\documentclass[%
 reprint,  
nofootinbib,
 amsmath,amssymb,
 aps,
 prd,
floatfix,
showkeys,
modulo,superscriptaddress
]{revtex4-2}

\usepackage[table,svgnames,dvipsnames]{xcolor}
\usepackage[normalem]{ulem}
\usepackage{aas_macros}
\usepackage{subfigure}
\usepackage{graphicx}
\usepackage{graphics}
\usepackage{float}
\usepackage{dcolumn}
\usepackage{bm}
\usepackage{cases}
\usepackage{booktabs}
\usepackage{comment}
\usepackage{multirow}
\usepackage{makecell}
\usepackage{siunitx}
\usepackage{tabularx}
\usepackage{xspace}
\usepackage{soul} 
\graphicspath{{figures/}}
\usepackage{fontawesome}
\usepackage{hyperref} 
\hypersetup{colorlinks=true,citecolor=blue,filecolor=blue,urlcolor=blue,linkcolor=blue}

\usepackage{orcidlink} 
\usepackage{hanging} 

\usepackage{adjustbox}
\usepackage{amsmath}
\usepackage[noabbrev]{cleveref}
\crefname{equation}{Eq.}{Eqs.}
\crefname{eqnarray}{Eq.}{Eqs.}
\crefname{section}{Section}{Sections}
\crefname{figure}{Figure}{Figures}
\crefname{table}{Table}{Tables}
\crefname{appendix}{Appendix}{Appendices}
\Crefname{figure}{Figure}{Figures}
\Crefname{equation}{Equation}{Equations}
\Crefname{section}{Section}{Sections}
\Crefname{table}{Table}{Tables}

\newcommand{\aperp}{\alpha_\perp}
\newcommand{\apar}{\alpha_\parallel}
\newcommand{\aiso}{\alpha_{\rm iso}}

\newcommand{\alap}{\alpha_{\rm AP}}

\newcommand{\hMpc}{h^{-1}\text{Mpc}}
\newcommand{\Mpc}{\text{Mpc}}

\newcommand{\ihMpc}{\,h{\rm Mpc}^{-1}}
\newcommand{\ihGpc}{\,h^{-1}{\rm Gpc}}

\newcommand{\rascalc}{{\sc RascalC}}

\newcommand{\abacussecond}{{\tt Abacus-2}}

\newcommand{\lrgxelg}{{\tt LRG$\times$ELG}}
\newcommand{\elgxqso}{{\tt ELG$\times$QSO}}
\newcommand{\lrgxqso}{{\tt LRG$\times$QSO}}
\newcommand{\lrgelg}{{\tt LRG$+$ELG}}

\newcommand{\elgo}{{\tt ELG1}}
\newcommand{\elgt}{{\tt ELG2}}
\newcommand{\elgs}{{\tt ELG}s}
\newcommand{\elg}{{\tt ELG}}
\newcommand{\lrgo}{{\tt LRG1}}
\newcommand{\lrgt}{{\tt LRG2}}
\newcommand{\lrgth}{{\tt LRG3}}
\newcommand{\lrg}{{\tt LRG}}
\newcommand{\lrgs}{{\tt LRG}s}
\newcommand{\bgs}{{\tt BGS}}
\newcommand{\qsoo}{{\tt QSO1}}
\newcommand{\qso}{{\tt QSO}}
\newcommand{\tleqo}{{\tt LEQ1}}
\newcommand{\tleqt}{{\tt LEQ2}}
\newcommand{\tleq}{{\tt LEQ}}
\newcommand{\eq}{{\tt EQ}}
\newcommand{\eqo}{{\tt EQ1}}
\newcommand{\eqt}{{\tt EQ2}}
\newcommand{\lpepq}{{\tt LRG+ELG+QSO}}
\newcommand{\pcs}{{\tt pcs2}}
\newcommand{\pow}{{\tt power3}}
\newcommand{\lcdm}{$\Lambda {\rm CDM}$}

\newcolumntype{Y}{>{\centering\arraybackslash}X}

\begin{document}

\title{A Unified Tracer Analysis of DESI DR2 Baryon Acoustic Oscillations}


\author{N.~Sanders\orcidlink{0009-0008-0020-2995}}
\affiliation{Department of Physics \& Astronomy, Ohio University, 139 University Terrace, Athens, OH 45701, USA}

\author{H.~Seo\orcidlink{0000-0002-6588-3508}}
\affiliation{Department of Physics \& Astronomy, Ohio University, 139 University Terrace, Athens, OH 45701, USA}

\author{M.~Rashkovetskyi\orcidlink{0000-0001-7144-2349}}
\affiliation{Center for Cosmology and AstroParticle Physics, The Ohio State University, 191 West Woodruff Avenue, Columbus, OH 43210, USA}
\affiliation{Department of Physics, The Ohio State University, 191 West Woodruff Avenue, Columbus, OH 43210, USA}
\affiliation{The Ohio State University, Columbus, 43210 OH, USA}

\author{U.~Andrade\orcidlink{0000-0002-4118-8236}}
\affiliation{Leinweber Center for Theoretical Physics, University of Michigan, 450 Church Street, Ann Arbor, Michigan 48109-1040, USA}
\affiliation{University of Michigan, 500 S. State Street, Ann Arbor, MI 48109, USA}

\author{D.~Valcin\orcidlink{0000-0003-0129-0620}}
\affiliation{University of California, Berkeley, 110 Sproul Hall \#5800 Berkeley, CA 94720, USA}

\author{E.~Paillas\orcidlink{0000-0002-4637-2868}}
\affiliation{Instituto de Estudios Astrof\'isicos, Facultad de Ingenier\'ia y Ciencias, Universidad Diego Portales, Av. Ej\'ercito Libertador 441, Santiago, Chile}
\affiliation{Steward Observatory, University of Arizona, 933 N. Cherry Avenue, Tucson, AZ 85721, USA}

\author{P.~McDonald\orcidlink{0000-0001-8346-8394}}
\affiliation{Lawrence Berkeley National Laboratory, 1 Cyclotron Road, Berkeley, CA 94720, USA}

\author{J.~Aguilar}
\affiliation{Lawrence Berkeley National Laboratory, 1 Cyclotron Road, Berkeley, CA 94720, USA}

\author{S.~Ahlen\orcidlink{0000-0001-6098-7247}}
\affiliation{Department of Physics, Boston University, 590 Commonwealth Avenue, Boston, MA 02215 USA}

\author{O.~Alves}
\affiliation{University of Michigan, 500 S. State Street, Ann Arbor, MI 48109, USA}

\author{E.~Armengaud\orcidlink{0000-0001-7600-5148}}
\affiliation{IRFU, CEA, Universit\'{e} Paris-Saclay, F-91191 Gif-sur-Yvette, France}

\author{A.~Aviles\orcidlink{0000-0001-5998-3986}}
\affiliation{Instituto Avanzado de Cosmolog\'{\i}a A.~C., San Marcos 11 - Atenas 202. Magdalena Contreras. Ciudad de M\'{e}xico C.~P.~10720, M\'{e}xico}
\affiliation{Instituto de Ciencias F\'{\i}sicas, Universidad Nacional Aut\'onoma de M\'exico, Av. Universidad s/n, Cuernavaca, Morelos, C.~P.~62210, M\'exico}

\author{F.~Beutler\orcidlink{0000-0003-0467-5438}}
\affiliation{Institute for Astronomy, University of Edinburgh, Royal Observatory, Blackford Hill, Edinburgh EH9 3HJ, UK}

\author{D.~Bianchi\orcidlink{0000-0001-9712-0006}}
\affiliation{Dipartimento di Fisica ``Aldo Pontremoli'', Universit\`a degli Studi di Milano, Via Celoria 16, I-20133 Milano, Italy}
\affiliation{INAF-Osservatorio Astronomico di Brera, Via Brera 28, 20122 Milano, Italy}

\author{D.~Brooks}
\affiliation{Department of Physics \& Astronomy, University College London, Gower Street, London, WC1E 6BT, UK}

\author{A.~Carnero Rosell\orcidlink{0000-0003-3044-5150}}
\affiliation{Departamento de Astrof\'{\i}sica, Universidad de La Laguna (ULL), E-38206, La Laguna, Tenerife, Spain}
\affiliation{Instituto de Astrof\'{\i}sica de Canarias, C/ V\'{\i}a L\'{a}ctea, s/n, E-38205 La Laguna, Tenerife, Spain}

\author{E.~Chaussidon\orcidlink{0000-0001-8996-4874}}
\affiliation{Lawrence Berkeley National Laboratory, 1 Cyclotron Road, Berkeley, CA 94720, USA}

\author{T.~Claybaugh}
\affiliation{Lawrence Berkeley National Laboratory, 1 Cyclotron Road, Berkeley, CA 94720, USA}

\author{S.~Cole\orcidlink{0000-0002-5954-7903}}
\affiliation{Institute for Computational Cosmology, Department of Physics, Durham University, South Road, Durham DH1 3LE, UK}

\author{A.~Cuceu\orcidlink{0000-0002-2169-0595}}
\affiliation{Lawrence Berkeley National Laboratory, 1 Cyclotron Road, Berkeley, CA 94720, USA}

\author{A.~de la Macorra\orcidlink{0000-0002-1769-1640}}
\affiliation{Instituto de F\'{\i}sica, Universidad Nacional Aut\'{o}noma de M\'{e}xico,  Circuito de la Investigaci\'{o}n Cient\'{\i}fica, Ciudad Universitaria, Cd. de M\'{e}xico  C.~P.~04510,  M\'{e}xico}

\author{Biprateep~Dey\orcidlink{0000-0002-5665-7912}}
\affiliation{Department of Astronomy \& Astrophysics, University of Toronto, Toronto, ON M5S 3H4, Canada}
\affiliation{Department of Physics \& Astronomy and Pittsburgh Particle Physics, Astrophysics, and Cosmology Center (PITT PACC), University of Pittsburgh, 3941 O'Hara Street, Pittsburgh, PA 15260, USA}

\author{Z.~Ding\orcidlink{0000-0002-3369-3718}}
\affiliation{University of Chinese Academy of Sciences, Nanjing 211135, People's Republic of China.}

\author{P.~Doel}
\affiliation{Department of Physics \& Astronomy, University College London, Gower Street, London, WC1E 6BT, UK}

\author{S.~Ferraro\orcidlink{0000-0003-4992-7854}}
\affiliation{Lawrence Berkeley National Laboratory, 1 Cyclotron Road, Berkeley, CA 94720, USA}
\affiliation{University of California, Berkeley, 110 Sproul Hall \#5800 Berkeley, CA 94720, USA}

\author{A.~Font-Ribera\orcidlink{0000-0002-3033-7312}}
\affiliation{Instituci\'{o} Catalana de Recerca i Estudis Avan\c{c}ats, Passeig de Llu\'{\i}s Companys, 23, 08010 Barcelona, Spain}
\affiliation{Institut de F\'{i}sica d’Altes Energies (IFAE), The Barcelona Institute of Science and Technology, Edifici Cn, Campus UAB, 08193, Bellaterra (Barcelona), Spain}

\author{D.~Forero-Sánchez\orcidlink{0000-0001-5957-332X}}
\affiliation{Institut de Ci\`encies del Cosmos (ICCUB), Universitat de Barcelona (UB), c. Mart\'i i Franqu\`es, 1, 08028 Barcelona, Spain.}

\author{J.~E.~Forero-Romero\orcidlink{0000-0002-2890-3725}}
\affiliation{Departamento de F\'isica, Universidad de los Andes, Cra. 1 No. 18A-10, Edificio Ip, CP 111711, Bogot\'a, Colombia}
\affiliation{Observatorio Astron\'omico, Universidad de los Andes, Cra. 1 No. 18A-10, Edificio H, CP 111711 Bogot\'a, Colombia}

\author{E.~Gaztañaga\orcidlink{0000-0001-9632-0815}}
\affiliation{Institut d'Estudis Espacials de Catalunya (IEEC), c/ Esteve Terradas 1, Edifici RDIT, Campus PMT-UPC, 08860 Castelldefels, Spain}
\affiliation{Institute of Cosmology and Gravitation, University of Portsmouth, Dennis Sciama Building, Portsmouth, PO1 3FX, UK}
\affiliation{Institute of Space Sciences, ICE-CSIC, Campus UAB, Carrer de Can Magrans s/n, 08913 Bellaterra, Barcelona, Spain}

\author{Satya~{Gontcho A Gontcho}\orcidlink{0000-0003-3142-233X}}
\affiliation{University of Virginia, Department of Astronomy, Charlottesville, VA 22904, USA}

\author{A.~X.~Gonzalez-Morales\orcidlink{0000-0003-4089-6924}}
\affiliation{Departamento de F\'{\i}sica, DCI-Campus Le\'{o}n, Universidad de Guanajuato, Loma del Bosque 103, Le\'{o}n, Guanajuato C.~P.~37150, M\'{e}xico}

\author{G.~Gutierrez}
\affiliation{Fermi National Accelerator Laboratory, PO Box 500, Batavia, IL 60510, USA}

\author{C.~Hahn\orcidlink{0000-0003-1197-0902}}
\affiliation{Department of Astronomy, University of Texas at Austin, 2515 Speedway, TX 78712, USA}

\author{H.~K.~Herrera-Alcantar\orcidlink{0000-0002-9136-9609}}
\affiliation{Institut d'Astrophysique de Paris. 98 bis boulevard Arago. 75014 Paris, France}
\affiliation{IRFU, CEA, Universit\'{e} Paris-Saclay, F-91191 Gif-sur-Yvette, France}

\author{K.~Honscheid\orcidlink{0000-0002-6550-2023}}
\affiliation{Center for Cosmology and AstroParticle Physics, The Ohio State University, 191 West Woodruff Avenue, Columbus, OH 43210, USA}
\affiliation{Department of Physics, The Ohio State University, 191 West Woodruff Avenue, Columbus, OH 43210, USA}
\affiliation{The Ohio State University, Columbus, 43210 OH, USA}

\author{S.~Juneau\orcidlink{0000-0002-0000-2394}}
\affiliation{NSF NOIRLab, 950 N. Cherry Ave., Tucson, AZ 85719, USA}

\author{T.~Karim\orcidlink{0000-0002-5652-8870}}
\affiliation{Department of Astronomy \& Astrophysics, University of Toronto, Toronto, ON M5S 3H4, Canada}

\author{D.~Kirkby\orcidlink{0000-0002-8828-5463}}
\affiliation{Department of Physics and Astronomy, University of California, Irvine, 92697, USA}

\author{A.~Kremin\orcidlink{0000-0001-6356-7424}}
\affiliation{Lawrence Berkeley National Laboratory, 1 Cyclotron Road, Berkeley, CA 94720, USA}

\author{O.~Lahav\orcidlink{0000-0002-1134-9035}}
\affiliation{Department of Physics \& Astronomy, University College London, Gower Street, London, WC1E 6BT, UK}

\author{C.~Lamman\orcidlink{0000-0002-6731-9329}}
\affiliation{The Ohio State University, Columbus, 43210 OH, USA}

\author{M.~Landriau\orcidlink{0000-0003-1838-8528}}
\affiliation{Lawrence Berkeley National Laboratory, 1 Cyclotron Road, Berkeley, CA 94720, USA}

\author{L.~Le~Guillou\orcidlink{0000-0001-7178-8868}}
\affiliation{Sorbonne Universit\'{e}, CNRS/IN2P3, Laboratoire de Physique Nucl\'{e}aire et de Hautes Energies (LPNHE), FR-75005 Paris, France}

\author{M.~Manera\orcidlink{0000-0003-4962-8934}}
\affiliation{Departament de F\'{i}sica, Serra H\'{u}nter, Universitat Aut\`{o}noma de Barcelona, 08193 Bellaterra (Barcelona), Spain}
\affiliation{Institut de F\'{i}sica d’Altes Energies (IFAE), The Barcelona Institute of Science and Technology, Edifici Cn, Campus UAB, 08193, Bellaterra (Barcelona), Spain}

\author{A.~Meisner\orcidlink{0000-0002-1125-7384}}
\affiliation{NSF NOIRLab, 950 N. Cherry Ave., Tucson, AZ 85719, USA}

\author{R.~Miquel}
\affiliation{Instituci\'{o} Catalana de Recerca i Estudis Avan\c{c}ats, Passeig de Llu\'{\i}s Companys, 23, 08010 Barcelona, Spain}
\affiliation{Institut de F\'{i}sica d’Altes Energies (IFAE), The Barcelona Institute of Science and Technology, Edifici Cn, Campus UAB, 08193, Bellaterra (Barcelona), Spain}

\author{J.~Moustakas\orcidlink{0000-0002-2733-4559}}
\affiliation{Department of Physics and Astronomy, Siena University, 515 Loudon Road, Loudonville, NY 12211, USA}

\author{A.~Muñoz-Gutiérrez}
\affiliation{Instituto de F\'{\i}sica, Universidad Nacional Aut\'{o}noma de M\'{e}xico,  Circuito de la Investigaci\'{o}n Cient\'{\i}fica, Ciudad Universitaria, Cd. de M\'{e}xico  C.~P.~04510,  M\'{e}xico}

\author{S.~Nadathur\orcidlink{0000-0001-9070-3102}}
\affiliation{Institute of Cosmology and Gravitation, University of Portsmouth, Dennis Sciama Building, Portsmouth, PO1 3FX, UK}

\author{J.~A.~Newman\orcidlink{0000-0001-8684-2222}}
\affiliation{Department of Physics \& Astronomy and Pittsburgh Particle Physics, Astrophysics, and Cosmology Center (PITT PACC), University of Pittsburgh, 3941 O'Hara Street, Pittsburgh, PA 15260, USA}

\author{H.~E.~Noriega\orcidlink{0000-0002-3397-3998}}
\affiliation{Instituto de Ciencias F\'{\i}sicas, Universidad Nacional Aut\'onoma de M\'exico, Av. Universidad s/n, Cuernavaca, Morelos, C.~P.~62210, M\'exico}
\affiliation{Instituto de F\'{\i}sica, Universidad Nacional Aut\'{o}noma de M\'{e}xico,  Circuito de la Investigaci\'{o}n Cient\'{\i}fica, Ciudad Universitaria, Cd. de M\'{e}xico  C.~P.~04510,  M\'{e}xico}

\author{N.~Palanque-Delabrouille\orcidlink{0000-0003-3188-784X}}
\affiliation{IRFU, CEA, Universit\'{e} Paris-Saclay, F-91191 Gif-sur-Yvette, France}
\affiliation{Lawrence Berkeley National Laboratory, 1 Cyclotron Road, Berkeley, CA 94720, USA}

\author{W.~J.~Percival\orcidlink{0000-0002-0644-5727}}
\affiliation{Department of Physics and Astronomy, University of Waterloo, 200 University Ave W, Waterloo, ON N2L 3G1, Canada}
\affiliation{Perimeter Institute for Theoretical Physics, 31 Caroline St. North, Waterloo, ON N2L 2Y5, Canada}
\affiliation{Waterloo Centre for Astrophysics, University of Waterloo, 200 University Ave W, Waterloo, ON N2L 3G1, Canada}

\author{F.~Prada\orcidlink{0000-0001-7145-8674}}
\affiliation{Instituto de Astrof\'{i}sica de Andaluc\'{i}a (CSIC), Glorieta de la Astronom\'{i}a, s/n, E-18008 Granada, Spain}

\author{I.~P\'erez-R\`afols\orcidlink{0000-0001-6979-0125}}
\affiliation{Departament de F\'isica, EEBE, Universitat Polit\`ecnica de Catalunya, c/Eduard Maristany 10, 08930 Barcelona, Spain}

\author{C.~Ravoux\orcidlink{0000-0002-3500-6635}}
\affiliation{Universit\'{e} Clermont-Auvergne, CNRS, LPCA, 63000 Clermont-Ferrand, France}

\author{A.~J.~Ross\orcidlink{0000-0002-7522-9083}}
\affiliation{Center for Cosmology and AstroParticle Physics, The Ohio State University, 191 West Woodruff Avenue, Columbus, OH 43210, USA}
\affiliation{Department of Astronomy, The Ohio State University, 4055 McPherson Laboratory, 140 W 18th Avenue, Columbus, OH 43210, USA}
\affiliation{The Ohio State University, Columbus, 43210 OH, USA}

\author{G.~Rossi}
\affiliation{Department of Physics and Astronomy, Sejong University, 209 Neungdong-ro, Gwangjin-gu, Seoul 05006, Republic of Korea}

\author{L.~Samushia\orcidlink{0000-0002-1609-5687}}
\affiliation{Abastumani Astrophysical Observatory, Tbilisi, GE-0179, Georgia}
\affiliation{Department of Physics, Kansas State University, 116 Cardwell Hall, Manhattan, KS 66506, USA}

\author{E.~Sanchez\orcidlink{0000-0002-9646-8198}}
\affiliation{CIEMAT, Avenida Complutense 40, E-28040 Madrid, Spain}

\author{C.~Saulder\orcidlink{0000-0002-0408-5633}}
\affiliation{Max Planck Institute for Extraterrestrial Physics, Gie\ss enbachstra\ss e 1, 85748 Garching, Germany}

\author{D.~Schlegel}
\affiliation{Lawrence Berkeley National Laboratory, 1 Cyclotron Road, Berkeley, CA 94720, USA}

\author{M.~Schubnell}
\affiliation{Department of Physics, University of Michigan, 450 Church Street, Ann Arbor, MI 48109, USA}
\affiliation{University of Michigan, 500 S. State Street, Ann Arbor, MI 48109, USA}

\author{J.~Silber\orcidlink{0000-0002-3461-0320}}
\affiliation{Lawrence Berkeley National Laboratory, 1 Cyclotron Road, Berkeley, CA 94720, USA}

\author{M.~Siudek\orcidlink{0000-0002-2949-2155}}
\affiliation{Institute of Space Sciences, ICE-CSIC, Campus UAB, Carrer de Can Magrans s/n, 08913 Bellaterra, Barcelona, Spain}
\affiliation{Instituto de Astrof\'{\i}sica de Canarias, C/ V\'{\i}a L\'{a}ctea, s/n, E-38205 La Laguna, Tenerife, Spain}

\author{G.~Tarl\'{e}\orcidlink{0000-0003-1704-0781}}
\affiliation{University of Michigan, 500 S. State Street, Ann Arbor, MI 48109, USA}

\author{M.~Vargas-Maga\~na\orcidlink{0000-0003-3841-1836}}
\affiliation{Instituto de F\'{\i}sica, Universidad Nacional Aut\'{o}noma de M\'{e}xico,  Circuito de la Investigaci\'{o}n Cient\'{\i}fica, Ciudad Universitaria, Cd. de M\'{e}xico  C.~P.~04510,  M\'{e}xico}

\author{B.~A.~Weaver}
\affiliation{NSF NOIRLab, 950 N. Cherry Ave., Tucson, AZ 85719, USA}

\begin{abstract}

We improve upon previous efforts to optimally combine overlapping galaxy samples in the Dark Energy Spectroscopic Instrument (DESI) baryon acoustic oscillation (BAO) analysis. By weighting each galaxy by its linear bias, overlapping galaxies are combined into a single, unified catalog, naturally avoiding double counting of cosmic volume and including all auto- and cross- information at the catalog level. Improvements over the previous effort include the addition of quasars out to $z=1.6$ to account for all overlapping DR2 tracers and redshift-dependent bias treatment to improve reconstruction. We report distance measurements using this unified tracer, and find them to be highly consistent with the baseline DR2 BAO analysis. We also test for tracer-dependent systematics within the DESI data, and find no evidence of tracer-dependent systematics within $0.8<z<1.6$. Finally, we take advantage of the unified tracer to rebin the analysis in redshift in order to more finely resolve the redshift-to-distance relation. Dynamical dark energy results on this finer redshift binning indicate that there is no missed feature in the expansion history in the redshifts $0.8<z<1.6$, and reproduces DESI's preference for an evolving dark energy equation of state.
\end{abstract}


\maketitle
\flushbottom
\tableofcontents

\section{Introduction}

Baryon Acoustic Oscillations (BAO) are the imprints of the primordial sound waves that propagated in the photon-baryon plasma of the pre-recombination universe. The competing effects of gravity and radiation pressure created pressure waves that propagated outward from the initial perturbation as spherical shells. The propagation halted near the epoch of recombination, when photons and baryons decoupled, leaving behind a characteristic spherical feature associated with the distance the waves traveled until that epoch, i.e., the sound horizon scale at decoupling.  After decoupling, baryons and dark matter behave similarly and grow together under gravity. This feature is observable in the distribution of total matter at late times, and therefore in the distribution of galaxies. In today's scale, the BAO feature that co-expanded with the Universe corresponds to a scale of $\sim 150\ \Mpc$. As the physics of the waves primarily depend on the physics of early cosmic time, which in turn depends on matter density, baryon density, and the density of the relativistic species, the sound horizon scale is precisely determined by the Cosmic Microwave Background, observed by, for example, the \textit{Planck} mission \citep{Planck-2018-overview}. With a known physical scale, the BAO feature observed in galaxy surveys serves as a standard ruler to measure the distances to various redshifts, revealing the precise nature of the accelerating expansion of the Universe.

The Dark Energy Spectroscopic Instrument (DESI) is a spectroscopic surveyor using robotically positioned fibers that operates at the Nicholas U. Mayall Telescope at Kitt Peak National Observatory on I'oligam Du'ag in the Tohono O'odham Nation \citep{DESI2022.KP1.Instr}. DESI simultaneously measures the spectra of nearly 5000 objects over a $\sim$3\textdegree\ field of view \citep{DESI2016b.Instr, Corrector.Miller.2023, FiberSystem.Poppett.2024}. Over the course of its planned eight-year survey, DESI is expected to observe 63 million spectroscopically-confirmed galaxies and quasars over about 17,000 ${\rm deg}^2$ of the sky \citep{SurveyOps.Schlafly.2023, Spectro.Pipeline.Guy.2023}. Over 18 million unique targets are publicly available as part of the first data release (DR1) \citep{DESI2024.I.DR1}. Early results from DR1 data have already placed exciting constraints on the nature of dark energy \citep{DESI2024.VII.KP7B}. The second data release (DR2) from DESI contains over 30 million galaxies and quasars, and is currently in a proprietary period within the DESI collaboration. Results from the analysis of BAO with DR2 data, in combination with data from the CMB and type Ia supernovae, have allowed for the tightest constraints on dynamical dark energy to date \citep{DESI.DR2.DR2}.

To cover a wide range of redshifts, the DESI analysis uses several galaxy types (tracers) across different redshift ranges, based on observational efficiencies. Different target types produce transition regions in redshift where one population gives way to another, as shown in \cref{fig:neff}. In particular, between $0.8 <z<1.1$, three target types, luminous red galaxies (\lrg), emission line galaxies (\elg), and quasars (\qso), occupy the same cosmic volume. One would want to maximize information extraction from all tracers while properly accounting for cross-covariances. \citep{KP4s5-Valcin} presented a method of constructing a combined tracer of \lrg\ and \elg\ for the DR1 analysis, the dominant two tracers in this redshift range, as the natural way to self-consistently combine their information. The same method was applied for the DESI DR2 data, and this combined catalog provided the most precise BAO measurements for DESI DR1 and DR2. This combined tracer was also used to test for tracer-dependent systematics, either observational or physical (e.g., the relative velocity effect), in the BAO measurements \citep{KP4s5-Valcin}.

The goal of this paper is to extend the approach of \cite{KP4s5-Valcin} and construct a single, unified catalog of all overlapping BAO tracers for the DR2 galaxy and quasar data. Specifically, we combine all three tracers over $0.8<z<1.1$:  \lrg, \elg, and \qso, and combine the two tracers over $1.1<z<1.6$: \elg\ and \qso. This is a self-consistent way to include all two-point cross-correlation information, as well as the cross-covariances of all auto- and cross- two-point statistics for the overlapping tracers, at the catalog level, allowing for a simple multi-tracer analysis. Additionally, by treating the overlapping tracers as independent, the baseline analysis double counts the cosmic volume where the overlap occurs. Unifying the overlapping tracers also naturally resolves this double counting.

A common technique in the analysis of the BAO as a standard ruler is density field reconstruction, which seeks to partially undo the smearing and displacement of the BAO feature due to non-linear structure growth, peculiar velocities, and galaxy bias \citep{2007ApJ...664..675E}. Reconstruction essentially solves the linearized continuity equation in reverse to obtain the field describing the displacements of galaxies from their initial linear positions. In our analysis, each tracer is quasi-optimally reweighted to form a single biased tracer of matter, albeit with a redshift-dependent bias.  Beyond the method in \citep{KP4s5-Valcin}, we account for the redshift dependence of the effective bias and the growth rate during reconstruction. The reconstruction process benefits from increased tracer densities \citep{KP4s3-Chen}, which is the additional advantage of combining overlapping tracers at the catalog level. Note that this is achieved with no significant increase in the complexity of the analysis pipeline, and only a marginal increase in the computation cost.

Once we ensure the stability of reconstruction across tracer boundaries within the unified catalog, a broader choice of redshift binning for the data analysis is allowed, no longer restricted by tracer transitions as it was previously, as well as a more flexible choice of redshift padding during density field reconstruction to reduce boundary effects. We present an example of this rebinning to check that the BAO measurements and the inferred cosmology results are robust to binning effects.

Additionally, we take the opportunity to test for tracer-dependent systematics in the DR2 data. In principle, since they trace the same underlying matter distribution, overlapping tracers should obtain consistent constraints on the BAO scale. A discrepancy could be indicative either of uncorrected systematics or a detection of new physics (eg. the relative velocity effect \citep{Dalal.RelV_2010, Beutler.RelV_2017}). Regardless, a discrepancy between overlapping tracers would be problematic for the unification procedure described, and thus must be checked as a validation test. We test the consistency in the BAO measurements, not just across the auto-correlations of the single and the combined tracers, but also across the cross-correlations between all pairs of the tracers available within the overlapping redshift ranges.

This paper is structured as follows. In \cref{sec:dr2}, we summarize the DESI Data Release 2, and in \cref{sec:mocks}, we summarize the mock data we used to validate our method.  In \cref{sec:methods}, we present the method of weighting and combining tracers, the two-point correlation function, redshift-rebinning, and the reconstruction details, as well as the covariance and the BAO fitting procedure. In \cref{sec:results}, we present the results: validation tests of the unified tracer method using the 25 DR2 mocks, followed by its application to the DR2 data. This section presents the BAO measurements of the unified tracer and those using the alternative redshift binning, followed by the resulting effects in the cosmological inference.  In \cref{sec:conclusion}, we present our conclusions.
\section{DESI Data Release 2} \label{sec:dr2}

The DESI Second Data Release (DR2) \citep{DESI.DR2.DR2} is the product of nearly three years of observations by DESI from 14 May 2021 to 9 April 2024. The instrument, mounted on the Nicholas U. Mayall Telescope at Kitt Peak National Observatory, simultaneously observes 5,000 spectra within the seven square degree field of view of the prime focus corrector \citep{Corrector.Miller.2023}. Robotic positioners \citep{FocalPlane.Silber.2023} align optical fibers \citep{FiberSystem.Poppett.2024} to the celestial coordinates of targets, which then record the spectra using ten bench-mounted spectrographs located in a climate-controlled enclosure. The data is treated by both spectroscopic reduction \citep{Spectro.Pipeline.Guy.2023} and redshift estimation \citep{Redrock.Bailey.2024} pipelines.

The DESI survey is divided into two observing programs: `bright time' and `dark time' based on observing conditions, with both programs having distinct target classes. The extragalactic sample from bright time is the bright galaxy survey (\bgs) \citep{BGS.TS.Hahn.2023}. The dark time extragalactic sample is comprised of luminous red galaxies (\lrg) \citep{LRG.TS.Zhou.2023}, emission line galaxies (\elg) \citep{ELG.TS.Raichoor.2023}, and quasars (\qso) \citep{QSO.TS.Chaussidon.2023}, which significantly overlap in redshift. In the primary DR2 BAO analysis \citep{DESI.DR2.DR2}, BAO is measured with the autocorrelation of members of each target class at $z<2.1$, with \lrg\ and \elg\ combined into one tracer in $0.8<z<1.1$ \citep{KP4s5-Valcin}. For $z>2.1$, BAO is measured with the autocorrelation of the Ly$\alpha$ forest absorption in the spectra of quasars, as well as the forest's cross-correlation with quasar positions \citep{DESI.DR2.BAO.lya}.

Throughout this work, we utilize the large-scale structure (LSS) catalogs denoted `Loa'. The LSS catalog pipeline is detailed in \citep{LSSCatalogs.Ross.2025, DESI2024.II.KP3}, and the choices unique to DR2 described in the DR2 BAO paper \citep{DESI.DR2.DR2}. We use the `v1.1' version of the catalogs, which are the same as those used in the baseline analysis.

In this work, we expand the DR2 BAO analysis in \cite{DESI.DR2.DR2} to naturally include all cross-correlations of the overlapping dark time galaxy tracers. The overlapping \lrg, \elg, and \qso\ are combined into a single catalog, as described in \cref{subsec:creation}. We then bin this unified tracer into two bins that we call \tleq\ and \eq. Additionally, the non-overlapping \qso\ at $z>1.6$ are placed in their own bin, which we label \qsoo. Note that DESI DR1 and DR2 approximated the \qso\ as an independent sample, although the sample shares the same cosmic volume as \lrg\ and \elg\ over $0.8<z<1.6$. This was considered a reasonable assumption given the dominance of shot noise in this sample. In our work, we explicitly avoid this double counting of this same cosmic volume, although we later show the covariance between \qso\ $0.8<z<1.6$ and other samples is indeed small. \cref{tab:tracers-info} shows information about both the baseline tracers, as well as the unified tracers. \cref{tab:tracers-info} shows that the double counting of \qso\ over $0.8<z<1.6$ in the DR2 baseline analysis corresponds to $V_{\rm eff} \sim 1.5\ \rm{Gpc}^3$, about half of the total $V_{\rm eff}$ of the entire \qso\ sample.

\begin{table}
    \centering
    \begin{tabular}{lcccc}
        \hline\hline
        Tracer & Redshfit Range & $z_{\rm eff}$ & $V_{\rm eff}\ (\rm{Gpc}^3)$ & $b$ \\
        \hline
        \bgs & $0.1<z<0.4$ & 0.295 & 3.8 & 1.5 \\
        \lrgo & $0.4<z<0.6$ & 0.510 & 4.9 & 2.0 \\
        \lrgt & $0.6<z<0.8$ & 0.706 & 7.6 & 2.0 \\
        \lrgelg & $0.8<z<1.1$ & 0.934 & 14.8 & 1.6 \\
        \hline
        \lrgth & $0.8<z<1.1$ & 0.922 & 9.8 & 2.0 \\
        \elgo & $0.8<z<1.1$ & 0.955 & 5.8 & 1.2 \\
        \elgt & $1.1<z<1.6$ & 1.321 & 8.3 & 1.2 \\
        \qso & $0.8<z<2.1$ & 1.484 & 2.7 & 2.1 \\
        \hline\hline
        \tleq & $0.8<z<1.1$ & 0.950 & 15.2 & --- \\
        \eq & $1.1<z<1.6$ & 1.329 & 10.1 & --- \\
        \qsoo & $1.6<z<2.1$ & 1.828 & 1.2 & --- \\
        \hline\hline
    \end{tabular}
    \caption{The redshift ranges of the DESI DR2 data tested in this paper. The upper 2 sections are tracers used in the default DR2 analysis \citep{DESI.DR2.DR2}. Among them, \lrgth, \elgo, \elgt, and part of \qso\ (middle section) are overlapping and are to be combined. \bgs, \lrgo, and \lrgt\ do not overlap, and are unchanged in this analysis, but still used for cosmology inference. The lower section shows the unified tracers constructed in this work, resulting from combining catalogs, as well as the high redshift \qso\ that do not overlap.
    \label{tab:tracers-info}}
\end{table}

\section{Mock Catalogs} \label{sec:mocks}

Mock catalogs are synthetic datasets created for validation purposes. They are designed to replicate observations, including clustering statistics, survey geometry, and observational effects. Mocks allow for the testing of our analysis pipeline by taking advantage of known cosmology and realistic noise.

We use the same set of mocks as the baseline DR2 BAO analysis in \citep{DESI.DR2.DR2}. These mocks are the \abacussecond{} DR2 mocks, built off the {\tt AbacusSummit} {\tt Planck} 2018-$\Lambda$CDM base simulations \citep{AbacusSummit}. {\tt AbacusSummit} halo catalogs are populated with galaxies following a halo occupation distribution (HOD) framework as described in \citep{EDR_HOD_ELG2023, EDR_HOD_LRGQSO2023} for the dark-time tracers considered in this work. Additionally, the mocks have had fiber assignment applied in a manner that mimics the real data by the `altmtl' procedure described in \citep{KP3s7-Lasker}. The mocks used for DR2 dark-time tracers are identical to DR1, save for updates to the footprint mask to reflect the DR2 geometry.

The 25 mock catalogs are constructed from 25 independently realized simulation boxes with side lengths of 2 $\ihGpc$ \citep{Y3.clust-s1.Andrade.2025}. The box must be replicated to fit the footprint of dark-time tracers, so the boxes are simply tiled $3\times 3\times 3$ to produce boxes of side lengths 6 $\ihGpc$. This box replication introduces artificial correlation into clustering measurements. However, mock results are still highly useful for testing biases in the measurement of BAO parameters and whether data results fall within the distribution of mocks.

\section{Methods} \label{sec:methods}

\subsection{Creation of Unified Catalog} \label{subsec:creation}

We construct a catalog containing all DESI galaxy tracers to naturally include all auto and cross information.
Following \cite{KP4s5-Valcin}, this is done by weighting each individual tracer by its constant bias, which is a simple and close proxy for the optimal weight.\footnote{Implementation of a truly optimal weight is non-trivial. We find in \cref{sec:alt-weights} that there is little expected gain in implementing an optimal weight.}
Then, these weighted catalogs are concatenated together to form a unified catalog.

\begin{figure}
    \centering
    \includegraphics[width=\linewidth]{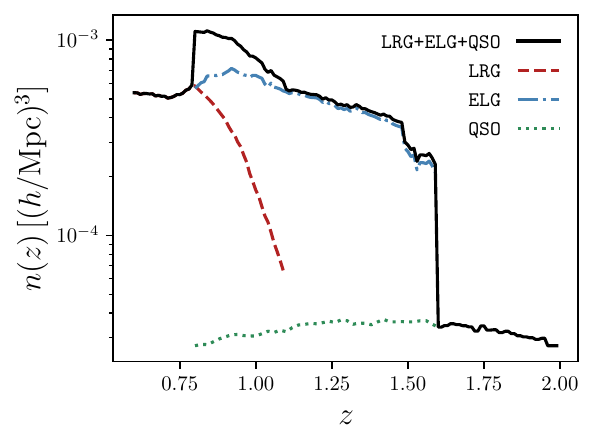}
    \caption{Effective tracer densities. \lpepq\ is the unified tracer of \lrg, \elg, and \qso. The combination of tracers into a unified catalog results in a higher-density sample, reducing shot noise and improving precision.}
    \label{fig:neff}
\end{figure}

The unified catalog has the obvious advantage of an increased sample density.
After accounting for the reweighting, the effective density of the unified tracer is
\begin{equation}
    \bar{n}_{\rm eff}(z) = \frac{\left[ \sum_i b_i \bar{n}_i(z) \right]^2}{\sum_i b_i^2 \bar{n}_i(z)} \equiv \frac{\sum_i b_i \bar{n}_i(z)}{b_{\rm eff}(z)},
    \label{eqn:neff}
\end{equation}
\noindent
where $b_i$ and $\bar{n}_i(z)$ are the constant bias and comoving number density of tracer $i$, respectively.
Here, $b_{\rm eff}(z)$ is the effective bias of the combined tracer, which normalizes the bias-weighted number density:
\begin{equation}
    b_{\rm eff} (z) = \frac{\sum_i b_i^2 \bar{n}_i(z)}{\sum_i b_i \bar{n}_i(z)}.
    \label{eqn:beff}
\end{equation}
\noindent
The effective density is shown in \cref{fig:neff}, along with the densities of the constituent single tracers. 
Only in the redshift range $0.8<z<1.6$ is there overlap between DESI tracers, so this procedure only applies there.
Thus, the analysis presented here will focus on just this range.
The density increase is most pronounced in $0.8<z<1.1$, because \lrg\ and \elg\ have similar densities.
The increase is much more modest in $1.1<z<1.6$ because \qso\ are significantly sparser than \elg.

The Feldman-Kaiser-Peacock (FKP) weight, defined to maximize the signal-to-noise \citep{FKP1994}, also must be updated to reflect $\bar{n}_{\rm eff}(z)$. The FKP weight is usually defined as
\begin{equation}
    W_{\rm FKP}(z)\;=\; \frac{1}{1 + \bar{n}_x(z)\, P_0},
\end{equation}
\noindent
where $\bar{n}_x(z)$ is the completeness uncorrected, ie. the raw, comoving number density and $P_0$ is an
effective constant approximation to the \emph{redshift-space monopole} amplitude of the
galaxy power spectrum at the BAO reference scale $k_\star = 0.14\,h\,\mathrm{Mpc}^{-1}$,
as it enters the variance of the FKP estimator.

Since we construct a catalog from multiple constituent tracers, which have different
biases and span a wide redshift range, a single constant value of $P_0$ is not
appropriate. Instead, for the purpose of defining the FKP weights only, we adopt a
redshift-dependent effective amplitude $P_0(z)$ defined by evaluating the expected
monopole at the fixed reference scale,
\begin{equation}
    P_0(z) \;\equiv\; P_{\ell=0}(k_\star,z)\Big|_{k_\star = 0.14\,h\,\mathrm{Mpc}^{-1}}.
\end{equation}
\noindent
In linear theory, this can be written as
\begin{equation}
    P_0(z) \;=\; \left[b_{\rm eff}^2(z) + \frac{2}{3}f(z)\,b_{\rm eff}(z)
    + \frac{1}{5}f^2(z)\right]\, P_m(k_\star,z),
\end{equation}
\noindent
where $P_m(k_\star,z)$ is the linear matter power spectrum evaluated at the BAO scale in the
fiducial cosmology and $f(z)$ is the linear growth rate. We stress that this prescription is used solely to set the relative FKP weighting as a function of redshift, and does not represent a modeling assumption for the measured multipoles.
The final FKP weights are then computed as
\begin{equation} \label{eqn:fkp}
    W_{\rm FKP}(z) \;=\; \frac{1}{1 + \bar{n}_{\rm eff}(z)\,P_0(z)}.
\end{equation}
Note that this is different from \cite{KP4s5-Valcin} which used a constant $P_0$ for each tracer for simplicity.

In summary, the galaxies are reweighted by their bias and an updated FKP weight reflecting the unified catalog's effective density:
\begin{equation}
    W_g = b \times W_{\rm fid} \times W_{\rm FKP}
\end{equation}
\noindent
where $b$ is the bias of the original catalog in \cref{tab:tracers-info} and $W_{\rm fid}$ is the WEIGHT column of the original catalog.

The random catalogs must also be consistently reweighted such that the (weighted) data-to-random ratio is the same between different tracers after the changes to the weights and FKP weights. Each random catalog is thus also reweighted by its data-to-random ratio,

\begin{equation}
    \begin{aligned}
    W_r = b \times W_{{\rm fid},r} \times W_{{\rm FKP},r} \times \frac{N_{\rm data}}{N_{\rm random}} \\
    N = \sum W_{\rm fid} \times W_{\rm FKP}.
    \end{aligned}
\end{equation}

While we assumed constant bias when combining tracers, bias, in fact, evolves for each tracer within its redshift range.
Across the redshift boundaries of different single tracers, effective bias jumps, while the underlying dark matter is continuous.
These boundaries exist because of cuts applied to the catalogs that cut out the redshifts where a given tracer has low number density.
A potential application of this method is to avoid such data cuts through unification of tracers.
However, we do not do this in this work in order to directly compare to the baeline DR2 results using the same data.

\begin{figure}
    \centering
    \includegraphics[width=\linewidth]{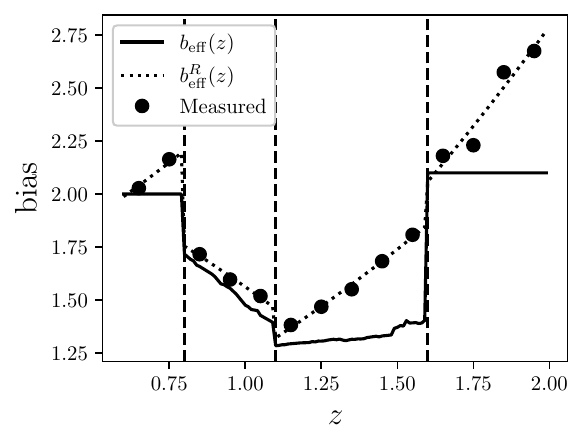}
    \caption{Effective bias of the unified catalog as a function of redshift. Black points show the measured effective bias in redshift bins of width $\Delta z = 0.1$, obtained by comparing the measured galaxy power spectrum monopole to the linear matter power spectrum from the fiducial cosmology. The solid line shows the effective bias computed directly from the unified catalog according to \cref{eqn:beff}, which exhibits discontinuities at the boundaries between different tracer populations. The discrepancy between the measured bias and \cref{eqn:beff} is caused by assuming a redshift-independent bias for the individual tracers. The dotted line, $b_{\rm eff}^{R}(z)$, shows a piecewise quadratic fit to the measured bias, which is used for density-field reconstruction to enforce a smoothly varying matter overdensity $\delta_m$ across tracer transitions and avoid spurious displacement fields.}
    \label{fig:beff}
\end{figure}

Accounting for the discontinuity in the bias (\cref{fig:beff}) is important for reconstruction because we want to ensure that the estimated underlying matter density $\delta_m$, which is used for calculating the displacement field during reconstruction, evolves smoothly across the redshift boundaries, as is physically expected.
For this purpose, the effective bias of the unified catalog was measured by measuring the power spectrum in redshift bins of width 0.1, and comparing it to the linear matter power spectrum computed from fiducial cosmology.
We then piecewise fit a quadratic to the measured bias.
This best fit, $b_{\rm eff}^R(z)$, is shown as the dotted line in \cref{fig:beff}.
The best fit will be used during reconstruction to ensure continuous $\delta_m$ over the tracer boundaries.

In \cref{fig:beff}, we note an apparent discrepancy between the effective bias $b_{\rm eff}(z)$ from \cref{eqn:beff} (solid line, i.e., the one used to construct the combined tracer through  \cref{eqn:neff} and \cref{eqn:fkp}), and the measured/resulting bias $b_{\rm eff}^{R}(z)$ (points, the one used in density field reconstruction).
The discrepancy arises because we assumed a redshift-independent bias to weight each individual tracer when calculating \cref{eqn:beff}, whereas in reality the bias of a given tracer evolves with redshift. 

Ideally, one could use an accurate redshift-dependent bias model to weight individual tracers in \cref{eqn:beff} so that it better matches the resulting $b_{\rm eff}^{R}(z)$.
We performed a case study with this weighting on the $1.1<z<1.6$ bin and found no significant difference in BAO scale measurements.
So, the simple case of constant bias was adopted for simplicity.

\subsection{Two-Point Correlation Function}

Following the standard procedures used in DESI \citep{DESI2024.II.KP3}, the two-point correlation function is computed using the Landy-Szalay estimator \citep{Landy1993}
\begin{equation}
    \xi(s, \mu) = \frac{DD(s, \mu) - 2DR(s, \mu) + RR(s, \mu)}{RR(s, \mu)}
    \label{eqn:LS-estimator}
\end{equation}
\noindent
where $s$ is the pair separation (measured in bins of $4\,h^{-1}\mathrm{Mpc}$) and $\mu \in [-1,1]$ is the cosine of the angle between the separation vector of the pair and the line of sight. Here $DD(s,\mu)$, $DR(s,\mu)$, and $RR(s,\mu)$ denote the appropriately normalized counts of data--data, data--random, and random--random pairs, respectively, after data and randoms are weighted as described in \cref{subsec:creation}. Various observational weights are also applied, accounting for imaging systematics, redshift failure, and survey completeness. The random catalog traces the survey selection function and geometry, and is constructed to be significantly denser than the data catalog to minimize shot noise. We use 5 random catalogs in calculating all correlation functions for unified tracers.
This is calculated using the publicly available code \textsc{pycorr}\footnote{\url{https://github.com/cosmodesi/pycorr}} \citep{pycorr}.
We then convert the output into Legendre multipoles, namely the monopole ($l=0$) and the quadrupole ($l=2$).

\subsection{Redshift Rebinning} \label{subsec:rebin}
The default DESI redshift binning used in DR1 and DR2 analyses was naturally chosen based on the tracer transitions, at the discontinuities shown in \cref{fig:beff}. Since the BAO feature is robust against galaxy bias/populations, ideally the redshift binning should not be limited by the population transition. With reconstruction, care needs to be given to ensure the stability of reconstruction across tracer boundaries within the unified catalog, such as by deriving the effective bias $b_{\rm eff}^R(z)$ in \cref{fig:beff}. 
This was not a concern for previous analyses because tracers were treated individually, so no bins cross a transition, and reconstruction was done in the same redshift range as pair counting and BAO fitting.

In \cref{subsec:recon-boundary}, in addition to the default redshift binning of DESI DR1 \citep{DESI2024.III.KP4} and DR2 \citep{DESI.DR2.DR2} as presented in \cref{tab:tracers-info}, we will also bin the unified tracer from redshifts $0.8<z<1.6$ into four equally spaced (in redshift, not in volume) bins, as summarized in \cref{tab:new-bin-summary}, an option that provides more redshift resolution for the cosmology inference.
We will demonstrate that this rebinning grants consistent results as compared to the baseline analysis and will also quantify any potential gain from better resolving the time dependence of the expansion history. 

\begin{table*}
    \centering
    \begin{tabular}{l|c|c|c|c|c}
        \hline
        \hline
        Tracer & Redshift Range & Reconstruction & $z_{\rm eff}$ & $V_{\rm eff}\ (\rm{Gpc}^3)$ & $P_0 (k=0.14)$ \\
         & & $z$ Range & & & \\
        \hline
        \tleqo & $0.8<z<1.0$ & $0.7<z<1.3$ & 0.903 & 10.2 & 5500 \\
        \tleqt & $1.0<z<1.2$ & $0.7<z<1.3$ & 1.095 & 6.8 & 3800 \\
        \eqo & $1.2<z<1.4$ & $1.1<z<1.7$ & 1.299 & 5.1 & 3300 \\
        \eqt & $1.4<z<1.6$ & $1.1<z<1.7$ & 1.490 & 4.2 & 3600 \\
        \qsoo & $1.6<z<2.1$ & $1.6<z<2.1$ & 1.828 & 1.2 & 5300 \\
        \hline
        \hline
    \end{tabular}
    \caption{New binning of unified catalog. This provides a finer redshift resolution to the cosmology inference, better resolving the expansion history.}
    \label{tab:new-bin-summary}
\end{table*}

\subsection{Reconstruction}

Nonlinear gravitational evolution in the late Universe both broadens the BAO feature and can introduce a subpercent-level shift in its apparent position/size towards a smaller scale, thereby degrading statistical precision and potentially biasing BAO distance measurements. Density-field reconstruction \citep{Eisenstein2007:astro-ph/0604362v1} is a widely used technique that mitigates the impact of large-scale nonlinear evolution by partially reversing bulk displacements and associated mode coupling, thereby sharpening the BAO feature and reducing systematic shifts in the
BAO scale.

Extensive analysis was performed in \cite{KP4s3-Chen,KP4s4-Paillas} exploring the application of different algorithms to DESI data, from which the iterative fast Fourier transform (iFFT) algorithm was chosen as the baseline method for DESI tracers.
We follow the standard choices for reconstruction parameters as the baseline DR2 analysis \citep{KP4s4-Paillas, Y3.clust-s1.Andrade.2025}. We use the \textbf{RecSym} convention, which does not remove redshift-space distortions when shifting the random catalogs, motivated by \cite{KP4s2-Chen}. We use a smoothing scale of $15 \hMpc$, with the exception of \qsoo\ (as well as the baseline \qso), where a smoothing scale of $30 \hMpc$ is used.

In contrast to the previous analyses in DR1 \citep{KP4s5-Valcin} and DR2 \cite{DESI.DR2.DR2}, we make the following changes to the reconstruction.

\begin{enumerate}
    \item Rather than using an average, constant $b_{\rm eff}$, we use $b_{\rm eff}^R(z)$ as measured according to \cref{subsec:creation}. This guarantees a smooth implied matter density over redshift boundaries. 
    \item Additionally, the growth rate is calculated from fiducial cosmology as a function of redshift, rather than being a constant calculated at $z_{\rm eff}$. 
    \item For the rebinned unified tracer, reconstruction is performed over a wider range of redshifts than the two-point correlation is measured, as specified in \cref{tab:new-bin-summary}. This `padding' is meant to avoid anomalous displacements near the boundaries in redshift. It was found in \cite{Y3.clust-s1.Andrade.2025} that padding has little effect on clustering measurements, though we apply it nonetheless out of caution, particularly given the smaller size of the new bins along the line of sight, which may increase the boundary effect on long wavelength modes.
\end{enumerate}
\noindent
These changes account for unique aspects of the unified tracer, such as the effective bias behaving in a qualitatively different way than single tracers (e.g., bias decreases with increasing redshift in $0.8<z<1.1$).

\subsection{Covariance Matrices}

We use the \rascalc{} code\footnote{\url{https://github.com/oliverphilcox/RascalC}} \citep{rascalC,2023MNRAS.524.3894R,KP4s7-Rashkovetskyi} to produce semi-analytical covariance matrices for the multipoles of the two-point correlation function, similarly to \cite{KP4s5-Valcin}.
Its covariance matrix model integrates the contributions of the non-Gaussian two-point correlation function measured from the data into three different terms, neglecting the three- and connected four-point functions and instead approximating their non-Gaussian effects with shot-noise rescaling.
A comprehensive review of these covariance matrices is available in \cite{KP4s7-Rashkovetskyi}.

We produce the covariance matrices for the combined tracers in the same manner as for single tracers, using the combined data and random catalogs and tuning the shot noise to the data correlation function jackknife covariance.
The covariance tuned on the data jackknife covariance is applied both to the data and mocks.
Validation of this approach is presented in \cref{sec:cov-as-multi}.
Additionally, in \cref{subsubsec:data-tracer-sys},  we measure the BAO scale from the cross-clustering between two tracers to test tracer-dependent systematics. We construct the auto-covariance matrix for such cross-clustering statistics, following Appendix A of \cite{KP4s7-Rashkovetskyi}, which tunes the shot noise for each tracer individually using the jackknife covariance of its autocorrelation function.

\subsection{BAO Fitting Model} \label{subsec:bao-model}

The procedure for fitting BAO for the DESI collaboration is discussed extensively in \cite{DESI2024.III.KP4,KP4s2-Chen}. Here we briefly summarize the procedure. Following the conventions of DESI DR1 and DR2, our baseline clustering statistic is the two-point correlation function in configuration space calculated after the observational coordinates are mapped to fiducial physical coordinates using the fiducial distance-redshift relation $D_M^{\rm fid}(z)$. The theory model is constructed in Fourier space, projected onto a Legendre basis, then converted to configuration space through a Hankel transform. 

The template-based theory model in Fourier space is
\begin{equation} \label{eqn:pk-model}
    P(k,\mu) = \mathcal{B}(k,\mu)P_{\rm nw}(k) + \mathcal{C}(k,\mu)P_{\rm w}(k) + \mathcal{D}(k),
\end{equation}
\noindent
where $P_{\rm nw}(k)$ and $P_{\rm w}(k)$ denote the smooth and BAO components of the linear matter power spectrum, as predicted from \textsc{class}\footnote{\url{https://github.com/lesgourg/class_public}} \citep{class-approximation-schemes} using our fiducial cosmology. The $\mathcal{C}(k,\mu)P_{\rm w}(k)$ term contains the BAO information, while $\mathcal{B}(k,\mu)P_{\rm nw}(k)$ models the smooth, broadband component using quasi-linear theory, and $\mathcal{D}(k)$ parametrically models additional nonlinearities and observational effects on the broadband.

What is measured in a BAO analysis is the observed BAO scale along and across the line of sight against the scale assumed in the template, expressed as the BAO dilation parameters $\apar$ and $\aperp$, respectively. For example, the BAO scale in the template is $r_d^{\rm fid}$, and the observed BAO scale across the line of sight in the fiducial comoving coordinates is $[r_d/D_M(z)] D_M^{\rm fid}(z)$, with $\aperp$ as the ratio of the former to the latter. Then, $\aperp$ or $\apar$ greater than unity means the observed BAO scale is smaller than the scale in the template, vice versa for less than unity.
Therefore the BAO parameters connect to cosmology through the Hubble parameter $H(z)$ and the comoving angular diameter distance $D_M(z)$:
\begin{equation}
    \apar = \frac{H^{\rm fid}(z) r_d^{\rm fid}}{H(z) r_d}, \quad \aperp = \frac{D_M(z) r_d^{\rm fid}}{D_M^{\rm fid}(z) r_d},
\end{equation}
\noindent
where the `fid' superscript denotes the quantities in the fiducial cosmology, and $r_d$ is the sound horizon scale at the end of the baryon drag epoch, i.e., the comoving size of the BAO feature.
In Fourier space, we fit for the $\alpha$'s that match the BAO feature in the template ($k^{\prime}$ and $\mu^{\prime}$) to the observed location ($k_{\rm obs}$ and $\mu_{\rm obs}$) through the BAO term $\mathcal{C}(k,\mu)P_{\rm w}(k)$:

\begin{equation}
    \begin{aligned}
        k^{\prime} = \frac{\alap^{1/3}}{\aiso}\left[1 + \mu_{\rm obs}^2 \left(\frac{1}{\alap^2} - 1\right) \right]^{1/2} k_{\rm obs} \\
        \mu^{\prime} = \frac{\mu_{\rm obs}}{\alap} \left[1 + \mu_{\rm obs}^2 \left(\frac{1}{\alap^2} - 1 \right) \right]^{-1/2}.
    \end{aligned}
\end{equation}
\noindent
Here $\aiso$ and $\alap$ are an alternative basis, a rotation of  $\aperp$ and $\apar$: $\aiso=(\apar \aperp^2)^{1/3}$ represents the ratio of the spherically averaged radius of the BAO feature between the template and what is observed, and $\alap=\apar/\aperp$ measures the anisotropic distortion introduced by assuming the fiducial grid cosmology $D_M^{\rm fid}(z)$. In this paper, we will mainly present the BAO measurements in terms of $\aiso$ and $\alap$ except for the final measurements. 
When fitting, flat priors are placed on $\aiso$ and $\alap$, with them being the primary parameters of interest in the model for cosmology inference.

We adopt the following form for $\mathcal{B}(k,\mu)$ to introduce anisotropic amplitude modulation due to redshift-space distortions to the quasi-linear theory smooth power spectrum component:

\begin{equation}
    \mathcal{B}(k,\mu) = \left(b_1 + f\mu^2 \right)^2F_{\rm fog}
\end{equation}
\noindent
where $F_{\rm fog} = \left(1 + \frac{1}{2}k^2\mu^2\Sigma_s^2 \right)$ accounts for the `Fingers of God' effect with a free smoothing scale parameter $\Sigma_s$. 

For the BAO component, $\mathcal{C}(k,\mu)$ is parameterized to account for the anisotropy in the amplitude as well as in the BAO damping due to redshift-space distortions:

\begin{equation}
    \mathcal{C}(k,\mu) = \left(b_1 + f\mu^2 \right)^2 \exp\left[-\frac{1}{2}k^2 \left(\mu^2\Sigma_{\parallel}^2 + (1-\mu^2)\Sigma_{\perp}^2 \right) \right]
\end{equation}
\noindent
where $\Sigma_{\parallel}$ and $\Sigma_{\perp}$ are free parameters modeling nonlinear damping of BAO along and perpendicular to the line of sight. The damping parameters are given Gaussian priors with values shown in \cref{tab:bao-priors}. Note that the process of reconstruction reduces the BAO damping parameters, as reconstruction is designed to partially undo the smearing of the BAO due to nonlinear structure growth.

Lastly, the $\mathcal{D}(k)$ term models deviations from linear theory in the broadband shape of the power spectrum multipoles. DESI has been parameterizing it using a basis of cubic splines separated by a defined scale $\Delta$,

\begin{equation} \label{eqn:spline-D}
    \mathcal{D}_{\ell}(k) = \sum_{n=-1}^{n_{\rm max}} a_{\ell,n}W_3\left(\frac{k}{\Delta} - n \right)
\end{equation}
\noindent
where $W_3$ is a piecewise cubic spline kernel \citep{CHANIOTIS2004253}. $\Delta$ is chosen such that $\mathcal{D}_{\ell}(k)$ cannot sample the BAO wiggles. By the Nyquist-Shannon sampling theorem, this sets a limit on $\Delta$ to be larger than half the BAO wavelength. Therefore $\Delta = 2\pi/r_{\rm d} \simeq 0.06\ihMpc$ is chosen. It was found in \citep{KP4s2-Chen} that within our range of fitting scales in configuration space, most of these functions quickly approach zero. Therefore, we exclude these terms, except for the $n=0,1$ terms of the quadrupole. Two additional nuisance parameters are introduced for each multipole to control large-scale systematics:

\begin{equation}
    \tilde{\mathcal{D}}_{\ell}(s) = b_{\ell,0} + b_{\ell,2} \left( \frac{sk_{\rm min}}{2\pi} \right)^2,
\end{equation}
\noindent
with $k_{\rm min}=0.02 \ihMpc$.

The model is fit to the two-point correlation function monopole and quadrupole, where the Fourier space model is converted to configuration space through a Hankel transform. Galaxy separations are restricted to $60 < s < 150 \ \hMpc$ to focus on the BAO feature \citep{Y3.clust-s1.Andrade.2025}. All tracers are fit to the correlation function of combined galactic caps (NGC and SGC).

\begin{table}
    \centering
    \begin{tabular}{lccc}
        \hline
        \hline
        Parameter & Reconstruction & Mean & Uncertainty \\
        \hline
        $\Sigma_\perp^{\rm fid}$ $[h^{-1} \rm Mpc]$ & Pre & 4.5 & 1.0 \\
        $\Sigma_\parallel^{\rm fid}$ $[h^{-1} \rm Mpc]$ & Pre & 9.0 & 2.0 \\
        $\Sigma_s^{\rm fid}$ $[h^{-1} \rm Mpc]$ & Pre & 2.0 & 2.0 \\
        \hline
        $\Sigma_\perp^{\rm fid}$ $[h^{-1} \rm Mpc]$ & Post & 3.0 & 1.0 \\
        $\Sigma_\parallel^{\rm fid}$ $[h^{-1} \rm Mpc]$ & Post & 6.0 & 2.0 \\
        $\Sigma_s^{\rm fid}$ $[h^{-1} \rm Mpc]$ & Post & 2.0 & 2.0 \\
        \hline
        \hline
    \end{tabular}
    \caption{Gaussian priors to damping parameters used for BAO fitting for all tracers unique to this work.}
    \label{tab:bao-priors}
\end{table}

\subsection{Cosmology Inference} \label{subsec:cosmo-methods}

We perform cosmological parameter inference using \texttt{Cobaya} \cite{Torrado:2021,Torrado:2019}, interfaced with \texttt{CAMB} \cite{LewisCAMB:2000}, and sample posteriors using Metropolis–Hastings MCMC. Parameter choices depend on the data combination considered, including BAO-only, BAO with external calibration, and joint BAO+CMB+SN analyses, and follow the same conventions as in the DESI key cosmology analysis \cite{DESI.DR2.DR2,DESI2024.VI.KP7A}. We calibrate the BAO ruler using the measurement of the baryon abundance from big bang nucleosynthesis (BBN) from \citep{Schoeneberg:2024}. For the CMB, we make use of the temperature, polarization, and cross power spectra from \textit{Planck}, the Atacama Cosmology Telescope (ACT) DR6, and the South Pole Telescope (SPT-3G), as well as all three surveys' lensing reconstructions \citep{Planck-2018-likelihoods, Madhavacheril:ACT-DR6, SPT-3G}. For type Ia supernovae, we use the Dark Energy Survey (DES) Dovekie likelihood \citep{DES-Dovekie}. Convergence and posterior summaries are assessed using standard Gelman–Rubin and effective sample size criteria, with results analyzed using \texttt{getdist} \cite{Lewis:2019xzd}.

\section{Results} \label{sec:results}
\subsection{Validation tests with DR2 mocks} \label{subsec:mock-results}

\subsubsection{Unified Tracer is Unbiased} \label{subsubsec:mock-unbias}

\begin{table*}
    \centering
    \begin{tabular}{l|l|c|c|c|c|c|r|c}
        \hline\hline
        Choice & Tracer & Redshift Range & $\langle \aiso^{\rm mock} \rangle$ & $\langle \sigma_{\rm iso}^{\rm mock} \rangle$ & $\langle \alap^{\rm mock} \rangle$ & $\langle \sigma_{\rm AP}^{\rm mock} \rangle$ & \multicolumn{1}{c|}{$\langle r \rangle$} & $\langle \chi^2 / {\rm dof} \rangle$ \\
        \hline
        & \tleq & $0.8<z<1.1$ & $1.0005 \pm 0.0013$ & $0.0053$ & $1.0018 \pm 0.0033$ & $0.0186$ & $0.062$ & $40.1/33$ \\
        Default zbin & \eq & $1.1<z<1.6$ & $0.9988 \pm 0.0015$ & $0.0063$ & $0.9952 \pm 0.0051$ & $0.0217$ & $-0.123$ & $37.1/33$ \\
        & \qsoo & $1.6<z<2.1$ & $0.9889 \pm 0.0056$ & $0.0193$ & $0.9888 \pm 0.0174$ & $0.0731$ & $-0.076$ & $34.4/33$ \\
        \hline\hline
        & \tleqo & $0.8<z<1.0$ & $0.9994 \pm 0.0012$ & $0.0059$ & $0.9987 \pm 0.0035$ & $0.0213$ & $0.125$ & $39.8/33$ \\
        Rebinned & \tleqt & $1.0<z<1.2$ & $0.9994 \pm 0.0021$ & $0.0090$ & $1.0099 \pm 0.0070$ & $0.0330$ & $0.040$ & $34.3/33$ \\
        & \eqo & $1.2<z<1.4$ & $0.9966 \pm 0.0021$ & $0.0098$ & $1.0000 \pm 0.0073$ & $0.0346$ & $-0.026$ & $34.5/33$ \\
        & \eqt & $1.4<z<1.6$ & $0.9988 \pm 0.0028$ & $0.0119$ & $0.9878 \pm 0.0082$ & $0.0410$ & $-0.040$ & $35.8/33$ \\
        \hline\hline
        & \lrgelg & $0.8<z<1.1$ & $1.0003 \pm 0.0012$ & 0.0052 & $1.0009 \pm 0.0032$ &  $0.0185$ & $0.072$ & $38.8/33$ \\
        DR2 Baseline & \elgt & $1.1<z<1.6$ & $1.0009 \pm 0.0019$ & $0.0077$ & $0.9966 \pm 0.0055$ & $0.0265$ & $-0.138$ & $24.6/33$ \\
        & \qso & $0.8<z<2.1$ & $0.9939 \pm 0.0026$ & $0.0116$ & $1.0030 \pm 0.0078$ & $0.0432$ & $-0.054$ & $29.1/33$ \\
        \hline\hline
    \end{tabular}
    \caption{BAO fits to the unified tracers in 25 mocks. The $\langle \alpha^{\rm mock} \rangle$ columns are the mean values measured from mocks, along with the dispersion within the mocks (divided by $\sqrt{25}$). The $\langle \sigma^{\rm mock} \rangle$ columns are the means of the errors from fits to each mock. The top block corresponds to the unified-tracer analysis, the middle block shows results obtained using the alternative redshift rebinning, and the bottom block is included for reference, corresponding to the baseline DR2 analysis used for comparison. We note a high mean $\chi^2$ for the lowest redshift bins in both binnings. This is also seen in the baseline results for \lrgelg; thus we do not find this concerning. Mocks for the unified tracers are consistent with unity within $2\sigma$ error on the mean, indicating they are non-biased tracers of BAO.}
    \label{tab:all-alphas-mock}
\end{table*}

\begin{figure}
    \subfigure{
        \centering
        \includegraphics[width=0.95\linewidth]{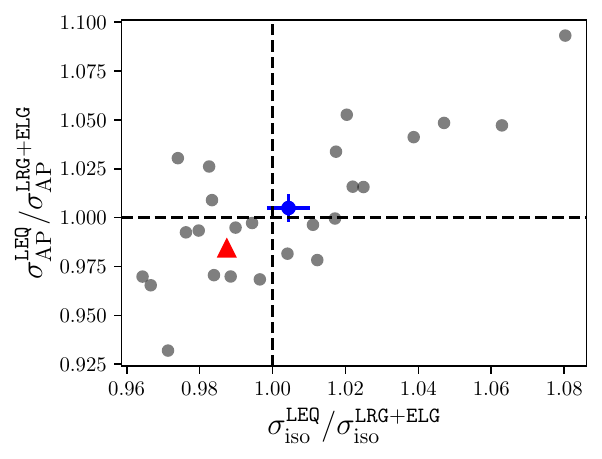}
        }
    \subfigure{
        \centering
        \includegraphics[width=0.95\linewidth]{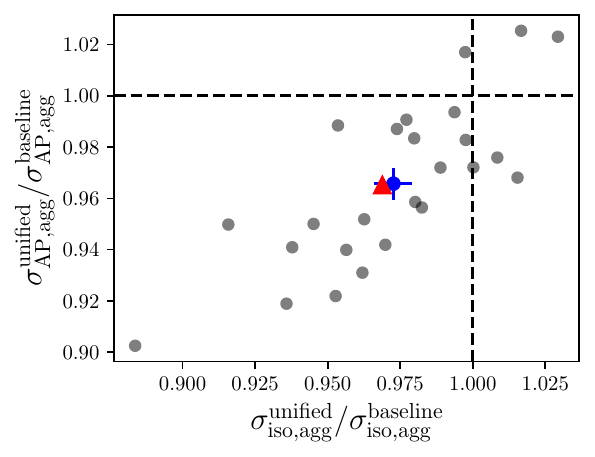}
        }
    \caption{Comparison of the precision in $\aiso$ and $\alap$. The top plot compares \tleq\ to \lrgelg\ in the redshift bin $0.8<z<1.1$. The bottom plot compares the aggregate precision \cref{eqn:agg-err} of the unified and the baseline analysis over $0.8<z<2.1$. Grey points are for 25 mocks, with their mean and dispersion in blue, and the DR2 data in red. No gain is found in the $0.8<z<1.1$ bin, but the unified tracer gains precision in aggregate across all bins, driven by gains in $1.1<z<1.6$}.
    \label{fig:err-gain}
\end{figure}

We first want to confirm that combining tracers into a unified catalog does not bias the BAO measurement.
We apply the unified catalog pipeline (\cref{sec:methods}) on 25 Abacus mocks.
Since the fiducial cosmology used for the redshift-to-distance conversion is the same as that used for mock production, measurements of BAO distance scales ($\alpha_{\rm iso}$ and $\alpha_{\rm AP}$) in mock catalogs should be equal to unity within statistical fluctuations.
Significant discrepancy from unity would indicate an uncorrected systematic bias within the unification procedure, since the constituent tracers alone are consistent with unity \citep{Y3.clust-s1.Andrade.2025}.

We present results from the mock analyses in \cref{tab:all-alphas-mock} (the unified tracer in the first three rows labeled `Default zbin'). We find that the mean BAO scale parameters measured from mocks, both $\langle \aiso^{\rm mock} \rangle$ and $\langle \alap^{\rm mock} \rangle$, are consistent with unity within the precision set by the ensemble of 25 mocks, i.e. within $2\sigma$, for \tleq, \eq, and the high-redshift \qsoo\ sample. This demonstrates that combining tracers into a unified catalog does not introduce a measurable bias in the BAO scale within the statistical precision of these mock tests.
As a measure of the goodness of the fits, the average $\chi^2$ per degree of freedom is near unity for all unified tracers, indicating no obvious issues with the covariances or modeling.
We can then compare \tleq\ to the baseline \lrgelg\ to check if there is any gain in precision due to the additional inclusion of \qso. The bottom three rows of \cref{tab:all-alphas-mock} present the baseline DR2 results \citep{DESI.DR2.DR2}. Additionally, a comparison of the distribution of errors in mocks between \tleq\ and \lrgelg\ is shown in the top panel of \cref{fig:err-gain}. The mocks indicate we expect no significant change in the precision of both BAO dilation parameters compared to the baseline DR2 choice, likely due to the subdominant contribution from \qso.
In \cref{subsec:data-results}, we will present the results of the DR2 data, in comparison to the mock results presented above. 

\subsubsection{Tracer-Dependent Systematics} \label{subsubsec:mock-systematics}

The unified tracer presents an opportunity to test for tracer-dependent systematics in DESI DR2. Tracer-dependent systematics could result either from uncorrected systematics within the data pipeline or indicative of new physics such as the relative velocity effect \citep{Dalal.RelV_2010, Beutler.RelV_2017}. This could manifest as a difference in the measured BAO scales between different tracers or combinations of tracers within the same cosmic volume.

\begin{figure*}
    \centering
    \includegraphics[width=0.75\linewidth]{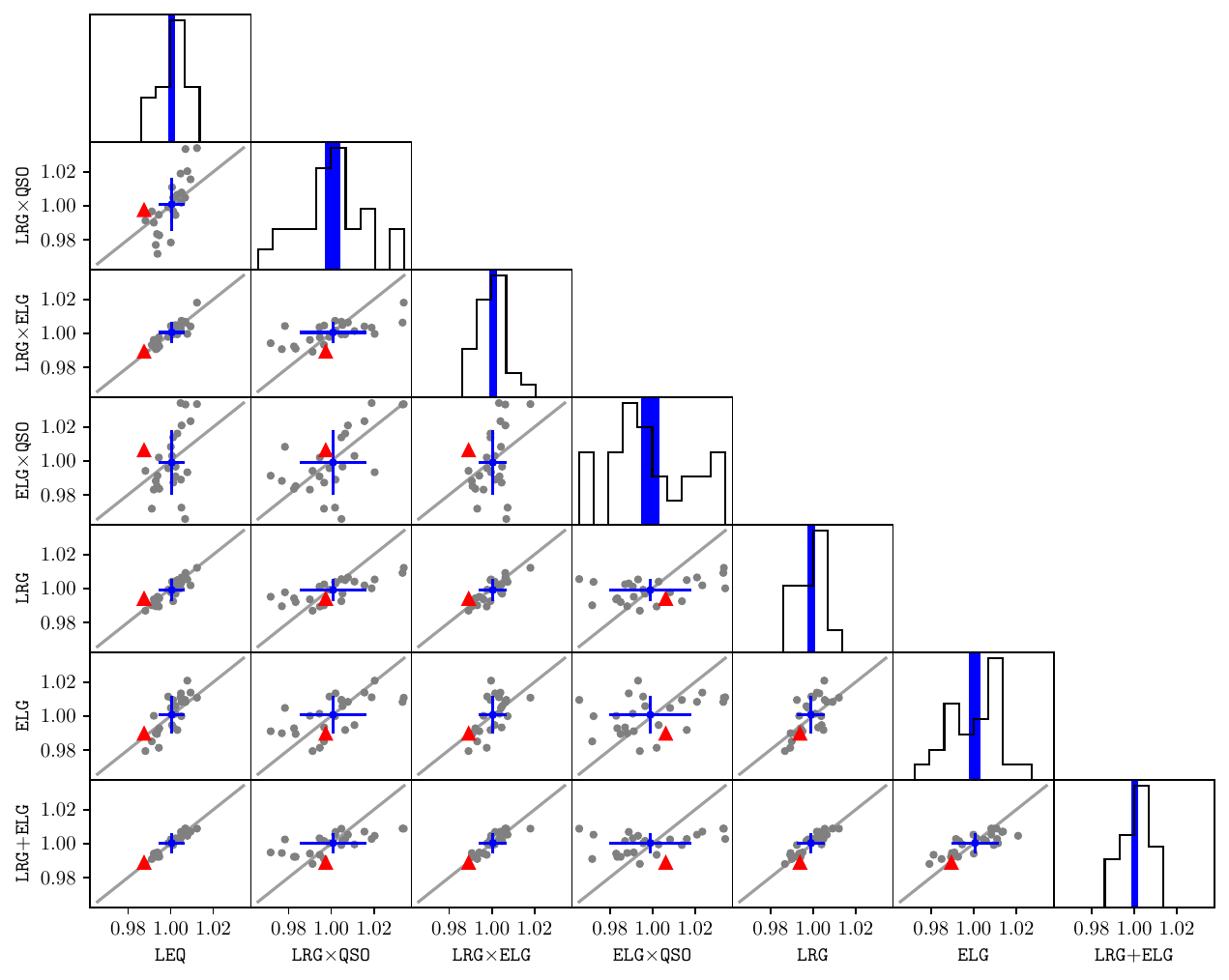}
    \caption{Systematic test of $\alpha_{\rm{iso}}$ fits to tracers in redshift $0.8<z<1.1$. Cross-correlations are denoted by $\times$. The grey points are best fit $\aiso$ values for the 25 mocks, with the mean and dispersion in mocks shown in blue. The red triangle points show the $\aiso$ values in DR2 data. Tracer-dependent systematics would manifest as the data falling further from the equality line than what is expected by the scatter in mocks where such systematics are not present. In no pairing of tracers is the data atypically far from the equality line, and thus no evidence of tracer-dependent systematics is observed.}
    \label{fig:qiso-scatter}
\end{figure*}

To test for systematics in the data, we must first verify that the catalog unification procedure does not introduce tracer-dependent biases. We do this by measuring BAO post-reconstruction for all possible tracers in redshift $0.8<z<1.1$ in the suite of 25 mocks. \cref{fig:qiso-scatter} summarizes the results of this test. The top diagonal subplots show that no tracer is inherently biased, repeating the information in \cref{tab:all-alphas-mock}. Each of the off-diagonal subplots shows the covariance between the $\aiso$ measurements for a pair of different BAO tracers constructed for $0.8<z<1.1$. The scatters from 25 mocks demonstrate that different tracers also tend to be correlated to varying degrees, with \tleq\ and \lrgelg\ being most tightly correlated, allowing variance cancellation and therefore a tight constraint on the differences. Tracer dependence will manifest as a significant offset from the diagonal line, and in all cases, the average offsets (the blue point with errors) are consistent with no relative bias. Quantitatively, this means that all pairs of tracers have a mean difference in BAO scales consistent with 0 within the dispersion of mocks, rescaled by $\sqrt{25}$ to reflect the error on the mean. The same test was performed for $\alap$ in \cref{sec:add-systematics} and again, we found no relative bias in mocks.

The conclusion from this mock test is that we do not expect our analysis to introduce artificial tracer-dependent systematics. The statistical fluctuations in the differences between different tracer BAO scales in mocks can now be used to quantify the level of detection of tracer-dependent systematics in the data. 

\subsection{The Unified Tracer in DR2 Data} \label{subsec:data-results}

\begin{table*}
    \centering
    \begin{tabular}{l|l|c|c|c|r|c}
        \hline\hline
        Choice & Tracer & Redshift Range & $\aiso$ & $\alap$ & \multicolumn{1}{c|}{$r$} & $\chi^2 / {\rm dof}$ \\
        \hline
        &  \tleq & $0.8<z<1.1$ & $0.9875 \pm 0.0045$ & $1.0248 \pm 0.0155$ & $-0.083$ & $33.9/33$ \\
        Default zbin & \eq & $1.1<z<1.6$ & $0.9902 \pm 0.0064$ & $1.0191 \pm 0.0213$ & $-0.140$ & $36.6/33$ \\
        & \qsoo & $1.6<z<2.1$ & $0.9962 \pm 0.0189$ & $0.9985 \pm 0.0693$ & $0.186$ & $42.1/33$ \\
        \hline\hline
        & \tleqo & $0.8<z<1.0$ & $0.9902 \pm 0.0053$ & $1.0207 \pm 0.0186$ & $-0.015$ & $29.4/33$ \\
        Rebinned & \tleqt & $1.0<z<1.2$ & $0.9855 \pm 0.0083$ & $1.0288 \pm 0.0310$ & $0.161$ & $44.5/33$ \\
        & \eqo$^*$ & $1.2<z<1.4$ & $0.9920 \pm 0.0114$ & $0.9999 \pm 0.0383$ & $-0.057$ & $51.1/33$ \\ 
        & \eqt & $1.4<z<1.6$ & $0.9879 \pm 0.0109$ & $1.0411 \pm 0.0413$ & $-0.043$ & $41.6/33$ \\
        \hline\hline 
        & \lrgelg & $0.8<z<1.1$ & $0.9886 \pm 0.0046$ & $1.0237 \pm 0.0157$ & $-0.099$ & $38.4/33$ \\
        DR2 Baseline & \elgt & $1.1<z<1.6$ & $0.9911 \pm 0.0071$ & $1.0257 \pm 0.0237$ & $-0.207$ & $42.4/33$ \\
        & \qso & $0.8<z<2.1$ & $1.0032 \pm 0.0153$ & $0.9885 \pm 0.0564$ & $0.009$ & $22.2/33$ \\
        \hline\hline
    \end{tabular}
    \caption{Compilation of the BAO fits to the unified tracer in the DR2 data. The $\alpha$ columns are the best fit values to data. The table layout and tracer definitions follow those in \cref{tab:all-alphas-mock}. $^*$The high $\chi^2/{\rm dof}$ for \eqo\ falls outside of the distribution of 25 mocks. We investigate this in detail in \cref{sec:broad-eq1}, and find it can be resolved by changing the modeling of the broadband term, with negligible effect on the best fit.
    }
    \label{tab:all-alphas-data}
\end{table*}

\subsubsection{Testing for Tracer-Dependent Systematics} \label{subsubsec:data-tracer-sys}

As we assured with the mocks that the unified tracers return unbiased BAO constraints, we now present the corresponding BAO measurements from the DR2 data in \cref{tab:all-alphas-data} (the first three rows). Unlike in the mock analyses, where agreement with the input cosmology can be explicitly tested, the data measurements cannot be required to agree with unity because the true cosmology is not known. Instead, we use the data to test for relative consistency between tracers, checking for possible tracer-dependent biases in the BAO scale parameters. Under the null hypothesis of no tracer-dependent systematics, any differences observed between tracers should be consistent with statistical noise, i.e., within the scatters of the mocks. 
The red points in \cref{fig:qiso-scatter} show the DR2 data  $\aiso$ fits over redshifts $0.8<z<1.1$. For all pairs of tracers, the data fall within the distribution of mocks in terms of the perpendicular distance from the diagonal lines, meaning that we find no evidence of tracer-dependent systematics. 
This systematic test is also done for $\alap$ and for both redshift ranges $0.8<z<1.1$ and $1.1<z<1.6$, which all support the same conclusion. These additional tests are presented in \cref{fig:qap-scatter,fig:tracer-sys-1.1-1.6} in \cref{sec:add-systematics}.

As a highlight of this point, we present a subset of tracer BAO parameter differences in \cref{tab:diff-subset}. For example, \tleq\ and \lrgelg\ agree on $\aiso$ within $\sim 0.1 \%$, less than $1\sigma$ as quantified by the standard deviation of the differences from the mocks. The most discrepant case of \lrg\ versus \tleq\ shows a difference at a significance of less than $2.5\sigma$, and falls within the full range of the 25 mocks. 
The conclusion is that we find no evidence of tracer-dependent systematics in DESI DR2 BAO data. For completeness, the significance of all tracer differences is presented in \cref{tab:all-diffs-0.8-1.1} in \cref{sec:add-systematics}.

\subsubsection{Precision of DR2 BAO Fits} \label{subsubsec:data-precision}

The red point in \cref{fig:err-gain} indicates the precision of the $\aiso$ and $\alap$ fits of \tleq\ in DR2 data as compared to \lrgelg. Consistent with expectations from mocks, there is no significant gain in precision.

We now compute an aggregate error, presented in \cref{tab:agg-err}, by summing the reciprocal of the squared errors following
\begin{equation} \label{eqn:agg-err}
\sigma_{\rm agg}^2 = \left[\sum_i \frac{1}{\sigma_i^2} \right]^{-1}
\end{equation}
\noindent
where $i$ sums over all redshift bins. This quantifies the overall precision of the BAO distance measurements across all bins. Note we only include bins in $0.8<z<2.1$ where the unified analysis differs from the baseline. The bottom panel of \cref{fig:err-gain} compares the aggregate errors between the unified and baseline analyses in both mocks and DR2 data. The mocks anticipate a $\sim 3\%$ gain in aggregate precision of both $\aiso$ and $\alap$ between the baseline and the Unified tracer (the first two rows of \cref{tab:agg-err}). The DR2 data also sees a $\sim 3\%$ gain over baseline. 
The small gain matches expectations, as the major difference from baseline is the addition of \qso\ into the unified tracer, which is subdominant to \lrg\ and \elg\ in terms of the number density. In fact, the true gain from the unified tracer must be slightly larger than the observed 3\%, since the unified tracer removes the double-counting of the \qso\ between $0.8<z<1.6$. That the net result is still positive, however small, implies that the benefit of the unified tracer, either from the cross-tracer information and/or the reduced shot noise, is enough to offset the loss.

\begin{table}
    \centering
    \begin{tabular}{l|r|r}
        \hline
        \hline
        Tracer & \multicolumn{1}{c|}{$\Delta \aiso \%$} & \multicolumn{1}{c}{$\Delta \alap \%$} \\
        \hline
        \lrgelg & $-0.12 \pm 0.17$ & $0.34 \pm 0.76$ \\
        \lrg & $-0.64 \pm 0.30$ & $2.79 \pm 1.69$ \\
        \elg & $-0.20 \pm 0.71$ & $-2.56 \pm 2.06$ \\
        \lrgxelg & $-0.27 \pm 0.32$ & $0.13 \pm 0.97$ \\
        \lrgxqso & $-0.99 \pm 1.16$ & $6.87 \pm 3.51$ \\
        \elgxqso & $-1.87 \pm 1.70$ & $3.21 \pm 4.28$ \\
        \hline
        \hline
    \end{tabular}
    \caption{Consistency in the BAO constraints across different BAO tracers from the DR2 data. The quoted values are the differences measured from the DR2 data between \tleq\ and other tracers in the redshift range $0.8<z<1.1$. The quoted errors are the $1\sigma$ standard deviations of 25 differences measured from the mocks for the corresponding pair. The data show no statistically significant difference, indicating no detection of tracer-dependent systematics. The significance of all pairs of tracers are presented in \cref{tab:all-diffs-0.8-1.1}.}
    \label{tab:diff-subset}
\end{table}

\begin{table}
    \centering
    \begin{tabular}{ccccc}
        \hline
        \hline
        Case & $\sigma_{\rm iso,agg} (\%)$ & $\sigma_{\rm AP,agg} (\%)$ & $\langle \sigma_{\rm iso,agg}^{\rm mock} \rangle (\%)$ & $\langle \sigma_{\rm AP,agg}^{\rm mock} \rangle (\%)$ \\
        \hline
        Baseline & 0.404 & 1.294 & 0.407 & 1.433 \\
        Unified & 0.395 & 1.250 & 0.395 & 1.388 \\
        Rebin & 0.400 & 1.375 & 0.405 & 1.453 \\
        \hline
        \hline
    \end{tabular}
    \caption{Aggregate errors of the DR2 data  (2nd-3rd columns) and average of mocks (4th-5th columns) in the baseline DR2 analysis, the unified tracer without rebinning, and the unified tracer with rebinning. Mocks indicate that the unified tracer nets a $\sim 3 \%$ improvement in the aggregate precision of both $\aiso$ and $\alap$, with similar results in the DR2 data. This gain is not realized in the rebinned case.}
    \label{tab:agg-err}
\end{table}

\subsection{Stability of reconstruction across the redshift boundaries} \label{subsec:recon-boundary}

With the unified tracer, we have more flexibility to choose different redshift bins. To rebin the unified tracer in redshift, we demonstrate that the BAO is not biased when reconstructed and measured across discontinuities in the effective bias due to survey design choices. There are 3 redshift boundaries between dark-time tracers due to the transition of samples: the \elg\ and \qso\ boundary at $z=0.8$, the \lrg\ boundary at $z=1.1$, and the \elg\ boundary at $z=1.6$. To test for biases in BAO measurements across boundaries, we define redshift bins that include the boundaries, and fit BAO to 25 mocks. We find the mean of the fits to $\aiso$ and $\alap$ to be consistent with unity (\cref{tab:boundaries}). We conclude from this that reconstruction and pair counting are stable and unbiased across the tracer redshift boundaries.

\begin{table*}
    \centering
    \begin{tabular}{c|c|c|c|c|c|c}
        \hline\hline
        Redshift Range & $\alpha_{\rm iso}$ & $\alpha_{\rm AP}$ & $\chi^2/{\rm dof}$ & $\langle \alpha_{\rm iso}^{\rm mock}\rangle$ & $\langle \alpha_{\rm AP}^{\rm mock}\rangle$ & $\langle \chi^2/{\rm dof} \rangle$ \\
        \hline
        $0.7<z<0.9$ & $0.9807 \pm 0.0057$ & $1.0142 \pm 0.0199$ & $31.7/33$ & $1.0005 \pm 0.0014$ & $1.0030 \pm 0.0044$ & $41.9/33$ \\
        $1.0<z<1.2$ & $0.9855 \pm 0.0083$ & $1.0288 \pm 0.0310$ & $44.5/33$ & $0.9996 \pm 0.0022$ & $1.0099 \pm 0.0076$ & $30.3/33$ \\
        $1.5<z<1.7$ & $0.9888 \pm 0.0170$ & $0.9315 \pm 0.0586$ & $42.0/33$ & $0.9969 \pm 0.0044$ & $0.9963 \pm 0.0189$ & $39.0/33$ \\
        \hline\hline
    \end{tabular}
    \caption{Stability of the BAO measurements across the redshift boundaries. We constructed catalogs crossing the boundaries and tested the post-reconstruction BAO measurements: the DR2 data (the 2nd and 3rd columns), in comparison to the mocks (the 5th and 6th columns), where we report the error on the mean of the mocks. The mean of the fits to 25 mocks is consistent with unity, indicating an unbiased measurement when crossing redshift boundaries.}
    \label{tab:boundaries}
\end{table*}

\subsection{The Rebinned Unified Tracer} \label{subsec:rebin-results}

\begin{figure*}
    \centering
    \includegraphics[width=\linewidth]{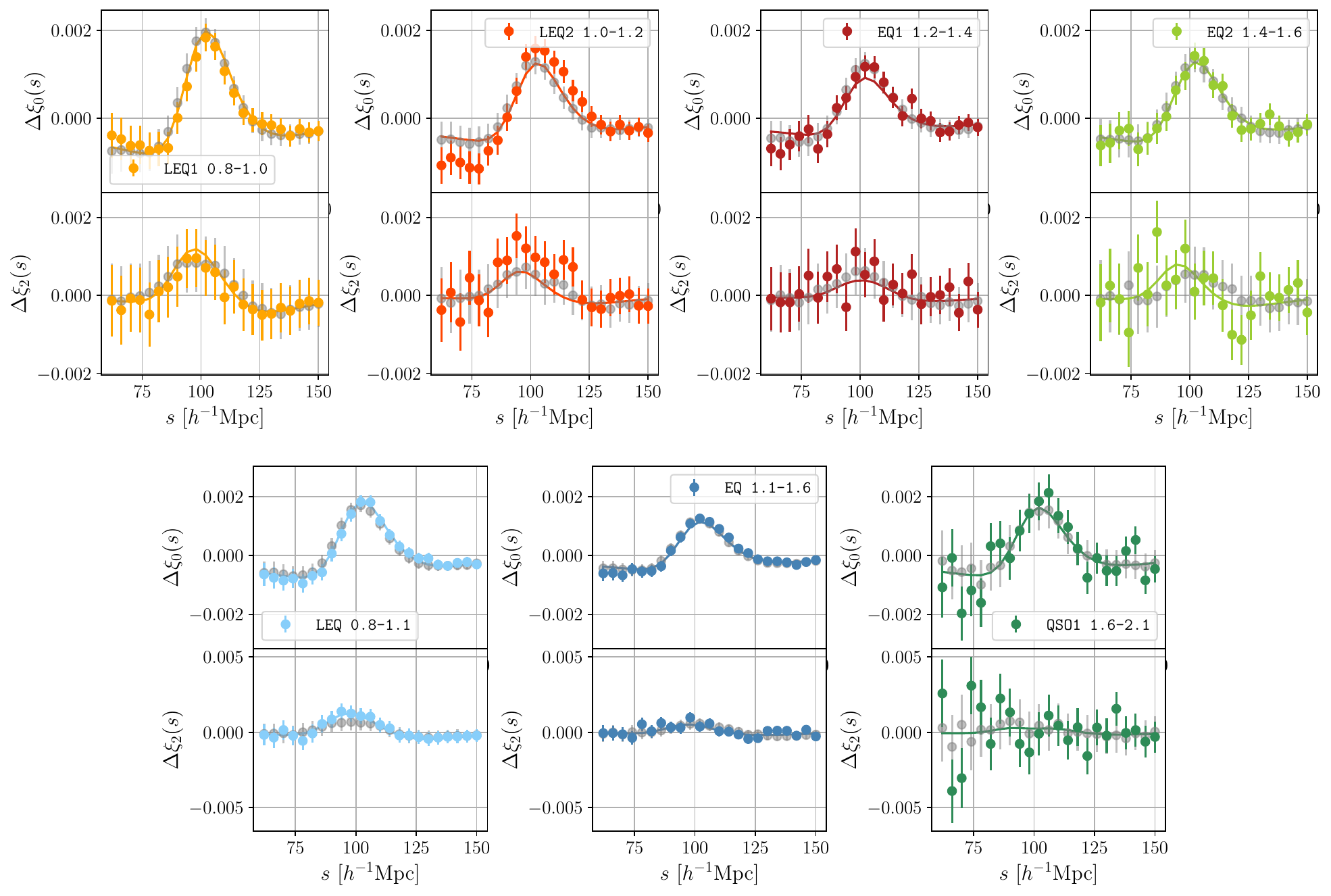}
    \caption{Two-point correlation function monopoles and quadrapoles of all unified tracers and the \qsoo\ bin, with the broadband subtracted off to display only the BAO feature. The top row is the rebinned unified tracers, and the bottom row is the unified tracer following the same redshift binning as the DR2 baseline analysis. The data values and best fits are the colored points and curves respectively, and the mean measurements in the mocks are shown in gray.}
    \label{fig:all-xis}
\end{figure*}

Having verified that reconstruction of the unified tracer is stable across tracer redshift boundaries, we are free to rebin the unified tracer as described in \cref{subsec:rebin}. We first test the new choice of bins on the mocks to ensure no bias is introduced in the BAO measurements, which the middle panel (`Rebinned') of \cref{tab:all-alphas-mock} confirms. \cref{fig:all-xis} shows the fits to the BAO feature of the DR2 data with respect to those of the mocks (with the rebinned unified tracers in the top row) and the middle panel of \cref{tab:all-alphas-data} presents the BAO measurements of the DR2 data. Clear BAO features are present in all bins. We note the visually poor fit to the data for \tleqt\ and \eqo. In the case of \eqo, the $\chi^2$ of the best fit falls outside the distribution of mocks (albeit marginally), failing a validation test. This is resolved by changing the model for the broadband term, discussed in detail in \cref{sec:broad-eq1}. Regardless, this broadband choice does not have significant effect on the best fit, and so results follow the fiducial model discussed in \cref{subsec:bao-model}.

The aggregate error on the new binning (including \qsoo) is also presented in the last row of \cref{tab:agg-err}. From mocks, no gain in precision for the BAO measurement is expected. The precision in DR2 data is consistent with the scatter in mocks, indicating the rebinned data includes equivalent information as the default binning. 
A loss of $\sim 6\%$ is observed for the aggregate precision in $\alap$. This is likely because the rebinned analysis is split into more redshift bins compared to baseline, so more galaxy pairs cross the boundaries between bins and are thus not counted.
We present this alternative binning only as a proof of concept, and future studies could investigate optimal binning choices, which fall outside the scope of this project.

\subsection{Adding Systematics} 

A variety of sources of systematic error are quantified, such as the fiducial cosmology \citep{KP4s9-Perez-Fernandez}, halo occupation  \citep{KP4s11-Garcia-Quintero, KP4s10-Mena-Fernandez}, and the theory model \citep{KP4s2-Chen}. These errors are added to the $\aiso$--$\alap$ covariance matrix, reflecting the increased uncertainty due to systematics. We inherit the same systematic error values as the baseline tracers in DR2 \citep{Y3.clust-s1.Andrade.2025}, with the respective $\aiso$ and $\alap$ \lrgelg\ values of $0.221\%$ and $0.329\%$ adopted by \tleq, \tleqo, and \tleqt; the \elgt\ values of $0.221\%$ and $0.224\%$ adopted by \eq, \eqo, and \eqt; and the \qso\ values of $0.221\%$ and $0.329\%$ adopted by \qsoo.

\begin{table*}[]
    \centering
    \begin{tabular}{l|c|c|c|r|c|c|c}
        \hline\hline
        Tracer & $z_{\rm eff}$ & $D_{\rm V}/r_{\rm d}$ & $D_{\rm M} /D_{\rm H}$ & $r_{\rm V, M/H}$ & $D_{\rm M}/r_{\rm d}$ & $D_{\rm H}/r_{\rm d}$ & $r_{\rm M,H}$ \\
        \hline
        \tleq & 0.950 & $19.915 \pm 0.102$ & $1.249 \pm 0.019$ & $-0.073$ & $21.817 \pm 0.158$ & $17.468 \pm 0.203$ & $-0.373$ \\
        \eq & 1.329 & $24.308 \pm 0.165$ & $1.976 \pm 0.041$ & $-0.131$ & $27.742 \pm 0.271$ & $14.042 \pm 0.220$ & $-0.404$ \\
        \qsoo & 1.828 & $28.462 \pm 0.544$ & $3.178 \pm 0.221$ & $0.184$ & $34.222 \pm 1.028$ & $10.769 \pm 0.539$ & $-0.451$ \\
        \hline
        \tleqo & 0.903 & $19.323 \pm 0.111$ & $1.174 \pm 0.022$ & $-0.014$ & $21.090 \pm 0.178$ & $17.965 \pm 0.247$ & $-0.404$ \\
        \tleqt & 1.095 & $21.696 \pm 0.189$ & $1.502 \pm 0.046$ & $0.155$ & $24.106 \pm 0.321$ & $16.051 \pm 0.358$ & $-0.443$ \\
        \eqo & 1.299 & $24.056 \pm 0.282$ & $1.951 \pm 0.075$ & $-0.056$ & $27.549 \pm 0.478$ & $14.121 \pm 0.397$ & $-0.414$ \\
        \eqt & 1.490 & $25.710 \pm 0.290$ & $2.271 \pm 0.090$ & $-0.042$ & $29.587 \pm 0.515$ & $13.029 \pm 0.383$ & $-0.501$ \\
        \hline\hline
    \end{tabular}
    \caption{BAO distance measurements, including contributions from the systematic error budget. $D_{\rm V}/r_{\rm d}$ is the average isotropic size of the BAO, $D_{\rm M}/D_{\rm H}$ the distortion of the BAO across and along the line-of-sight due to the Alcock-Paczynski effect, and $D_{\rm M}/r_{\rm d}$ and $D_{\rm H}/r_{\rm d}$ are the distances across and along the line-of sight, respectively. Here, $r_{\rm M,H}$ is the correlation between the perpendicular and parallel components of the BAO. }
    \label{tab:distances}
\end{table*}

\subsection{Hubble Diagrams and Cosmology Results} \label{sec:cosmo-results}

\begin{figure*}
    \centering
    \includegraphics[width=\linewidth]{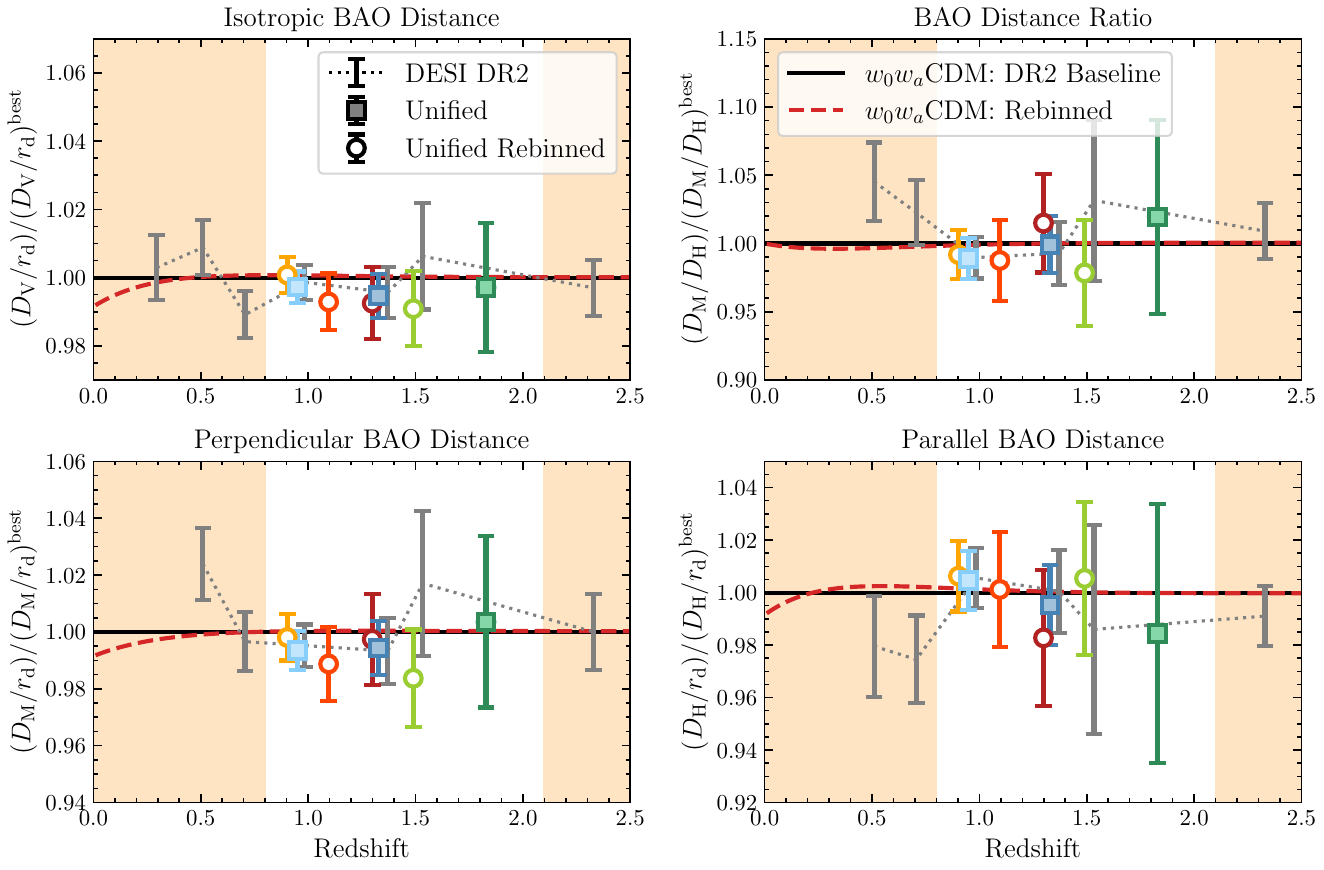}
    \caption{Hubble Diagram of DESI DR2 with the baseline analysis shown as the gray error bars, the unified tracer as the square points, and the rebinned unified tracer as the circle points. The highest redshift colored point corresponds to the \qsoo\ bin, and is common between both binnings. The shaded regions are redshifts where no tracers overlap and the unified tracer analysis is identical to the baseline analysis. The baseline points in the non-shaded region have been shifted by $\Delta z=0.05$ to improve readability. All distances are plotted with respect to the best fit $w_0w_a$ cosmology to the baseline data. Both binnings demonstrate strong agreement with the baseline distance measurements.}
    \label{fig:hubble}
\end{figure*}

\begin{table*}
    \centering
    \begin{tabular}{llcccc}
        \hline
        \hline
        Model/Dataset &  BAO & $\Omega_{\rm m}$ & $H_0$ & $w_0$ & $w_a$ \\
        \hline
        ${\bf Flat \ \Lambda {\rm \bf CDM}}$ & & & & \\
        DESI & Baseline & $0.2975 \pm 0.0086$ & --- & --- & --- \\
        DESI & Unified & $0.2981\pm 0.0086$ & --- & --- & --- \\
        DESI & Rebin & $0.2998\pm 0.0089$ & --- & --- & --- \\
        \hline
        DESI+BBN (no Lya) & Baseline & $0.3001 \pm 0.0126$ & $68.60 \pm 0.69$ & --- & --- \\
        DESI+BBN (no Lya) & Unified & $0.3021\pm 0.0124$ & $68.78\pm 0.67$ & --- & --- \\
        DESI+BBN (no Lya) & Rebin & $0.3069 \pm 0.0137$ & $68.87 \pm 0.73$ & --- & --- \\
        \hline
        DESI+CMB+SN & Baseline & $0.3054 \pm 0.0033$ & $68.03 \pm 0.24$ & --- & --- \\
        DESI+CMB+SN & Unified & $0.3036 \pm 0.0033$ & $68.16\pm 0.24$ & --- & --- \\
        DESI+CMB+SN & Rebin & $0.3042\pm 0.0032$ & $68.12\pm 0.24$ & --- & --- \\
        \hline\hline
        ${\bf Flat \ w_0w_a {\rm \bf CDM}}$ & & & & & \\
        DESI+CMB+SN & Baseline & $0.3135 \pm 0.0053$ & $67.38 \pm 0.54$ & $-0.820 \pm 0.055$ & $-0.66_{-0.19}^{+0.22}$ \\
        DESI+CMB+SN & Unified & $0.3125 \pm 0.0053$ & $67.45 \pm 0.55$ & $-0.803 \pm 0.054$ & $-0.73_{-0.19}^{+0.21}$ \\
        DESI+CMB+SN & Rebin & $0.3129 \pm 0.0053$ & $67.41 \pm 0.55$ & $-0.807 \pm 0.055$ & $-0.71_{-0.20}^{+0.22}$ \\
        \hline
        \hline
    \end{tabular}
    \caption{Best fit cosmology results for a variety of combinations of datasets and cosmologies. Results for the baseline, unified, and rebinned BAO analyses are presented. In all cases, both unified tracer binnings are extremely consistent with the baseline.}
    \label{tab:cosmology}
\end{table*}

\begin{figure}
    \subfigure{
        \centering
        \includegraphics[width=0.95\linewidth]{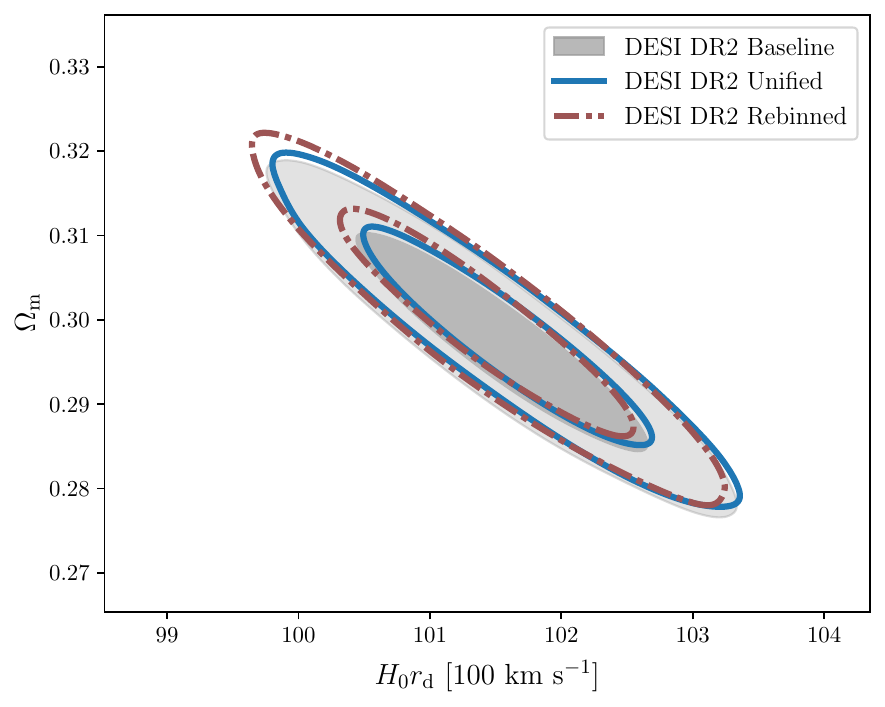}
        }
    \subfigure{
        \centering
        \includegraphics[width=0.95\linewidth]{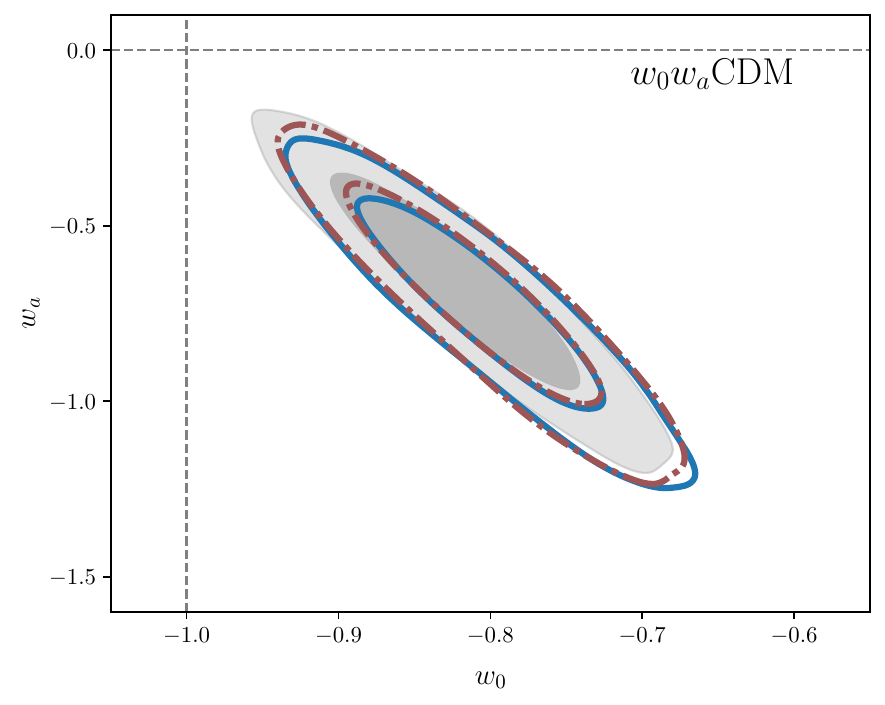}
        }
    \caption{Consistency in cosmology inference between the DR2 baseline, the unified tracer, and the rebinned unified tracer. The top panel shows the BAO-only constraints on $\Omega_{\rm m}$ and $H_0r_{\rm d}$ in \lcdm\ and the bottom panel shows the constraints on $w_0$ and $w_a$ from the combination of BAO, CMB, and supernovae. Both cases demonstrate that both binnings of the unified tracer are highly consistent with the baseline analysis.}
    \label{fig:cosmo-contours}
\end{figure}

We test the impact of the unified tracer on the cosmological parameters. We first check if the inferred cosmology from the unified tracer is consistent with the baseline DR2 BAO results \cite{DESI.DR2.DR2}. We also check whether the rebinning, with the boundary bin, produces a consistent cosmology best fit. Additionally, we check the relative precision on cosmological inference for both binnings. Despite a similar aggregate precision in BAO for the rebinning, the increased resolution in redshift space may improve the best fit by better tracking the expansion history. \cref{tab:distances} shows the BAO measurements in $D_{\rm M}/r_{\rm d}$, $D_{\rm H}/r_{\rm d}$, $D_{\rm V}/r_{\rm d}$, and $D_{\rm M}/D_{\rm H}$, after including systematics.

\cref{fig:hubble} shows the distance measurements of the unified tracer as compared to the baseline DR2 analysis. Both the default and new binning of the unified tracer infer expansion histories that are consistent with the baseline DR2 and with one another. In the default binning case, the expansion history looks nearly identical to the baseline, as expected. 
The greater redshift resolution of the rebinned case captures the distance at $z\approx1.5$ slightly differently, with the \eqt\ point being about $1\sigma$ lower than the baseline \qso\ point at about the same $z_{\rm eff}$ for $D_{\rm V}$, $D_{\rm M}/D_{\rm H}$, and $D_{M}$.

For both binning choices, we perform cosmological parameter inference. Results for a variety of cosmologies and dataset combinations (explained in \cref{subsec:cosmo-methods}) are presented in \cref{tab:cosmology}. Overall, we find the results for both binnings to be consistent with both the baseline DR2 analysis and each other. In \cref{fig:cosmo-contours}, we show comparisons of the constraints on $\Omega_m$ vs $H_0r_{\rm d}$ assuming flat $\Lambda$CDM cosmology (top) and the $w_0w_a$ constraint when extending to flat $w_0w_a$CDM cosmology (bottom), for the baseline, unified, and rebinned analyses. The three choices are all extremely consistent. We conclude that both the unified tracer procedure and rebinning the analysis return consistent cosmological interpretations of the BAO measurement. 
The consistency of the rebinned analysis in particular implies that the coarse binning of the baseline analysis in the redshifts $0.8<z<1.6$ does not miss a potential feature of dynamical dark energy between the \lrgelg\ and \elgt\ points.

We compute the significance of preference for $w_0w_a\rm{CDM}$ over \lcdm\ via Wilks' theorem \citep{Wilks:1938dza}. Since \lcdm\ is contained within $w_0w_a\rm{CDM}$, $\Delta \chi^2_{\rm MAP}$ should follow a $\chi^2$ distribution with two degrees of freedom under the assumption of \lcdm. We then translate $\Delta \chi^2_{\rm MAP}$ into the corresponding frequentest significance for a 1D Gaussian distribution. We find preference for evolving dark energy to be $3.2\sigma$ for the baseline case, $3.4\sigma$ for the unified case, and $3.3\sigma$ for the rebinned case, reproducing consistent preference for evolving dark energy.

\section{Conclusion} \label{sec:conclusion}

We have tested a method for combining all overlapping BAO galaxy tracers into a single, unified tracer for the DESI DR2 data over $0.8<z<1.6$. It is designed to coherently include all information from all available tracers, such as the cross-tracer correlations that are ignored in the baseline analysis, at the catalog level, and to reduce shot noise by increasing tracer density during density field reconstruction. Ensuring the stability of the reconstruction across the tracer boundaries, we demonstrate that this single unified tracer allows flexibility in the redshift binning.

The analysis in this paper largely follows that of the baseline DR2 BAO analysis \citep{DESI.DR2.DR2}, with a change to the catalog definition, extending the technique established in \citep{KP4s5-Valcin}. 
Improvements on \citep{KP4s5-Valcin} include the addition of \qso\ to the unified tracer, extending out to $z=1.6$ to include all overlapping tracers. Consequently, it avoids the issue of double counting the same cosmic volume between \qso\ and other tracers over $0.8<z<1.6$ that results from treating those \qso\ as independent in the DESI DR1 and DR2 baseline analyses. Additional improvements are the redshift-dependent FKP weights, a redshift-dependent bias used for more accurate reconstruction, and an additional analysis that leverages the properties of the unified tracer to rebin the analysis in redshift.

Using catalogs of mock data, we verify that the method of constructing the unified tracer does not bias BAO measurement. Once the method is confirmed, utilizing the unified tracer as well as all possible auto and cross-combinations of the single tracers over the same redshift range, we tested for consistency against tracer-dependent systematics in the DESI DR2 BAO measurements. We find that the BAO measurements from the different tracers are consistent, at most a $\sim2.5\sigma$ discrepancy with respect to the expected dispersion derived from the mocks, demonstrating no tracer-dependent systematics in the DESI DR2 BAO measurement either due to uncorrected observational systematics or due to new physics.
Furthermore, by adopting the redshift-dependent bias during the density field reconstruction, we demonstrate that the unified tracer can be stably reconstructed across redshift boundaries of different galaxy tracers. This consequently allows more flexibility in redshift binning than the DR2 baseline choice.

We investigate the effect of the unified tracer on the precision of the BAO measurements. We find a $\sim3\%$ improvement over the DR2 baseline \citep{DESI.DR2.DR2} on the aggregate precision across the redshift range we tested ($0.8<z<1.6$).
While the gain is small, the net gain being still positive implies that the benefit of the unified tracer, either from the cross-tracer information or the reduced shot noise, is enough to offset the loss due to avoiding the double-counting of \qso\ in $0.8<z<1.6$.
In the case of the rebinning, the aggregate precision was also fairly consistent with that of the baseline and with that of the unified tracer with the default binning. In detail, a $\sim 6\%$ loss was observed for $\alap$ compared to the unified tracer with the default binning, which could be because increasing the number of bins excludes galaxy pairs that cross the bin boundaries.

Finally, we check that both binning choices of the unified tracer are consistent with the DR2 baseline in terms of distance measurements and inferred cosmology. In terms of the distance measurements, the effective redshifts additionally probed by the two binning choices of the unified tracer show a consistent expansion history between redshifts of the baseline data points. That is, the baseline analysis does not seem to miss a redshift-dependent feature in the expansion history between the redshift $0.8<z<1.6$. This conclusion is further supported by the excellent agreement in dynamical dark energy constraints between the baseline and the rebinned analyses.

Overall, we find the unified tracer to be extremely consistent with the baseline analysis in all respects, while marginally improving the precision of the BAO measurement. Additionally, the freedom and consistency in rebinning opens the door to future investigations of optimal binning choices.

\section{Data Availability}

The data used in this analysis will be made public in the DESI Data Release 2 (details at \url{https://data.desi.lbl.gov/doc/releases/}).
The data points corresponding to the figures in this paper are available at \citep{data-zenodo}.

\acknowledgments
NS and H-JS acknowledge support from the U.S. Department of Energy, Office of Science, Office of High Energy Physics under grant No. DE-SC0023241. UA acknowledges support from the Leinweber Institute for Theoretical Physics at the University of Michigan Postdoctoral Research Fellowship and DOE Grant No. DE-SC009193.

This material is based upon work supported by the U.S. Department of Energy (DOE), Office of Science, Office of High-Energy Physics, under Contract No. DE–AC02–05CH11231, and by the National Energy Research Scientific Computing Center, a DOE Office of Science User Facility under the same contract. Additional support for DESI was provided by the U.S. National Science Foundation (NSF), Division of Astronomical Sciences under Contract No. AST-0950945 to the NSF’s National Optical-Infrared Astronomy Research Laboratory; the Science and Technology Facilities Council of the United Kingdom; the Gordon and Betty Moore Foundation; the Heising-Simons Foundation; the French Alternative Energies and Atomic Energy Commission (CEA); the Secretariat of Science, Humanities, Technology and Innovation (SECIHTI) of Mexico; the Ministry of Science, Innovation and Universities of Spain (MICIU/AEI/10.13039/501100011033), and by the DESI Member Institutions: \url{https://www.desi.lbl.gov/collaborating-institutions}. Any opinions, findings, and conclusions or recommendations expressed in this material are those of the author(s) and do not necessarily reflect the views of the U. S. National Science Foundation, the U. S. Department of Energy, or any of the listed funding agencies.

The authors are honored to be permitted to conduct scientific research on I'oligam Du'ag (Kitt Peak), a mountain with particular significance to the Tohono O’odham Nation.

\bibliographystyle{JHEP}
\bibliography{DESI_supporting_papers, biblio, Y1KP7a_references}

@ARTICLE{DESI2022.KP1.Instr,
       author = {{DESI Collaboration} and {Abareshi}, B. and {Aguilar}, J. and {Ahlen}, S. and {Alam}, Shadab and {Alexander}, David M. and {Alfarsy}, R. and {Allen}, L. and {Allende Prieto}, C. and {Alves}, O. and {Ameel}, J. and {Armengaud}, E. and {Asorey}, J. and {Aviles}, Alejandro and {Bailey}, S. and {Balaguera-Antol{\'\i}nez}, A. and {Ballester}, O. and {Baltay}, C. and {Bault}, A. and {Beltran}, S.~F. and {Benavides}, B. and {BenZvi}, S. and {Berti}, A. and {Besuner}, R. and {Beutler}, Florian and {Bianchi}, D. and {Blake}, C. and {Blanc}, P. and {Blum}, R. and {Bolton}, A. and {Bose}, S. and {Bramall}, D. and {Brieden}, S. and {Brodzeller}, A. and {Brooks}, D. and {Brownewell}, C. and {Buckley-Geer}, E. and {Cahn}, R.~N. and {Cai}, Z. and {Canning}, R. and {Capasso}, R. and {Carnero Rosell}, A. and {Carton}, P. and {Casas}, R. and {Castander}, F.~J. and {Cervantes-Cota}, J.~L. and {Chabanier}, S. and {Chaussidon}, E. and {Chuang}, C. and {Circosta}, C. and {Cole}, S. and {Cooper}, A.~P. and {da Costa}, L. and {Cousinou}, M. -C. and {Cuceu}, A. and {Davis}, T.~M. and {Dawson}, K. and {de la Cruz-Noriega}, R. and {de la Macorra}, A. and {de Mattia}, A. and {Della Costa}, J. and {Demmer}, P. and {Derwent}, M. and {Dey}, A. and {Dey}, B. and {Dhungana}, G. and {Ding}, Z. and {Dobson}, C. and {Doel}, P. and {Donald-McCann}, J. and {Donaldson}, J. and {Douglass}, K. and {Duan}, Y. and {Dunlop}, P. and {Edelstein}, J. and {Eftekharzadeh}, S. and {Eisenstein}, D.~J. and {Enriquez-Vargas}, M. and {Escoffier}, S. and {Evatt}, M. and {Fagrelius}, P. and {Fan}, X. and {Fanning}, K. and {Fawcett}, V.~A. and {Ferraro}, S. and {Ereza}, J. and {Flaugher}, B. and {Font-Ribera}, A. and {Forero-Romero}, J.~E. and {Frenk}, C.~S. and {Fromenteau}, S. and {G{\"a}nsicke}, B.~T. and {Garcia-Quintero}, C. and {Garrison}, L. and {Gazta{\~n}aga}, E. and {Gerardi}, F. and {Gil-Mar{\'\i}n}, H. and {Gontcho a Gontcho}, S. and {Gonzalez-Morales}, Alma X. and {Gonzalez-de-Rivera}, G. and {Gonzalez-Perez}, V. and {Gordon}, C. and {Graur}, O. and {Green}, D. and {Grove}, C. and {Gruen}, D. and {Gutierrez}, G. and {Guy}, J. and {Hahn}, C. and {Harris}, S. and {Herrera}, D. and {Herrera-Alcantar}, Hiram K. and {Honscheid}, K. and {Howlett}, C. and {Huterer}, D. and {Ir{\v{s}}i{\v{c}}}, V. and {Ishak}, M. and {Jelinsky}, P. and {Jiang}, L. and {Jimenez}, J. and {Jing}, Y.~P. and {Joyce}, R. and {Jullo}, E. and {Juneau}, S. and {Kara{\c{c}}ayl{\i}}, N.~G. and {Karamanis}, M. and {Karcher}, A. and {Karim}, T. and {Kehoe}, R. and {Kent}, S. and {Kirkby}, D. and {Kisner}, T. and {Kitaura}, F. and {Koposov}, S.~E. and {Kov{\'a}cs}, A. and {Kremin}, A. and {Krolewski}, Alex and {L'Huillier}, B. and {Lahav}, O. and {Lambert}, A. and {Lamman}, C. and {Lan}, Ting-Wen and {Landriau}, M. and {Lane}, S. and {Lang}, D. and {Lange}, J.~U. and {Lasker}, J. and {Le Guillou}, L. and {Leauthaud}, A. and {Le Van Suu}, A. and {Levi}, Michael E. and {Li}, T.~S. and {Magneville}, C. and {Manera}, M. and {Manser}, Christopher J. and {Marshall}, B. and {Martini}, Paul and {McCollam}, W. and {McDonald}, P. and {Meisner}, Aaron M. and {Mena-Fern{\'a}ndez}, J. and {Meneses-Rizo}, J. and {Mezcua}, M. and {Miller}, T. and {Miquel}, R. and {Montero-Camacho}, P. and {Moon}, J. and {Moustakas}, J. and {Mueller}, E. and {Mu{\~n}oz-Guti{\'e}rrez}, Andrea and {Myers}, Adam D. and {Nadathur}, S. and {Najita}, J. and {Napolitano}, L. and {Neilsen}, E. and {Newman}, Jeffrey A. and {Nie}, J.~D. and {Ning}, Y. and {Niz}, G. and {Norberg}, P. and {Noriega}, Hern{\'a}n E. and {O'Brien}, T. and {Obuljen}, A. and {Palanque-Delabrouille}, N. and {Palmese}, A. and {Zhiwei}, P. and {Pappalardo}, D. and {PENG}, X. and {Percival}, W.~J. and {Perruchot}, S. and {Pogge}, R. and {Poppett}, C. and {Porredon}, A. and {Prada}, F. and {Prochaska}, J. and {Pucha}, R. and {P{\'e}rez-Fern{\'a}ndez}, A. and {P{\'e}rez-R{\`a}fols}, I. and {Rabinowitz}, D. and {Raichoor}, A. and {Ramirez-Solano}, S. and {Ram{\'\i}rez-P{\'e}rez}, C{\'e}sar and {Ravoux}, C. and {Reil}, K. and {Rezaie}, M. and {Rocher}, A. and {Rockosi}, C. and {Roe}, N.~A. and {Roodman}, A. and {Ross}, A.~J. and {Rossi}, G. and {Ruggeri}, R. and {Ruhlmann-Kleider}, V. and {Sabiu}, C.~G. and {Safonova}, S. and {Said}, K. and {Saintonge}, A. and {Salas Catonga}, Javier and {Samushia}, L. and {Sanchez}, E. and {Saulder}, C. and {Schaan}, E. and {Schlafly}, E. and {Schlegel}, D. and {Schmoll}, J. and {Scholte}, D. and {Schubnell}, M. and {Secroun}, A. and {Seo}, H. and {Serrano}, S. and {Sharples}, Ray M. and {Sholl}, Michael J. and {Silber}, Joseph Harry and {Silva}, D.~R. and {Sirk}, M. and {Siudek}, M. and {Smith}, A. and {Sprayberry}, D. and {Staten}, R. and {Stupak}, B. and {Tan}, T. and {Tarl{\'e}}, Gregory and {Tie}, Suk Sien and {Tojeiro}, R. and {Ure{\~n}a-L{\'o}pez}, L.~A. and {Valdes}, F. and {Valenzuela}, O. and {Valluri}, M. and {Vargas-Maga{\~n}a}, M. and {Verde}, L. and {Walther}, M. and {Wang}, B. and {Wang}, M.~S. and {Weaver}, B.~A. and {Weaverdyck}, C. and {Wechsler}, R. and {Wilson}, Michael J. and {Yang}, J. and {Yu}, Y. and {Yuan}, S. and {Y{\`e}che}, Christophe and {Zhang}, H. and {Zhang}, K. and {Zhao}, Cheng and {Zhou}, Rongpu and {Zhou}, Zhimin and {Zou}, H. and {Zou}, J. and {Zou}, S. and {Zu}, Y. and {DESI Collaboration}},
        title = "{Overview of the Instrumentation for the Dark Energy Spectroscopic Instrument}",
      journal = {\aj},
         year = 2022,
        month = nov,
       volume = {164},
       number = {5},
          eid = {207},
        pages = {207},
          doi = {10.3847/1538-3881/ac882b},
archivePrefix = {arXiv},
       eprint = {2205.10939},
 primaryClass = {astro-ph.IM},
       adsurl = {https://ui.adsabs.harvard.edu/abs/2022AJ....164..207A}
}

@ARTICLE{DESI2024.I.DR1,
       author = {{DESI Collaboration} and {Abdul-Karim}, M  and {Adame}, A.~G. and {Aguado}, D. and {Aguilar}, J. and {Ahlen}, S. and {Alam}, S. and {Aldering}, G. and {Alexander}, D.~M. and {Alfarsy}, R. and {Allen}, L. and {Allende Prieto}, C. and {Alves}, O. and {Anand}, A. and {Andrade}, U. and {Armengaud}, E. and {Avila}, S. and {Aviles}, A. and {Awan}, H. and {Bailey}, S. and {Baleato Lizancos}, A. and {Ballester}, O. and {Bault}, A. and {Bautista}, J. and {BenZvi}, S. and {Beraldo e Silva}, L. and {Bermejo-Climent}, J.~R. and {Beutler}, F. and {Bianchi}, D. and {Blake}, C. and {Blum}, R. and {Bolton}, A.~S. and {Bonici}, M. and {Brieden}, S. and {Brodzeller}, A. and {Brooks}, D. and {Buckley-Geer}, E. and {Burtin}, E. and {Canning}, R. and {Carnero Rosell}, A. and {Carr}, A. and {Carrilho}, P. and {Casas}, L. and {Castander}, F.~J. and {Cereskaite}, R. and {Cervantes-Cota}, J.~L. and {Chaussidon}, E. and {Chaves-Montero}, J. and {Chen}, S. and {Chen}, X. and {Claybaugh}, T. and {Cole}, S. and {Cooper}, A.~P. and {Cousinou}, M. -C. and {Cuceu}, A. and {Davis}, T.~M. and {Dawson}, K.~S. and {de Belsunce}, R. and {de la Cruz}, R. and {de la Macorra}, A. and {de Mattia}, A. and {Deiosso}, N. and {Della Costa}, J. and {Demina}, R. and {Demirbozan}, U. and {DeRose}, J. and {Dey}, A. and {Dey}, B. and {Ding}, J. and {Ding}, Z. and {Doel}, P. and {Douglass}, K. and {Dowicz}, M. and {Ebina}, H. and {Edelstein}, J. and {Eisenstein}, D.~J. and {Elbers}, W. and {Emas}, N. and {Escoffier}, S. and {Fagrelius}, P. and {Fan}, X. and {Fanning}, K. and {Fawcett}, V.~A. and {Fern{\'a}ndez-Garc{\'\i}a}, E. and {Ferraro}, S. and {Findlay}, N. and {Font-Ribera}, A. and {Forero-Romero}, J.~E. and {Forero-S{\'a}nchez}, D. and {Frenk}, C.~S. and {G{\"a}nsicke}, B.~T. and {Galbany}, L. and {Garc{\'\i}a-Bellido}, J. and {Garcia-Quintero}, C. and {Garrison}, L.~H. and {Gazta{\~n}aga}, E. and {Gil-Mar{\'\i}n}, H. and {Gnedin}, O.~Y. and {Gontcho}, S. Gontcho A and {Gonzalez-Morales}, A.~X. and {Gonzalez-Perez}, V. and {Gordon}, C. and {Graur}, O. and {Green}, D. and {Gruen}, D. and {Gsponer}, R. and {Guandalin}, C. and {Gutierrez}, G. and {Guy}, J. and {Hahn}, C. and {Han}, J.~J. and {Han}, J. and {He}, S. and {Herrera-Alcantar}, H.~K. and {Honscheid}, K. and {Hou}, J. and {Howlett}, C. and {Huterer}, D. and {Ir{\v{s}}i{\v{c}}}, V. and {Ishak}, M. and {Jacques}, A. and {Jimenez}, J. and {Jing}, Y.~P. and {Joachimi}, B. and {Joudaki}, S. and {Joyce}, R. and {Jullo}, E. and {Juneau}, S. and {Kara{\c{c}}ayl{\i}}, N.~G. and {Karim}, T. and {Kehoe}, R. and {Kent}, S. and {Khederlarian}, A. and {Kirkby}, D. and {Kisner}, T. and {Kitaura}, F. -S. and {Kizhuprakkat}, N. and {Kong}, H. and {Koposov}, S.~E. and {Kremin}, A. and {Krolewski}, A. and {Lahav}, O. and {Lai}, Y. and {Lamman}, C. and {Lan}, T. -W. and {Landriau}, M. and {Lang}, D. and {Lange}, J.~U. and {Lasker}, J. and {Le Goff}, J.~M. and {Le Guillou}, L. and {Leauthaud}, A. and {Levi}, M.~E. and {Li}, S. and {Li}, T.~S. and {Lodha}, K. and {Lokken}, M. and {Luo}, Y. and {Magneville}, C. and {Manera}, M. and {Manser}, C.~J. and {Margala}, D. and {Martini}, P. and {Maus}, M. and {McCullough}, J. and {McDonald}, P. and {Medina}, G.~E. and {Medina-Varela}, L. and {Meisner}, A. and {Mena-Fern{\'a}ndez}, J. and {Menegas}, A. and {Mezcua}, M. and {Miquel}, R. and {Montero-Camacho}, P. and {Moon}, J. and {Moustakas}, J. and {Mu{\~n}oz-Guti{\'e}rrez}, A. and {Mu{\~n}oz-Santos}, D. and {Myers}, A.~D. and {Myles}, J. and {Nadathur}, S. and {Najita}, J. and {Napolitano}, L. and {Newman}, J.~A. and {Nikakhtar}, F. and {Nikutta}, R. and {Niz}, G. and {Noriega}, H.~E. and {Padmanabhan}, N. and {Paillas}, E. and {Palanque-Delabrouille}, N. and {Palmese}, A. and {Pan}, J. and {Pan}, Z. and {Parkinson}, D. and {Peacock}, J.~A.  and {Percival}, W.~J. and {P{\'e}rez-Fern{\'a}ndez}, A. and {P{\'e}rez-R{\`a}fols}, I. and {Peterson}, P.},
        title = "{Data Release 1 of the Dark Energy Spectroscopic Instrument}",
      journal = {arXiv e-prints},
         year = 2025,
        month = mar,
          eid = {arXiv:2503.14745},
        pages = {arXiv:2503.14745},
          doi = {10.48550/arXiv.2503.14745},
archivePrefix = {arXiv},
       eprint = {2503.14745},
 primaryClass = {astro-ph.CO},
       adsurl = {https://ui.adsabs.harvard.edu/abs/2025arXiv250314745D}
}

@ARTICLE{DESI2024.II.KP3,
       author = {{DESI Collaboration} and {Adame}, A.~G. and {Aguilar}, J. and {Ahlen}, S. and {Alam}, S. and {Alexander}, D.~M. and {Alvarez}, M. and {Alves}, O. and {Anand}, A. and {Andrade}, U. and {Armengaud}, E. and {Avila}, S. and {Aviles}, A. and {Awan}, H. and {Bailey}, S. and {Baltay}, C. and {Bault}, A. and {Behera}, J. and {BenZvi}, S. and {Beutler}, F. and {Bianchi}, D. and {Blake}, C. and {Blum}, R. and {Brieden}, S. and {Brodzeller}, A. and {Brooks}, D. and {Brown}, Z. and {Buckley-Geer}, E. and {Burtin}, E. and {Calderon}, R. and {Canning}, R. and {Carnero Rosell}, A. and {Cereskaite}, R. and {Cervantes-Cota}, J.~L. and {Chabanier}, S. and {Chaussidon}, E. and {Chaves-Montero}, J. and {Chen}, S. and {Chen}, X. and {Claybaugh}, T. and {Cole}, S. and {Cuceu}, A. and {Davis}, T.~M. and {Dawson}, K. and {de la Macorra}, A. and {de Mattia}, A. and {Deiosso}, N. and {Demina}, R. and {Dey}, A. and {Dey}, B. and {Ding}, Z. and {Doel}, P. and {Edelstein}, J. and {Eftekharzadeh}, S. and {Eisenstein}, D.~J. and {Elliott}, A. and {Fagrelius}, P. and {Fanning}, K. and {Ferraro}, S. and {Ereza}, J. and {Findlay}, N. and {Flaugher}, B. and {Font-Ribera}, A. and {Forero-S{\'a}nchez}, D. and {Forero-Romero}, J.~E. and {Frenk}, C.~S. and {Garcia-Quintero}, C. and {Gazta{\~n}aga}, E. and {Gil-Mar{\'\i}n}, H. and {Gontcho}, S. Gontcho A. and {Gonzalez-Morales}, A.~X. and {Gonzalez-Perez}, V. and {Gordon}, C. and {Green}, D. and {Gruen}, D. and {Gsponer}, R. and {Gutierrez}, G. and {Guy}, J. and {Hadzhiyska}, B. and {Hahn}, C. and {Hanif}, M.~M.~S. and {Herrera-Alcantar}, H.~K. and {Honscheid}, K. and {Hou}, J. and {Howlett}, C. and {Huterer}, D. and {Ir{\v{s}}i{\v{c}}}, V. and {Ishak}, M. and {Juneau}, S. and {Kara{\c{c}}ayl{\i}}, N.~G. and {Kehoe}, R. and {Kent}, S. and {Kirkby}, D. and {Kitaura}, F. -S. and {Kong}, H. and {Kremin}, A. and {Krolewski}, A. and {Lai}, Y. and {Lan}, T. -W. and {Landriau}, M. and {Lang}, D. and {Lasker}, J. and {Le Goff}, J.~M. and {Le Guillou}, L. and {Leauthaud}, A. and {Levi}, M.~E. and {Li}, T.~S. and {Lodha}, K. and {Magneville}, C. and {Manera}, M. and {Margala}, D. and {Martini}, P. and {Maus}, M. and {McDonald}, P. and {Medina-Varela}, L. and {Meisner}, A. and {Mena-Fern{\'a}ndez}, J. and {Miquel}, R. and {Moon}, J. and {Moore}, S. and {Moustakas}, J. and {Mudur}, N. and {Mueller}, E. and {Mu{\~n}oz-Guti{\'e}rrez}, A. and {Myers}, A.~D. and {Nadathur}, S. and {Napolitano}, L. and {Neveux}, R. and {Newman}, J.~A. and {Nguyen}, N.~M. and {Nie}, J. and {Niz}, G. and {Noriega}, H.~E. and {Padmanabhan}, N. and {Paillas}, E. and {Palanque-Delabrouille}, N. and {Pan}, J. and {Penmetsa}, S. and {Percival}, W.~J. and {Pieri}, M.~M. and {Pinon}, M. and {Poppett}, C. and {Porredon}, A. and {Prada}, F. and {P{\'e}rez-Fern{\'a}ndez}, A. and {P{\'e}rez-R{\`a}fols}, I. and {Rabinowitz}, D. and {Raichoor}, A. and {Ram{\'\i}rez-P{\'e}rez}, C. and {Ramirez-Solano}, S. and {Rashkovetskyi}, M. and {Ravoux}, C. and {Rezaie}, M. and {Rich}, J. and {Rocher}, A. and {Rockosi}, C. and {Roe}, N.~A. and {Rosado-Marin}, A. and {Ross}, A.~J. and {Rossi}, G. and {Ruggeri}, R. and {Ruhlmann-Kleider}, V. and {Samushia}, L. and {Sanchez}, E. and {Saulder}, C. and {Schlafly}, E.~F. and {Schlegel}, D. and {Scholte}, D. and {Schubnell}, M. and {Seo}, H. and {Sharples}, R. and {Silber}, J. and {Slosar}, A. and {Smith}, A. and {Sprayberry}, D. and {Tan}, T. and {Tarl{\'e}}, G. and {Trusov}, S. and {Vaisakh}, R. and {Valcin}, D. and {Valdes}, F. and {Vargas-Maga{\~n}a}, M. and {Verde}, L. and {Walther}, M. and {Wang}, B. and {Wang}, M.~S. and {Weaver}, B.~A. and {Weaverdyck}, N. and {Wechsler}, R.~H. and {Weinberg}, D.~H. and {White}, M. and {Wilson}, M.~J. and {Yu}, J. and {Yu}, Y. and {Yuan}, S. and {Y{\`e}che}, C. and {Zaborowski}, E.~A. and {Zarrouk}, P. and {Zhang}, H. and {Zhao}, C. and {Zhao}, R.},
        title = "{DESI 2024 II: sample definitions, characteristics, and two-point clustering statistics}",
      journal = {\jcap},
         year = 2025,
        month = jul,
       volume = {2025},
       number = {7},
          eid = {017},
        pages = {017},
          doi = {10.1088/1475-7516/2025/07/017},
archivePrefix = {arXiv},
       eprint = {2411.12020},
 primaryClass = {astro-ph.CO},
       adsurl = {https://ui.adsabs.harvard.edu/abs/2025JCAP...07..017A}
}

@ARTICLE{DESI2024.III.KP4,
       author = {{DESI Collaboration} and {Adame}, A.~G. and {Aguilar}, J. and {Ahlen}, S. and {Alam}, S. and {Alexander}, D.~M. and {Alvarez}, M. and {Alves}, O. and {Anand}, A. and {Andrade}, U. and {Armengaud}, E. and {Avila}, S. and {Aviles}, A. and {Awan}, H. and {Bailey}, S. and {Baltay}, C. and {Bault}, A. and {Behera}, J. and {BenZvi}, S. and {Beutler}, F. and {Bianchi}, D. and {Blake}, C. and {Blum}, R. and {Brieden}, S. and {Brodzeller}, A. and {Brooks}, D. and {Buckley-Geer}, E. and {Burtin}, E. and {Calderon}, R. and {Canning}, R. and {Carnero Rosell}, A. and {Cereskaite}, R. and {Cervantes-Cota}, J.~L. and {Chabanier}, S. and {Chaussidon}, E. and {Chaves-Montero}, J. and {Chen}, S. and {Chen}, X. and {Claybaugh}, T. and {Cole}, S. and {Cuceu}, A. and {Davis}, T.~M. and {Dawson}, K. and {de la Macorra}, A. and {de Mattia}, A. and {Deiosso}, N. and {Dey}, A. and {Dey}, B. and {Ding}, Z. and {Doel}, P. and {Edelstein}, J. and {Eftekharzadeh}, S. and {Eisenstein}, D.~J. and {Elliott}, A. and {Fagrelius}, P. and {Fanning}, K. and {Ferraro}, S. and {Ereza}, J. and {Findlay}, N. and {Flaugher}, B. and {Font-Ribera}, A. and {Forero-S{\'a}nchez}, D. and {Forero-Romero}, J.~E. and {Garcia-Quintero}, C. and {Gazta{\~n}aga}, E. and {Gil-Mar{\'\i}n}, H. and {Gontcho a Gontcho}, S. and {Gonzalez-Morales}, A.~X. and {Gonzalez-Perez}, V. and {Gordon}, C. and {Green}, D. and {Gruen}, D. and {Gsponer}, R. and {Gutierrez}, G. and {Guy}, J. and {Hadzhiyska}, B. and {Hahn}, C. and {Hanif}, M.~M.~S. and {Herrera-Alcantar}, H.~K. and {Honscheid}, K. and {Howlett}, C. and {Huterer}, D. and {Ir{\v{s}}i{\v{c}}}, V. and {Ishak}, M. and {Juneau}, S. and {Kara{\c{c}}ayl{\i}}, N.~G. and {Kehoe}, R. and {Kent}, S. and {Kirkby}, D. and {Kong}, H. and {Kremin}, A. and {Krolewski}, A. and {Lai}, Y. and {Lan}, T. -W. and {Landriau}, M. and {Lang}, D. and {Lasker}, J. and {Le Goff}, J.~M. and {Le Guillou}, L. and {Leauthaud}, A. and {Levi}, M.~E. and {Li}, T.~S. and {Linder}, E. and {Lodha}, K. and {Magneville}, C. and {Manera}, M. and {Margala}, D. and {Martini}, P. and {Maus}, M. and {McDonald}, P. and {Medina-Varela}, L. and {Meisner}, A. and {Mena-Fern{\'a}ndez}, J. and {Miquel}, R. and {Moon}, J. and {Moore}, S. and {Moustakas}, J. and {Mueller}, E. and {Mu{\~n}oz-Guti{\'e}rrez}, A. and {Myers}, A.~D. and {Nadathur}, S. and {Napolitano}, L. and {Neveux}, R. and {Newman}, J.~A. and {Nguyen}, N.~M. and {Nie}, J. and {Niz}, G. and {Noriega}, H.~E. and {Padmanabhan}, N. and {Paillas}, E. and {Palanque-Delabrouille}, N. and {Pan}, J. and {Penmetsa}, S. and {Percival}, W.~J. and {Pieri}, M.~M. and {Pinon}, M. and {Poppett}, C. and {Porredon}, A. and {Prada}, F. and {P{\'e}rez-Fern{\'a}ndez}, A. and {P{\'e}rez-R{\`a}fols}, I. and {Rabinowitz}, D. and {Raichoor}, A. and {Ram{\'\i}rez-P{\'e}rez}, C. and {Ramirez-Solano}, S. and {Rashkovetskyi}, M. and {Ravoux}, C. and {Rezaie}, M. and {Rich}, J. and {Rocher}, A. and {Rockosi}, C. and {Roe}, N.~A. and {Rosado-Marin}, A. and {Ross}, A.~J. and {Rossi}, G. and {Ruggeri}, R. and {Ruhlmann-Kleider}, V. and {Samushia}, L. and {Sanchez}, E. and {Saulder}, C. and {Schlafly}, E.~F. and {Schlegel}, D. and {Schubnell}, M. and {Seo}, H. and {Sharples}, R. and {Silber}, J. and {Slosar}, A. and {Smith}, A. and {Sprayberry}, D. and {Swanson}, J. and {Tan}, T. and {Tarl{\'e}}, G. and {Trusov}, S. and {Vaisakh}, R. and {Valcin}, D. and {Valdes}, F. and {Vargas-Maga{\~n}a}, M. and {Verde}, L. and {Walther}, M. and {Wang}, B. and {Wang}, M.~S. and {Weaver}, B.~A. and {Weaverdyck}, N. and {Wechsler}, R.~H. and {Weinberg}, D.~H. and {White}, M. and {Wilson}, M.~J. and {Yu}, J. and {Yu}, Y. and {Yuan}, S. and {Y{\`e}che}, C. and {Zaborowski}, E.~A. and {Zarrouk}, P. and {Zhang}, H. and {Zhao}, C. and {Zhao}, R. and {Zhou}, R. and {Zou}, H. and {DESI Collaboration}},
        title = "{DESI 2024 III: baryon acoustic oscillations from galaxies and quasars}",
      journal = {\jcap},
         year = 2025,
        month = apr,
       volume = {2025},
       number = {4},
          eid = {012},
        pages = {012},
          doi = {10.1088/1475-7516/2025/04/012},
archivePrefix = {arXiv},
       eprint = {2404.03000},
 primaryClass = {astro-ph.CO},
       adsurl = {https://ui.adsabs.harvard.edu/abs/2025JCAP...04..012A}
}

@ARTICLE{DESI2024.VI.KP7A,
       author = {{DESI Collaboration} and {Adame}, A.~G. and {Aguilar}, J. and {Ahlen}, S. and {Alam}, S. and {Alexander}, D.~M. and {Alvarez}, M. and {Alves}, O. and {Anand}, A. and {Andrade}, U. and {Armengaud}, E. and {Avila}, S. and {Aviles}, A. and {Awan}, H. and {Bahr-Kalus}, B. and {Bailey}, S. and {Baltay}, C. and {Bault}, A. and {Behera}, J. and {BenZvi}, S. and {Bera}, A. and {Beutler}, F. and {Bianchi}, D. and {Blake}, C. and {Blum}, R. and {Brieden}, S. and {Brodzeller}, A. and {Brooks}, D. and {Buckley-Geer}, E. and {Burtin}, E. and {Calderon}, R. and {Canning}, R. and {Carnero Rosell}, A. and {Cereskaite}, R. and {Cervantes-Cota}, J.~L. and {Chabanier}, S. and {Chaussidon}, E. and {Chaves-Montero}, J. and {Chen}, S. and {Chen}, X. and {Claybaugh}, T. and {Cole}, S. and {Cuceu}, A. and {Davis}, T.~M. and {Dawson}, K. and {de la Macorra}, A. and {de Mattia}, A. and {Deiosso}, N. and {Dey}, A. and {Dey}, B. and {Ding}, Z. and {Doel}, P. and {Edelstein}, J. and {Eftekharzadeh}, S. and {Eisenstein}, D.~J. and {Elliott}, A. and {Fagrelius}, P. and {Fanning}, K. and {Ferraro}, S. and {Ereza}, J. and {Findlay}, N. and {Flaugher}, B. and {Font-Ribera}, A. and {Forero-S{\'a}nchez}, D. and {Forero-Romero}, J.~E. and {Frenk}, C.~S. and {Garcia-Quintero}, C. and {Gazta{\~n}aga}, E. and {Gil-Mar{\'\i}n}, H. and {Gontcho a Gontcho}, S. and {Gonzalez-Morales}, A.~X. and {Gonzalez-Perez}, V. and {Gordon}, C. and {Green}, D. and {Gruen}, D. and {Gsponer}, R. and {Gutierrez}, G. and {Guy}, J. and {Hadzhiyska}, B. and {Hahn}, C. and {Hanif}, M.~M.~S. and {Herrera-Alcantar}, H.~K. and {Honscheid}, K. and {Howlett}, C. and {Huterer}, D. and {Ir{\v{s}}i{\v{c}}}, V. and {Ishak}, M. and {Juneau}, S. and {Kara{\c{c}}ayl{\i}}, N.~G. and {Kehoe}, R. and {Kent}, S. and {Kirkby}, D. and {Kremin}, A. and {Krolewski}, A. and {Lai}, Y. and {Lan}, T. -W. and {Landriau}, M. and {Lang}, D. and {Lasker}, J. and {Le Goff}, J.~M. and {Le Guillou}, L. and {Leauthaud}, A. and {Levi}, M.~E. and {Li}, T.~S. and {Linder}, E. and {Lodha}, K. and {Magneville}, C. and {Manera}, M. and {Margala}, D. and {Martini}, P. and {Maus}, M. and {McDonald}, P. and {Medina-Varela}, L. and {Meisner}, A. and {Mena-Fern{\'a}ndez}, J. and {Miquel}, R. and {Moon}, J. and {Moore}, S. and {Moustakas}, J. and {Mueller}, E. and {Mu{\~n}oz-Guti{\'e}rrez}, A. and {Myers}, A.~D. and {Nadathur}, S. and {Napolitano}, L. and {Neveux}, R. and {Newman}, J.~A. and {Nguyen}, N.~M. and {Nie}, J. and {Niz}, G. and {Noriega}, H.~E. and {Padmanabhan}, N. and {Paillas}, E. and {Palanque-Delabrouille}, N. and {Pan}, J. and {Penmetsa}, S. and {Percival}, W.~J. and {Pieri}, M.~M. and {Pinon}, M. and {Poppett}, C. and {Porredon}, A. and {Prada}, F. and {P{\'e}rez-Fern{\'a}ndez}, A. and {P{\'e}rez-R{\`a}fols}, I. and {Rabinowitz}, D. and {Raichoor}, A. and {Ram{\'\i}rez-P{\'e}rez}, C. and {Ramirez-Solano}, S. and {Rashkovetskyi}, M. and {Ravoux}, C. and {Rezaie}, M. and {Rich}, J. and {Rocher}, A. and {Rockosi}, C. and {Roe}, N.~A. and {Rosado-Marin}, A. and {Ross}, A.~J. and {Rossi}, G. and {Ruggeri}, R. and {Ruhlmann-Kleider}, V. and {Samushia}, L. and {Sanchez}, E. and {Saulder}, C. and {Schlafly}, E.~F. and {Schlegel}, D. and {Schubnell}, M. and {Seo}, H. and {Shafieloo}, A. and {Sharples}, R. and {Silber}, J. and {Slosar}, A. and {Smith}, A. and {Sprayberry}, D. and {Tan}, T. and {Tarl{\'e}}, G. and {Taylor}, P. and {Trusov}, S. and {Ure{\~n}a-L{\'o}pez}, L.~A. and {Vaisakh}, R. and {Valcin}, D. and {Valdes}, F. and {Vargas-Maga{\~n}a}, M. and {Verde}, L. and {Walther}, M. and {Wang}, B. and {Wang}, M.~S. and {Weaver}, B.~A. and {Weaverdyck}, N. and {Wechsler}, R.~H. and {Weinberg}, D.~H. and {White}, M. and {Yu}, J. and {Yu}, Y. and {Yuan}, S. and {Y{\`e}che}, C. and {Zaborowski}, E.~A. and {Zarrouk}, P. and {Zhang}, H. and {Zhao}, C. and {Zhao}, R. and {Zhou}, R. and {Zhuang}, T.},
        title = "{DESI 2024 VI: cosmological constraints from the measurements of baryon acoustic oscillations}",
      journal = {\jcap},
         year = 2025,
        month = feb,
       volume = {2025},
       number = {2},
          eid = {021},
        pages = {021},
          doi = {10.1088/1475-7516/2025/02/021},
archivePrefix = {arXiv},
       eprint = {2404.03002},
 primaryClass = {astro-ph.CO},
       adsurl = {https://ui.adsabs.harvard.edu/abs/2025JCAP...02..021A}
}

@ARTICLE{DESI2024.VII.KP7B,
       author = {{DESI Collaboration} and {Adame}, A.~G. and {Aguilar}, J. and {Ahlen}, S. and {Alam}, S. and {Alexander}, D.~M. and {Allende Prieto}, C. and {Alvarez}, M. and {Alves}, O. and {Anand}, A. and {Andrade}, U. and {Armengaud}, E. and {Avila}, S. and {Aviles}, A. and {Awan}, H. and {Bahr-Kalus}, B. and {Bailey}, S. and {Baltay}, C. and {Bault}, A. and {Behera}, J. and {BenZvi}, S. and {Beutler}, F. and {Bianchi}, D. and {Blake}, C. and {Blum}, R. and {Bonici}, M. and {Brieden}, S. and {Brodzeller}, A. and {Brooks}, D. and {Buckley-Geer}, E. and {Burtin}, E. and {Calderon}, R. and {Canning}, R. and {Carnero Rosell}, A. and {Cereskaite}, R. and {Cervantes-Cota}, J.~L. and {Chabanier}, S. and {Chaussidon}, E. and {Chaves-Montero}, J. and {Chebat}, D. and {Chen}, S. and {Chen}, X. and {Claybaugh}, T. and {Cole}, S. and {Cuceu}, A. and {Davis}, T.~M. and {Dawson}, K. and {de la Macorra}, A. and {de Mattia}, A. and {Deiosso}, N. and {Dey}, A. and {Dey}, B. and {Ding}, Z. and {Doel}, P. and {Edelstein}, J. and {Eftekharzadeh}, S. and {Eisenstein}, D.~J. and {Elbers}, W. and {Elliott}, A. and {Fagrelius}, P. and {Fanning}, K. and {Ferraro}, S. and {Ereza}, J. and {Findlay}, N. and {Flaugher}, B. and {Font-Ribera}, A. and {Forero-S{\'a}nchez}, D. and {Forero-Romero}, J.~E. and {Frenk}, C.~S. and {Garcia-Quintero}, C. and {Garrison}, L.~H. and {Gazta{\~n}aga}, E. and {Gil-Mar{\'\i}n}, H. and {Gontcho}, S. Gontcho A. and {Gonzalez-Morales}, A.~X. and {Gonzalez-Perez}, V. and {Gordon}, C. and {Green}, D. and {Gruen}, D. and {Gsponer}, R. and {Gutierrez}, G. and {Guy}, J. and {Hadzhiyska}, B. and {Hahn}, C. and {Hanif}, M.~M.~S. and {Herrera-Alcantar}, H.~K. and {Honscheid}, K. and {Howlett}, C. and {Huterer}, D. and {Ir{\v{s}}i{\v{c}}}, V. and {Ishak}, M. and {Joyce}, R. and {Juneau}, S. and {Kara{\c{c}}ayl{\i}}, N.~G. and {Kehoe}, R. and {Kent}, S. and {Kirkby}, D. and {Kong}, H. and {Koposov}, S.~E. and {Kremin}, A. and {Krolewski}, A. and {Lahav}, O. and {Lai}, Y. and {Lan}, T. -W. and {Landriau}, M. and {Lang}, D. and {Lasker}, J. and {Le Goff}, J.~M. and {Le Guillou}, L. and {Leauthaud}, A. and {Levi}, M.~E. and {Li}, T.~S. and {Lodha}, K. and {Magneville}, C. and {Manera}, M. and {Margala}, D. and {Martini}, P. and {Matthewson}, W. and {Maus}, M. and {McDonald}, P. and {Medina-Varela}, L. and {Meisner}, A. and {Mena-Fern{\'a}ndez}, J. and {Miquel}, R. and {Moon}, J. and {Moore}, S. and {Moustakas}, J. and {Mudur}, N. and {Mueller}, E. and {Mu{\~n}oz-Guti{\'e}rrez}, A. and {Myers}, A.~D. and {Nadathur}, S. and {Napolitano}, L. and {Neveux}, R. and {Newman}, J.~A. and {Nguyen}, N.~M. and {Nie}, J. and {Niz}, G. and {Noriega}, H.~E. and {Padmanabhan}, N. and {Paillas}, E. and {Palanque-Delabrouille}, N. and {Pan}, J. and {Penmetsa}, S. and {Percival}, W.~J. and {Pieri}, M.~M. and {Pinon}, M. and {Poppett}, C. and {Porredon}, A. and {Prada}, F. and {P{\'e}rez-Fern{\'a}ndez}, A. and {P{\'e}rez-R{\`a}fols}, I. and {Rabinowitz}, D. and {Raichoor}, A. and {Ram{\'\i}rez-P{\'e}rez}, C. and {Ramirez-Solano}, S. and {Rashkovetskyi}, M. and {Ravoux}, C. and {Rezaie}, M. and {Rich}, J. and {Rocher}, A. and {Rockosi}, C. and {Roe}, N.~A. and {Rosado-Marin}, A. and {Ross}, A.~J. and {Rossi}, G. and {Ruggeri}, R. and {Ruhlmann-Kleider}, V. and {Samushia}, L. and {Sanchez}, E. and {Saulder}, C. and {Schlafly}, E.~F. and {Schlegel}, D. and {Schubnell}, M. and {Seo}, H. and {Shafieloo}, A. and {Sharples}, R. and {Silber}, J. and {Slosar}, A. and {Smith}, A. and {Sprayberry}, D. and {Tan}, T. and {Tarl{\'e}}, G. and {Taylor}, P. and {Trusov}, S. and {Vaisakh}, R. and {Valcin}, D. and {Valdes}, F. and {Valogiannis}, G. and {Vargas-Maga{\~n}a}, M. and {Verde}, L. and {Walther}, M. and {Wang}, B. and {Wang}, M.~S. and {Weaver}, B.~A. and {Weaverdyck}, N. and {Wechsler}, R.~H. and {Weinberg}, D.~H. and {White}, M. and {Wilson}, M.~J. and {Yi}, L.},
        title = "{DESI 2024 VII: cosmological constraints from the full-shape modeling of clustering measurements}",
      journal = {\jcap},
         year = 2025,
        month = jul,
       volume = {2025},
       number = {7},
          eid = {028},
        pages = {028},
          doi = {10.1088/1475-7516/2025/07/028},
archivePrefix = {arXiv},
       eprint = {2411.12022},
 primaryClass = {astro-ph.CO},
       adsurl = {https://ui.adsabs.harvard.edu/abs/2025JCAP...07..028A}
}

@ARTICLE{KP3s7-Lasker,
       author = {{Lasker}, J. and {Carnero Rosell}, A. and {Myers}, A.~D. and {Ross}, A.~J. and {Bianchi}, D. and {Hanif}, M.~M.~S. and {Kehoe}, R. and {de Mattia}, A. and {Napolitano}, L. and {Percival}, W.~J. and {Staten}, R. and {Aguilar}, J. and {Ahlen}, S. and {Bigwood}, L. and {Brooks}, D. and {Claybaugh}, T. and {Cole}, S. and {de la Macorra}, A. and {Ding}, Z. and {Doel}, P. and {Fanning}, K. and {Forero-Romero}, J.~E. and {Gazta{\~n}aga}, E. and {Gontcho A Gontcho}, S. and {Gutierrez}, G. and {Honscheid}, K. and {Howlett}, C. and {Juneau}, S. and {Kremin}, A. and {Landriau}, M. and {Le Guillou}, L. and {Levi}, M.~E. and {Manera}, M. and {Meisner}, A. and {Miquel}, R. and {Moustakas}, J. and {Mueller}, E. and {Nie}, J. and {Niz}, G. and {Oh}, M. and {Palanque-Delabrouille}, N. and {Poppett}, C. and {Prada}, F. and {Rezaie}, M. and {Rossi}, G. and {Sanchez}, E. and {Schlegel}, D. and {Schubnell}, M. and {Seo}, H. and {Sprayberry}, D. and {Tarl{\'e}}, G. and {Vargas-Maga{\~n}a}, M. and {Weaver}, B.~A. and {Wilson}, Michael J. and {Zheng}, Y. and {DESI Collaboration}},
        title = "{Production of alternate realizations of DESI fiber assignment for unbiased clustering measurement in data and simulations}",
      journal = {\jcap},
         year = 2025,
        month = jan,
       volume = {2025},
       number = {1},
          eid = {127},
        pages = {127},
          doi = {10.1088/1475-7516/2025/01/127},
archivePrefix = {arXiv},
       eprint = {2404.03006},
 primaryClass = {astro-ph.CO},
       adsurl = {https://ui.adsabs.harvard.edu/abs/2025JCAP...01..127L}
}

@ARTICLE{2023MNRAS.524.3894R,
       author = {{Rashkovetskyi}, Michael and {Eisenstein}, Daniel J. and {Aguilar}, Jessica Nicole and {Brooks}, David and {Claybaugh}, Todd and {Cole}, Shaun and {Dawson}, Kyle and {de la Macorra}, Axel and {Doel}, Peter and {Fanning}, Kevin and {Font-Ribera}, Andreu and {Forero-Romero}, Jaime E. and {Gontcho A Gontcho}, Satya and {Hahn}, ChangHoon and {Honscheid}, Klaus and {Kehoe}, Robert and {Kisner}, Theodore and {Landriau}, Martin and {Levi}, Michael and {Manera}, Marc and {Miquel}, Ramon and {Moon}, Jeongin and {Nadathur}, Seshadri and {Nie}, Jundan and {Poppett}, Claire and {Ross}, Ashley J. and {Rossi}, Graziano and {Sanchez}, Eusebio and {Saulder}, Christoph and {Schubnell}, Michael and {Seo}, Hee-Jong and {Tarle}, Gregory and {Valcin}, David and {Weaver}, Benjamin Alan and {Zhao}, Cheng and {Zhou}, Zhimin and {Zou}, Hu},
        title = "{Validation of semi-analytical, semi-empirical covariance matrices for two-point correlation function for early DESI data}",
      journal = {\mnras},
         year = 2023,
        month = sep,
       volume = {524},
       number = {3},
        pages = {3894-3911},
          doi = {10.1093/mnras/stad2078},
archivePrefix = {arXiv},
       eprint = {2306.06320},
 primaryClass = {astro-ph.CO},
       adsurl = {https://ui.adsabs.harvard.edu/abs/2023MNRAS.524.3894R}
}

@ARTICLE{KP4s2-Chen,
       author = {{Chen}, S. -F. and {Howlett}, C. and {White}, M. and {McDonald}, P. and {Ross}, A.~J. and {Seo}, H. -J. and {Padmanabhan}, N. and {Aguilar}, J. and {Ahlen}, S. and {Alam}, S. and {Alves}, O. and {Andrade}, U. and {Blum}, R. and {Brooks}, D. and {Chen}, X. and {Cole}, S. and {Dawson}, K. and {de la Macorra}, A. and {Dey}, A. and {Ding}, Z. and {Doel}, P. and {Ferraro}, S. and {Font-Ribera}, A. and {Forero-S{\'a}nchez}, D. and {Forero-Romero}, J.~E. and {Garcia-Quintero}, C. and {Gazta{\~n}aga}, E. and {Gontcho}, S.~G.~A. and {Hanif}, M.~M.~S. and {Honscheid}, K. and {Kisner}, T. and {Kremin}, A. and {Lambert}, A. and {Landriau}, M. and {Levi}, M.~E. and {Manera}, M. and {Meisner}, A. and {Mena-Fern{\'a}ndez}, J. and {Miquel}, R. and {Munoz-Gutierrez}, A. and {Paillas}, E. and {Palanque-Delabrouille}, N. and {Percival}, W.~J. and {P{\'e}rez-Fern{\'a}ndez}, A. and {Prada}, F. and {Rashkovetskyi}, M. and {Rezaie}, M. and {Rosado-Marin}, A. and {Rossi}, G. and {Ruggeri}, R. and {Sanchez}, E. and {Schlegel}, D. and {Silber}, J. and {Tarl{\'e}}, G. and {Vargas-Maga{\~n}a}, M. and {Weaver}, B.~A. and {Yu}, J. and {Yuan}, S. and {Zhou}, R. and {Zhou}, Z.},
        title = "{Baryon acoustic oscillation theory and modelling systematics for the DESI 2024 results}",
      journal = {\mnras},
         year = 2024,
        month = oct,
       volume = {534},
       number = {1},
        pages = {544-574},
          doi = {10.1093/mnras/stae2090},
archivePrefix = {arXiv},
       eprint = {2402.14070},
 primaryClass = {astro-ph.CO},
       adsurl = {https://ui.adsabs.harvard.edu/abs/2024MNRAS.534..544C}
}

@ARTICLE{KP4s3-Chen,
       author = {{Chen}, X. and {Ding}, Z. and {Paillas}, E. and {Nadathur}, S. and {Seo}, H. and {Chen}, S. and {Padmanabhan}, N. and {White}, M. and {de Mattia}, A. and {McDonald}, P. and {Ross}, A.~J. and {Variu}, A. and {Carnero Rosell}, A. and {Hadzhiyska}, B. and {Hanif}, M.~M. S and {Forero-S{\'a}nchez}, D. and {Ahlen}, S. and {Alves}, O. and {Andrade}, U. and {BenZvi}, S. and {Bianchi}, D. and {Brooks}, D. and {Chaussidon}, E. and {Claybaugh}, T. and {de la Macorra}, A. and {Dey}, Biprateep and {Fanning}, K. and {Ferraro}, S. and {Font-Ribera}, A. and {Forero-Romero}, J.~E. and {Garcia-Quintero}, C. and {Gazta{\~n}aga}, E. and {Gontcho}, S. Gontcho A and {Gutierrez}, G. and {Hahn}, C. and {Honscheid}, K. and {Juneau}, S. and {Kehoe}, R. and {Kirkby}, D. and {Kisner}, T. and {Kremin}, A. and {Levi}, M.~E. and {Meisner}, A. and {Mena-Fern{\'a}ndez}, J. and {Miquel}, R. and {Moustakas}, J. and {Mu{\~n}oz-Guti{\'e}rrez}, A. and {Nikakhtar}, F. and {Palanque-Delabrouille}, N. and {Percival}, W.~J. and {Prada}, F. and {P{\'e}rez-R{\`a}fols}, I. and {Rashkovetskyi}, M. and {Rossi}, G. and {Ruggeri}, R. and {Sanchez}, E. and {Saulder}, C. and {Schlegel}, D. and {Schubnell}, M. and {Smith}, A. and {Sprayberry}, D. and {Tarl{\'e}}, G. and {Valcin}, D. and {Vargas-Maga{\~n}a}, M. and {Weaver}, B.~A. and {Yuan}, S. and {Zhou}, R.},
        title = "{Extensive analysis of reconstruction algorithms for DESI 2024 baryon acoustic oscillations}",
      journal = {arXiv e-prints},
         year = 2024,
        month = nov,
          eid = {arXiv:2411.19738},
        pages = {arXiv:2411.19738},
          doi = {10.48550/arXiv.2411.19738},
archivePrefix = {arXiv},
       eprint = {2411.19738},
 primaryClass = {astro-ph.CO},
       adsurl = {https://ui.adsabs.harvard.edu/abs/2024arXiv241119738C}
}

@ARTICLE{KP4s4-Paillas,
       author = {{Paillas}, E. and {Ding}, Z. and {Chen}, X. and {Seo}, H. and {Padmanabhan}, N. and {de Mattia}, A. and {Ross}, A.~J. and {Nadathur}, S. and {Howlett}, C. and {Aguilar}, J. and {Ahlen}, S. and {Alves}, O. and {Andrade}, U. and {Brooks}, D. and {Buckley-Geer}, E. and {Burtin}, E. and {Chen}, S. and {Claybaugh}, T. and {Cole}, S. and {Dawson}, K. and {de la Macorra}, A. and {Dey}, Arjun and {Doel}, P. and {Fanning}, K. and {Ferraro}, S. and {Forero-Romero}, J.~E. and {Garcia-Quintero}, C. and {Gazta{\~n}aga}, E. and {Gil-Mar{\'\i}n}, H. and {Gontcho}, S. Gontcho A. and {Gutierrez}, G. and {Hahn}, C. and {Hanif}, M.~M.~S. and {Honscheid}, K. and {Ishak}, M. and {Kehoe}, R. and {Kremin}, A. and {Landriau}, M. and {Le Guillou}, L. and {Levi}, M.~E. and {Manera}, M. and {Martini}, P. and {Medina-Varela}, L. and {Meisner}, A. and {Mena-Fern{\'a}ndez}, J. and {Miquel}, R. and {Moustakas}, J. and {Mueller}, E. and {Mu{\~n}oz-Guti{\'e}rrez}, A. and {Myers}, A.~D. and {Newman}, J.~A. and {Nie}, J. and {Niz}, G. and {Palanque-Delabrouille}, N. and {Percival}, W.~J. and {Poppett}, C. and {Prada}, F. and {P{\'e}rez-Fern{\'a}ndez}, A. and {Rashkovetskyi}, M. and {Rezaie}, M. and {Rosado-Marin}, A. and {Rossi}, G. and {Ruggeri}, R. and {Sanchez}, E. and {Saulder}, C. and {Schlafly}, E.~F. and {Schlegel}, D. and {Schubnell}, M. and {Sprayberry}, D. and {Tarl{\'e}}, G. and {Valcin}, D. and {Vargas-Maga{\~n}a}, M. and {Yu}, J. and {Yuan}, S. and {Zhou}, R. and {Zou}, H.},
        title = "{Optimal reconstruction of baryon acoustic oscillations for DESI 2024}",
      journal = {\jcap},
         year = 2025,
        month = jan,
       volume = {2025},
       number = {1},
          eid = {142},
        pages = {142},
          doi = {10.1088/1475-7516/2025/01/142},
archivePrefix = {arXiv},
       eprint = {2404.03005},
 primaryClass = {astro-ph.CO},
       adsurl = {https://ui.adsabs.harvard.edu/abs/2025JCAP...01..142P}
}

@article{KP4s5-Valcin,
doi = {10.1088/1475-7516/2026/04/058},
url = {https://doi.org/10.1088/1475-7516/2026/04/058},
year = {2026},
month = {apr},
publisher = {IOP Publishing},
volume = {2026},
number = {04},
pages = {058},
author = {Valcin, D. and Rashkovetskyi, M. and Seo, H. and Beutler, F. and McDonald, P. and de Mattia, A. and Rosado-Marín, A.J. and Ross, A.J. and Padmanabhan, N. and Aguilar, J. and Ahlen, S. and Andrade, U. and Bianchi, D. and Brooks, D. and Chaussidon, E. and Chen, S. and Chen, X. and Claybaugh, T. and Cuceu, A. and Dawson, K.S. and de la Macorra, A. and Dey, Biprateep and Ding, Z. and Doel, P. and Ferraro, S. and Font-Ribera, A. and Forero-Romero, J.E. and Gaztañaga, E. and Gontcho A Gontcho, S. and Gutierrez, G. and Hahn, C. and Honscheid, K. and Howlett, C. and Ishak, M. and Kehoe, R. and Kirkby, D. and Kisner, T. and Kremin, A. and Lahav, O. and Lambert, A. and Landriau, M. and Levi, M.E. and Manera, M. and Meisner, A. and Mena-Fernández, J. and Miquel, R. and Moustakas, J. and Nadathur, S. and Paillas, E. and Palanque-Delabrouille, N. and Percival, W.J. and Prada, F. and Pérez-Ràfols, I. and Rossi, G. and Ruggeri, R. and Samushia, L. and Sanchez, E. and Saulder, C. and Schlegel, D. and Schubnell, M. and Silber, J. and Sprayberry, D. and Tarlé, G. and Weaver, B.A. and Yu, J. and Zhou, R. and Zou, H.},
title = {Combined tracer analysis for DESI 2024 BAO},
journal = {Journal of Cosmology and Astroparticle Physics}
}

@ARTICLE{KP4s7-Rashkovetskyi,
       author = {{Rashkovetskyi}, M. and {Forero-S{\'a}nchez}, D. and {de Mattia}, A. and {Eisenstein}, D.~J. and {Padmanabhan}, N. and {Seo}, H. and {Ross}, A.~J. and {Aguilar}, J. and {Ahlen}, S. and {Alves}, O. and {Andrade}, U. and {Brooks}, D. and {Burtin}, E. and {Chen}, X. and {Claybaugh}, T. and {Cole}, S. and {de la Macorra}, A. and {Ding}, Z. and {Doel}, P. and {Fanning}, K. and {Ferraro}, S. and {Font-Ribera}, A. and {Forero-Romero}, J.~E. and {Garcia-Quintero}, C. and {Gil-Mar{\'\i}n}, H. and {Gontcho A Gontcho}, S. and {Gonzalez-Morales}, A.~X. and {Gutierrez}, G. and {Honscheid}, K. and {Howlett}, C. and {Juneau}, S. and {Kremin}, A. and {Le Guillou}, L. and {Manera}, M. and {Medina-Varela}, L. and {Mena-Fern{\'a}ndez}, J. and {Miquel}, R. and {Mueller}, E. and {Mu{\~n}oz-Guti{\'e}rrez}, A. and {Myers}, A.~D. and {Nie}, J. and {Niz}, G. and {Paillas}, E. and {Percival}, W.~J. and {Poppett}, C. and {P{\'e}rez-Fern{\'a}ndez}, A. and {Rezaie}, M. and {Rosado-Marin}, A. and {Rossi}, G. and {Ruggeri}, R. and {Sanchez}, E. and {Saulder}, C. and {Schlegel}, D. and {Schubnell}, M. and {Sprayberry}, D. and {Tarl{\'e}}, G. and {Weaver}, B.~A. and {Yu}, J. and {Zhao}, C. and {Zou}, H.},
        title = "{Semi-analytical covariance matrices for two-point correlation function for DESI 2024 data}",
      journal = {\jcap},
         year = 2025,
        month = jan,
       volume = {2025},
       number = {1},
          eid = {145},
        pages = {145},
          doi = {10.1088/1475-7516/2025/01/145},
archivePrefix = {arXiv},
       eprint = {2404.03007},
 primaryClass = {astro-ph.CO},
       adsurl = {https://ui.adsabs.harvard.edu/abs/2025JCAP...01..145R}
}

@ARTICLE{KP4s9-Perez-Fernandez,
       author = {{P{\'e}rez-Fern{\'a}ndez}, A. and {Medina-Varela}, L. and {Ruggeri}, R. and {Vargas-Maga{\~n}a}, M. and {Seo}, H. and {Padmanabhan}, N. and {Ishak}, M. and {Aguilar}, J. and {Ahlen}, S. and {Alam}, S. and {Alves}, O. and {Andrade}, U. and {Brieden}, S. and {Brooks}, D. and {Carnero Rosell}, A. and {Chen}, X. and {Claybaugh}, T. and {Cole}, S. and {Dawson}, K. and {de la Macorra}, A. and {de Mattia}, A. and {Dey}, Arjun and {Ding}, Z. and {Doel}, P. and {Fanning}, K. and {Garcia-Quintero}, C. and {Gazta{\~n}aga}, E. and {Gontcho}, S. Gontcho A. and {Gutierrez}, G. and {Honscheid}, K. and {Juneau}, S. and {Kirkby}, D. and {Kisner}, T. and {Lambert}, A. and {Landriau}, M. and {Lasker}, J. and {Le Guillou}, L. and {Manera}, M. and {Martini}, P. and {Meisner}, A. and {Mena-Fern{\'a}ndez}, J. and {Miquel}, R. and {Moustakas}, J. and {Myers}, A.~D. and {Nadathur}, S. and {Newman}, J.~A. and {Niz}, G. and {Paillas}, E. and {Palanque-Delabrouille}, N. and {Percival}, W.~J. and {Poppett}, C. and {Prada}, F. and {Rashkovetskyi}, M. and {Rocher}, A. and {Rossi}, G. and {Sanchez}, A. and {Sanchez}, E. and {Schubnell}, M. and {Sprayberry}, D. and {Tarl{\'e}}, G. and {Valcin}, D. and {Weaver}, B.~A. and {Yu}, J. and {Zou}, H.},
        title = "{Fiducial-cosmology-dependent systematics for the DESI 2024 BAO analysis}",
      journal = {\jcap},
         year = 2025,
        month = jan,
       volume = {2025},
       number = {1},
          eid = {144},
        pages = {144},
          doi = {10.1088/1475-7516/2025/01/144},
archivePrefix = {arXiv},
       eprint = {2406.06085},
 primaryClass = {astro-ph.CO},
       adsurl = {https://ui.adsabs.harvard.edu/abs/2025JCAP...01..144P}
}

@ARTICLE{KP4s10-Mena-Fernandez,
       author = {{Mena-Fern{\'a}ndez}, J. and {Garcia-Quintero}, C. and {Yuan}, S. and {Hadzhiyska}, B. and {Alves}, O. and {Rashkovetskyi}, M. and {Seo}, H. and {Padmanabhan}, N. and {Nadathur}, S. and {Howlett}, C. and {Alam}, S. and {Rocher}, A. and {Ross}, A.~J. and {Sanchez}, E. and {Ishak}, M. and {Aguilar}, J. and {Ahlen}, S. and {Andrade}, U. and {BenZvi}, S. and {Brooks}, D. and {Burtin}, E. and {Chen}, S. and {Chen}, X. and {Claybaugh}, T. and {Cole}, S. and {de la Macorra}, A. and {de Mattia}, A. and {Dey}, Arjun and {Dey}, B. and {Ding}, Z. and {Doel}, P. and {Fanning}, K. and {Forero-Romero}, J.~E. and {Gazta{\~n}aga}, E. and {Gil-Mar{\'\i}n}, H. and {Gontcho}, S. Gontcho A. and {Gutierrez}, G. and {Guy}, J. and {Hahn}, C. and {Honscheid}, K. and {Juneau}, S. and {Kremin}, A. and {Landriau}, M. and {Le Guillou}, L. and {Levi}, M.~E. and {Manera}, M. and {Martini}, P. and {Medina-Varela}, L. and {Meisner}, A. and {Miquel}, R. and {Moustakas}, J. and {Mueller}, E. and {Mu{\~n}oz-Guti{\'e}rrez}, A. and {Myers}, A.~D. and {Newman}, J.~A. and {Nie}, J. and {Niz}, G. and {Paillas}, E. and {Palanque-Delabrouille}, N. and {Percival}, W.~J. and {Poppett}, C. and {P{\'e}rez-Fern{\'a}ndez}, A. and {Rosado-Marin}, A. and {Rossi}, G. and {Ruggeri}, R. and {Saulder}, C. and {Schlegel}, D. and {Schubnell}, M. and {Sprayberry}, D. and {Tarl{\'e}}, G. and {Vargas-Maga{\~n}a}, M. and {Weaver}, B.~A. and {Yu}, J. and {Zhang}, H. and {Zou}, H.},
        title = "{HOD-dependent systematics for luminous red galaxies in the DESI 2024 BAO analysis}",
      journal = {\jcap},
         year = 2025,
        month = jan,
       volume = {2025},
       number = {1},
          eid = {133},
        pages = {133},
          doi = {10.1088/1475-7516/2025/01/133},
archivePrefix = {arXiv},
       eprint = {2404.03008},
 primaryClass = {astro-ph.CO},
       adsurl = {https://ui.adsabs.harvard.edu/abs/2025JCAP...01..133M}
}

@ARTICLE{KP4s11-Garcia-Quintero,
       author = {{Garcia-Quintero}, C. and {Mena-Fern{\'a}ndez}, J. and {Rocher}, A. and {Yuan}, S. and {Hadzhiyska}, B. and {Alves}, O. and {Rashkovetskyi}, M. and {Seo}, H. and {Padmanabhan}, N. and {Nadathur}, S. and {Howlett}, C. and {Ishak}, M. and {Medina-Varela}, L. and {McDonald}, P. and {Ross}, A.~J. and {Xie}, Y. and {Chen}, X. and {Bera}, A. and {Aguilar}, J. and {Ahlen}, S. and {Andrade}, U. and {BenZvi}, S. and {Brooks}, D. and {Burtin}, E. and {Chen}, S. and {Claybaugh}, T. and {Cole}, S. and {de la Macorra}, A. and {de Mattia}, A. and {Dey}, A. and {Dey}, B. and {Ding}, Z. and {Doel}, P. and {Fanning}, K. and {Forero-Romero}, J.~E. and {Gazta{\~n}aga}, E. and {Gil-Mar{\'\i}n}, H. and {Gontcho}, S. Gontcho A. and {Gutierrez}, G. and {Guy}, J. and {Hahn}, C. and {Honscheid}, K. and {Kremin}, A. and {Landriau}, M. and {Le Guillou}, L. and {Levi}, M.~E. and {Manera}, M. and {Martini}, P. and {Meisner}, A. and {Miquel}, R. and {Moustakas}, J. and {Mueller}, E. and {Mu{\~n}oz-Guti{\'e}rrez}, A. and {Myers}, A.~D. and {Newman}, J.~A. and {Nie}, J. and {Niz}, G. and {Paillas}, E. and {Palanque-Delabrouille}, N. and {Percival}, W.~J. and {Poppett}, C. and {P{\'e}rez-Fern{\'a}ndez}, A. and {Rosado-Marin}, A. and {Rossi}, G. and {Ruggeri}, R. and {Sanchez}, E. and {Schlegel}, D. and {Schubnell}, M. and {Sprayberry}, D. and {Tarl{\'e}}, G. and {Vargas-Maga{\~n}a}, M. and {Weaver}, B.~A. and {Yu}, J. and {Zhang}, H. and {Zhou}, R. and {Zou}, H.},
        title = "{HOD-dependent systematics in Emission Line Galaxies for the DESI 2024 BAO analysis}",
      journal = {\jcap},
         year = 2025,
        month = jan,
       volume = {2025},
       number = {1},
          eid = {132},
        pages = {132},
          doi = {10.1088/1475-7516/2025/01/132},
archivePrefix = {arXiv},
       eprint = {2404.03009},
 primaryClass = {astro-ph.CO},
       adsurl = {https://ui.adsabs.harvard.edu/abs/2025JCAP...01..132G}
}

@ARTICLE{DESI2016b.Instr,
       author = {{DESI Collaboration} and {Aghamousa}, Amir and {Aguilar}, Jessica and {Ahlen}, Steve and {Alam}, Shadab and {Allen}, Lori E. and {Allende Prieto}, Carlos and {Annis}, James and {Bailey}, Stephen and {Balland}, Christophe and {Ballester}, Otger and {Baltay}, Charles and {Beaufore}, Lucas and {Bebek}, Chris and {Beers}, Timothy C. and {Bell}, Eric F. and {Bernal}, Jos{\'e} Luis and {Besuner}, Robert and {Beutler}, Florian and {Blake}, Chris and {Bleuler}, Hannes and {Blomqvist}, Michael and {Blum}, Robert and {Bolton}, Adam S. and {Briceno}, Cesar and {Brooks}, David and {Brownstein}, Joel R. and {Buckley-Geer}, Elizabeth and {Burden}, Angela and {Burtin}, Etienne and {Busca}, Nicolas G. and {Cahn}, Robert N. and {Cai}, Yan-Chuan and {Cardiel-Sas}, Laia and {Carlberg}, Raymond G. and {Carton}, Pierre-Henri and {Casas}, Ricard and {Castander}, Francisco J. and {Cervantes-Cota}, Jorge L. and {Claybaugh}, Todd M. and {Close}, Madeline and {Coker}, Carl T. and {Cole}, Shaun and {Comparat}, Johan and {Cooper}, Andrew P. and {Cousinou}, M. -C. and {Crocce}, Martin and {Cuby}, Jean-Gabriel and {Cunningham}, Daniel P. and {Davis}, Tamara M. and {Dawson}, Kyle S. and {de la Macorra}, Axel and {De Vicente}, Juan and {Delubac}, Timoth{\'e}e and {Derwent}, Mark and {Dey}, Arjun and {Dhungana}, Govinda and {Ding}, Zhejie and {Doel}, Peter and {Duan}, Yutong T. and {Ealet}, Anne and {Edelstein}, Jerry and {Eftekharzadeh}, Sarah and {Eisenstein}, Daniel J. and {Elliott}, Ann and {Escoffier}, St{\'e}phanie and {Evatt}, Matthew and {Fagrelius}, Parker and {Fan}, Xiaohui and {Fanning}, Kevin and {Farahi}, Arya and {Farihi}, Jay and {Favole}, Ginevra and {Feng}, Yu and {Fernandez}, Enrique and {Findlay}, Joseph R. and {Finkbeiner}, Douglas P. and {Fitzpatrick}, Michael J. and {Flaugher}, Brenna and {Flender}, Samuel and {Font-Ribera}, Andreu and {Forero-Romero}, Jaime E. and {Fosalba}, Pablo and {Frenk}, Carlos S. and {Fumagalli}, Michele and {Gaensicke}, Boris T. and {Gallo}, Giuseppe and {Garcia-Bellido}, Juan and {Gaztanaga}, Enrique and {Pietro Gentile Fusillo}, Nicola and {Gerard}, Terry and {Gershkovich}, Irena and {Giannantonio}, Tommaso and {Gillet}, Denis and {Gonzalez-de-Rivera}, Guillermo and {Gonzalez-Perez}, Violeta and {Gott}, Shelby and {Graur}, Or and {Gutierrez}, Gaston and {Guy}, Julien and {Habib}, Salman and {Heetderks}, Henry and {Heetderks}, Ian and {Heitmann}, Katrin and {Hellwing}, Wojciech A. and {Herrera}, David A. and {Ho}, Shirley and {Holland}, Stephen and {Honscheid}, Klaus and {Huff}, Eric and {Hutchinson}, Timothy A. and {Huterer}, Dragan and {Hwang}, Ho Seong and {Illa Laguna}, Joseph Maria and {Ishikawa}, Yuzo and {Jacobs}, Dianna and {Jeffrey}, Niall and {Jelinsky}, Patrick and {Jennings}, Elise and {Jiang}, Linhua and {Jimenez}, Jorge and {Johnson}, Jennifer and {Joyce}, Richard and {Jullo}, Eric and {Juneau}, St{\'e}phanie and {Kama}, Sami and {Karcher}, Armin and {Karkar}, Sonia and {Kehoe}, Robert and {Kennamer}, Noble and {Kent}, Stephen and {Kilbinger}, Martin and {Kim}, Alex G. and {Kirkby}, David and {Kisner}, Theodore and {Kitanidis}, Ellie and {Kneib}, Jean-Paul and {Koposov}, Sergey and {Kovacs}, Eve and {Koyama}, Kazuya and {Kremin}, Anthony and {Kron}, Richard and {Kronig}, Luzius and {Kueter-Young}, Andrea and {Lacey}, Cedric G. and {Lafever}, Robin and {Lahav}, Ofer and {Lambert}, Andrew and {Lampton}, Michael and {Landriau}, Martin and {Lang}, Dustin and {Lauer}, Tod R. and {Le Goff}, Jean-Marc and {Le Guillou}, Laurent and {Le Van Suu}, Auguste and {Lee}, Jae Hyeon and {Lee}, Su-Jeong and {Leitner}, Daniela and {Lesser}, Michael and {Levi}, Michael E. and {L'Huillier}, Benjamin and {Li}, Baojiu and {Liang}, Ming and {Lin}, Huan and {Linder}, Eric and {Loebman}, Sarah R. and {Luki{\'c}}, Zarija and {Ma}, Jun and {MacCrann}, Niall and {Magneville}, Christophe and {Makarem}, Laleh and {Manera}, Marc and {Manser}, Christopher J. and {Marshall}, Robert and {Martini}, Paul and {Massey}, Richard and {Matheson}, Thomas and {McCauley}, Jeremy and {McDonald}, Patrick and {McGreer}, Ian D. and {Meisner}, Aaron and {Metcalfe}, Nigel and {Miller}, Timothy N. and {Miquel}, Ramon and {Moustakas}, John and {Myers}, Adam and {Naik}, Milind and {Newman}, Jeffrey A. and {Nichol}, Robert C. and {Nicola}, Andrina and {Nicolati da Costa}, Luiz and {Nie}, Jundan and {Niz}, Gustavo and {Norberg}, Peder and {Nord}, Brian and {Norman}, Dara and {Nugent}, Peter and {O'Brien}, Thomas and {Oh}, Minji and {Olsen}, Knut A.~G.},
        title = "{The DESI Experiment Part II: Instrument Design}",
      journal = {arXiv e-prints},
         year = 2016,
        month = oct,
          eid = {arXiv:1611.00037},
        pages = {arXiv:1611.00037},
          doi = {10.48550/arXiv.1611.00037},
archivePrefix = {arXiv},
       eprint = {1611.00037},
 primaryClass = {astro-ph.IM},
       adsurl = {https://ui.adsabs.harvard.edu/abs/2016arXiv161100037D}
}

@ARTICLE{FocalPlane.Silber.2023,
       author = {{Silber}, Joseph Harry and {Fagrelius}, Parker and {Fanning}, Kevin and {Schubnell}, Michael and {Aguilar}, Jessica Nicole and {Ahlen}, Steven and {Ameel}, Jon and {Ballester}, Otger and {Baltay}, Charles and {Bebek}, Chris and {Benton Beard}, Dominic and {Besuner}, Robert and {Cardiel-Sas}, Laia and {Casas}, Ricard and {Castander}, Francisco Javier and {Claybaugh}, Todd and {Dobson}, Carl and {Duan}, Yutong and {Dunlop}, Patrick and {Edelstein}, Jerry and {Emmet}, William T. and {Elliott}, Ann and {Evatt}, Matthew and {Gershkovich}, Irena and {Guy}, Julien and {Harris}, Stu and {Heetderks}, Henry and {Heetderks}, Ian and {Honscheid}, Klaus and {Illa}, Jose Maria and {Jelinsky}, Patrick and {Jelinsky}, Sharon R. and {Jimenez}, Jorge and {Karcher}, Armin and {Kent}, Stephen and {Kirkby}, David and {Kneib}, Jean-Paul and {Lambert}, Andrew and {Lampton}, Mike and {Leitner}, Daniela and {Levi}, Michael and {McCauley}, Jeremy and {Meisner}, Aaron and {Miller}, Timothy N. and {Miquel}, Ramon and {Mundet}, Juli{\'a} and {Poppett}, Claire and {Rabinowitz}, David and {Reil}, Kevin and {Roman}, David and {Schlegel}, David and {Serrano}, Santiago and {Van Shourt}, William and {Sprayberry}, David and {Tarl{\'e}}, Gregory and {Tie}, Suk Sien and {Weaverdyck}, Curtis and {Zhang}, Kai and {Azzaro}, Marco and {Bailey}, Stephen and {Becerril}, Santiago and {Blackwell}, Tami and {Bouri}, Mohamed and {Brooks}, David and {Buckley-Geer}, Elizabeth and {Castro}, Jose Pe{\~n}ate and {Derwent}, Mark and {Dey}, Arjun and {Dhungana}, Govinda and {Doel}, Peter and {Eisenstein}, Daniel J. and {Fahim}, Nasib and {Garcia-Bellido}, Juan and {Gazta{\~n}aga}, Enrique and {A Gontcho}, Satya Gontcho and {Gutierrez}, Gaston and {H{\"o}rler}, Philipp and {Kehoe}, Robert and {Kisner}, Theodore and {Kremin}, Anthony and {Kronig}, Luzius and {Landriau}, Martin and {Le Guillou}, Laurent and {Martini}, Paul and {Moustakas}, John and {Palanque-Delabrouille}, Nathalie and {Peng}, Xiyan and {Percival}, Will and {Prada}, Francisco and {Allende Prieto}, Carlos and {de Rivera}, Guillermo Gonzalez and {Sanchez}, Eusebio and {Sanchez}, Justo and {Sharples}, Ray and {Soares-Santos}, Marcelle and {Schlafly}, Edward and {Weaver}, Benjamin Alan and {Zhou}, Zhimin and {Zhu}, Yaling and {Zou}, Hu and {DESI Collaboration}},
        title = "{The Robotic Multiobject Focal Plane System of the Dark Energy Spectroscopic Instrument (DESI)}",
      journal = {\aj},
         year = 2023,
        month = jan,
       volume = {165},
       number = {1},
          eid = {9},
        pages = {9},
          doi = {10.3847/1538-3881/ac9ab1},
archivePrefix = {arXiv},
       eprint = {2205.09014},
 primaryClass = {astro-ph.IM},
       adsurl = {https://ui.adsabs.harvard.edu/abs/2023AJ....165....9S}
}

@ARTICLE{Corrector.Miller.2023,
       author = {{Miller}, Timothy N. and {Doel}, Peter and {Gutierrez}, Gaston and {Besuner}, Robert and {Brooks}, David and {Gallo}, Giuseppe and {Heetderks}, Henry and {Jelinsky}, Patrick and {Kent}, Stephen M. and {Lampton}, Michael and {Levi}, Michael E. and {Liang}, Ming and {Meisner}, Aaron and {Sholl}, Michael J. and {Silber}, Joseph Harry and {Sprayberry}, David and {Aguilar}, Jessica Nicole and {de la Macorra}, Axel and {Eisenstein}, Daniel and {Fanning}, Kevin and {Font-Ribera}, Andreu and {Gazta{\~n}aga}, Enrique and {Gontcho A Gontcho}, Satya and {Honscheid}, Klaus and {Jimenez}, Jorge and {Joyce}, Dick and {Kehoe}, Robert and {Kisner}, Theodore and {Kremin}, Anthony and {Landriau}, Martin and {Le Guillou}, Laurent and {Magneville}, Christophe and {Martini}, Paul and {Miquel}, Ramon and {Moustakas}, John and {Nie}, Jundan and {Percival}, Will and {Poppett}, Claire and {Prada}, Francisco and {Rossi}, Graziano and {Schlegel}, David and {Schubnell}, Michael and {Seo}, Hee-Jong and {Sharples}, Ray and {Tarl{\'e}}, Gregory and {Vargas-Maga{\~n}a}, Mariana and {Zhou}, Zhimin and {the DESI Collaboration}},
        title = "{The Optical Corrector for the Dark Energy Spectroscopic Instrument}",
      journal = {\aj},
         year = 2024,
        month = aug,
       volume = {168},
       number = {2},
          eid = {95},
        pages = {95},
          doi = {10.3847/1538-3881/ad45fe},
archivePrefix = {arXiv},
       eprint = {2306.06310},
 primaryClass = {astro-ph.IM},
       adsurl = {https://ui.adsabs.harvard.edu/abs/2024AJ....168...95M}
}

@ARTICLE{FiberSystem.Poppett.2024,
       author = {{Poppett}, Claire and {Tyas}, Luke and {Aguilar}, J. and {Bebek}, Christopher and {Bramall}, D. and {Claybaugh}, T. and {Edelstein}, J. and {Fagrelius}, P. and {Heetderks}, H. and {Jelinsky}, P. and {Jelinsky}, S. and {Lafever}, Robin and {Lambert}, A. and {Lampton}, M. and {Levi}, Michael E. and {Martini}, P. and {Rockosi}, C. and {Schmoll}, J. and {Sharples}, Ray M. and {Sirk}, Martin and {Wishnow}, Edward and {Yu}, Jiaxi and {Ahlen}, S. and {Bault}, A. and {BenZvi}, S. and {Brooks}, D. and {Cole}, S. and {de la Macorra}, A. and {Dey}, Arjun and {Doel}, P. and {Fanning}, K. and {Font-Ribera}, A. and {Forero-Romero}, J.~E. and {Gazta{\~n}aga}, E. and {Gontcho A Gontcho}, S. and {Gonzalez-Morales}, A.~X. and {Hahn}, C. and {Honscheid}, K. and {Jimenez}, J. and {Juneau}, S. and {Kirkby}, D. and {Kremin}, A. and {Landriau}, M. and {Le Guillou}, L. and {Manera}, M. and {Meisner}, A. and {Miquel}, R. and {Moustakas}, J. and {Mueller}, E. and {Mu{\~n}oz-Guti{\'e}rrez}, A. and {Myers}, A.~D. and {Nie}, J. and {Niz}, G. and {Palanque-Delabrouille}, N. and {Percival}, W.~J. and {Prada}, F. and {Rabinowitz}, D. and {Rezaie}, M. and {Rossi}, G. and {Sanchez}, E. and {Schlafly}, Edward F. and {Schlegel}, D. and {Schubnell}, M. and {Seo}, H. and {Sprayberry}, D. and {Tarl{\'e}}, G. and {Vargas-Maga{\~n}a}, M. and {Weaver}, B.~A. and {Zhou}, R.},
        title = "{Overview of the Fiber System for the Dark Energy Spectroscopic Instrument}",
      journal = {\aj},
         year = 2024,
        month = dec,
       volume = {168},
       number = {6},
          eid = {245},
        pages = {245},
          doi = {10.3847/1538-3881/ad76a4},
       adsurl = {https://ui.adsabs.harvard.edu/abs/2024AJ....168..245P}
}

@ARTICLE{Spectro.Pipeline.Guy.2023,
       author = {{Guy}, J. and {Bailey}, S. and {Kremin}, A. and {Alam}, Shadab and {Alexander}, D.~M. and {Allende Prieto}, C. and {BenZvi}, S. and {Bolton}, A.~S. and {Brooks}, D. and {Chaussidon}, E. and {Cooper}, A.~P. and {Dawson}, K. and {de la Macorra}, A. and {Dey}, A. and {Dey}, Biprateep and {Dhungana}, G. and {Eisenstein}, D.~J. and {Font-Ribera}, A. and {Forero-Romero}, J.~E. and {Gazta{\~n}aga}, E. and {Gontcho A Gontcho}, S. and {Green}, D. and {Honscheid}, K. and {Ishak}, M. and {Kehoe}, R. and {Kirkby}, D. and {Kisner}, T. and {Koposov}, Sergey E. and {Lan}, Ting-Wen and {Landriau}, M. and {Le Guillou}, L. and {Levi}, Michael E. and {Magneville}, C. and {Manser}, Christopher J. and {Martini}, P. and {Meisner}, Aaron M. and {Miquel}, R. and {Moustakas}, J. and {Myers}, Adam D. and {Newman}, Jeffrey A. and {Nie}, Jundan and {Palanque-Delabrouille}, N. and {Percival}, W.~J. and {Poppett}, C. and {Prada}, F. and {Raichoor}, A. and {Ravoux}, C. and {Ross}, A.~J. and {Schlafly}, E.~F. and {Schlegel}, D. and {Schubnell}, M. and {Sharples}, Ray M. and {Tarl{\'e}}, Gregory and {Weaver}, B.~A. and {Y{\'e}che}, Christophe and {Zhou}, Rongpu and {Zhou}, Zhimin and {Zou}, H.},
        title = "{The Spectroscopic Data Processing Pipeline for the Dark Energy Spectroscopic Instrument}",
      journal = {\aj},
         year = 2023,
        month = apr,
       volume = {165},
       number = {4},
          eid = {144},
        pages = {144},
          doi = {10.3847/1538-3881/acb212},
archivePrefix = {arXiv},
       eprint = {2209.14482},
 primaryClass = {astro-ph.IM},
       adsurl = {https://ui.adsabs.harvard.edu/abs/2023AJ....165..144G}
}

@software{Redrock.Bailey.2024,
       author = {{Bailey}, Stephen and {Schlegel}, David},
        title = "{redrock: Redshift fitting for spectroperfectionism}",
 howpublished = {Astrophysics Source Code Library, record ascl:2606.024},
         year = 2026,
        month = jun,
          eid = {ascl:2606.024},
archivePrefix = {ascl},
       eprint = {2606.024},
       adsurl = {https://ui.adsabs.harvard.edu/abs/2026ascl.soft06024B}
}

@ARTICLE{SurveyOps.Schlafly.2023,
       author = {{Schlafly}, Edward F. and {Kirkby}, David and {Schlegel}, David J. and {Myers}, Adam D. and {Raichoor}, Anand and {Dawson}, Kyle and {Aguilar}, Jessica and {Allende Prieto}, Carlos and {Bailey}, Stephen and {BenZvi}, Segev and {Bermejo-Climent}, Jose and {Brooks}, David and {de la Macorra}, Axel and {Dey}, Arjun and {Doel}, Peter and {Fanning}, Kevin and {Font-Ribera}, Andreu and {Forero-Romero}, Jaime E. and {Garc{\'\i}a-Bellido}, Juan and {Gontcho A Gontcho}, Satya and {Guy}, Julien and {Hahn}, ChangHoon and {Honscheid}, Klaus and {Ishak}, Mustapha and {Juneau}, St{\'e}phanie and {Kehoe}, Robert and {Kisner}, Theodore and {Kremin}, Anthony and {Landriau}, Martin and {Lang}, Dustin A. and {Lasker}, James and {Levi}, Michael E. and {Magneville}, Christophe and {Manser}, Christopher J. and {Martini}, Paul and {Meisner}, Aaron M. and {Miquel}, Ramon and {Moustakas}, John and {Newman}, Jeffrey A. and {Nie}, Jundan and {Palanque-Delabrouille}, Nathalie. and {Percival}, Will J. and {Poppett}, Claire and {Rockosi}, Constance and {Ross}, Ashley J. and {Rossi}, Graziano and {Tarl{\'e}}, Gregory and {Weaver}, Benjamin A. and {Y{\`e}che}, Christophe and {Zhou}, Rongpu and {DESI Collaboration}},
        title = "{Survey Operations for the Dark Energy Spectroscopic Instrument}",
      journal = {\aj},
         year = 2023,
        month = dec,
       volume = {166},
       number = {6},
          eid = {259},
        pages = {259},
          doi = {10.3847/1538-3881/ad0832},
archivePrefix = {arXiv},
       eprint = {2306.06309},
 primaryClass = {astro-ph.CO},
       adsurl = {https://ui.adsabs.harvard.edu/abs/2023AJ....166..259S}
}

@ARTICLE{LSSCatalogs.Ross.2025,
       author = {{Ross}, A.~J. and {Aguilar}, J. and {Ahlen}, S. and {Alam}, S. and {Anand}, A. and {Bailey}, S. and {Bianchi}, D. and {Brieden}, S. and {Brooks}, D. and {Burtin}, E. and {Carnero Rosell}, A. and {Chaussidon}, E. and {Claybaugh}, T. and {Cole}, S. and {Dawson}, K. and {de la Macorra}, A. and {de Mattia}, A. and {Dey}, A. and {Dey}, B. and {Doel}, P. and {Fanning}, K. and {Ferraro}, S. and {Ereza}, J. and {Font-Ribera}, A. and {Forero-Romero}, J.~E. and {Gazta{\~n}aga}, E. and {Gil-Mar{\'\i}n}, H. and {Gontcho A Gontcho}, S. and {Gonzalez-Morales}, A.~X. and {Guy}, J. and {Hahn}, C. and {Heydenreich}, S. and {Honscheid}, K. and {Howlett}, C. and {Ishak}, M. and {Karim}, T. and {Kirkby}, D. and {Kisner}, T. and {Kong}, H. and {Kremin}, A. and {Krolewski}, A. and {Lambert}, A. and {Landriau}, M. and {Lasker}, J. and {Guillou}, L.~L. and {Levi}, M.~E. and {Manera}, M. and {Martini}, P. and {McDonald}, P. and {Meisner}, A. and {Miquel}, R. and {Moon}, J. and {Moustakas}, J. and {Mu{\~n}oz-Guti{\'e}rrez}, A. and {Myers}, A.~D. and {Nadathur}, S. and {Napolitano}, L. and {Newman}, J.~A. and {Nie}, J. and {Niz}, G. and {Palanque-Delabrouille}, N. and {Percival}, W.~J. and {Poppett}, C. and {Prada}, F. and {Raichoor}, A. and {Ravoux}, C. and {Rezaie}, M. and {Rosado-Marin}, A. and {Rossi}, G. and {Samushia}, L. and {Sanchez}, E. and {Schlafly}, E.~F. and {Schlegel}, D. and {Seo}, H. and {Smith}, A. and {Sprayberry}, D. and {Tarl{\'e}}, G. and {Valcin}, D. and {Vargas-Maga{\~n}a}, M. and {Weaver}, B.~A. and {Wilson}, M.~J. and {Yu}, J. and {Zarrouk}, P. and {Zhao}, C. and {Zhou}, R. and {Zou}, H.},
        title = "{The construction of large-scale structure catalogs for the Dark Energy Spectroscopic Instrument}",
      journal = {\jcap},
         year = 2025,
        month = jan,
       volume = {2025},
       number = {1},
          eid = {125},
        pages = {125},
          doi = {10.1088/1475-7516/2025/01/125},
archivePrefix = {arXiv},
       eprint = {2405.16593},
 primaryClass = {astro-ph.CO},
       adsurl = {https://ui.adsabs.harvard.edu/abs/2025JCAP...01..125R}
}

@ARTICLE{BGS.TS.Hahn.2023,
       author = {{Hahn}, ChangHoon and {Wilson}, Michael J. and {Ruiz-Macias}, Omar and {Cole}, Shaun and {Weinberg}, David H. and {Moustakas}, John and {Kremin}, Anthony and {Tinker}, Jeremy L. and {Smith}, Alex and {Wechsler}, Risa H. and {Ahlen}, Steven and {Alam}, Shadab and {Bailey}, Stephen and {Brooks}, David and {Cooper}, Andrew P. and {Davis}, Tamara M. and {Dawson}, Kyle and {Dey}, Arjun and {Dey}, Biprateep and {Eftekharzadeh}, Sarah and {Eisenstein}, Daniel J. and {Fanning}, Kevin and {Forero-Romero}, Jaime E. and {Frenk}, Carlos S. and {Gazta{\~n}aga}, Enrique and {Gontcho A Gontcho}, Satya and {Guy}, Julien and {Honscheid}, Klaus and {Ishak}, Mustapha and {Juneau}, St{\'e}phanie and {Kehoe}, Robert and {Kisner}, Theodore and {Lan}, Ting-Wen and {Landriau}, Martin and {Le Guillou}, Laurent and {Levi}, Michael E. and {Magneville}, Christophe and {Martini}, Paul and {Meisner}, Aaron and {Myers}, Adam D. and {Nie}, Jundan and {Norberg}, Peder and {Palanque-Delabrouille}, Nathalie and {Percival}, Will J. and {Poppett}, Claire and {Prada}, Francisco and {Raichoor}, Anand and {Ross}, Ashley J. and {Safonova}, Sasha and {Saulder}, Christoph and {Schlafly}, Eddie and {Schlegel}, David and {Sierra-Porta}, David and {Tarle}, Gregory and {Weaver}, Benjamin A. and {Y{\`e}che}, Christophe and {Zarrouk}, Pauline and {Zhou}, Rongpu and {Zhou}, Zhimin and {Zou}, Hu},
        title = "{The DESI Bright Galaxy Survey: Final Target Selection, Design, and Validation}",
      journal = {\aj},
         year = 2023,
        month = jun,
       volume = {165},
       number = {6},
          eid = {253},
        pages = {253},
          doi = {10.3847/1538-3881/accff8},
archivePrefix = {arXiv},
       eprint = {2208.08512},
 primaryClass = {astro-ph.CO},
       adsurl = {https://ui.adsabs.harvard.edu/abs/2023AJ....165..253H}
}

@ARTICLE{LRG.TS.Zhou.2023,
       author = {{Zhou}, Rongpu and {Dey}, Biprateep and {Newman}, Jeffrey A. and {Eisenstein}, Daniel J. and {Dawson}, K. and {Bailey}, S. and {Berti}, A. and {Guy}, J. and {Lan}, Ting-Wen and {Zou}, H. and {Aguilar}, J. and {Ahlen}, S. and {Alam}, Shadab and {Brooks}, D. and {de la Macorra}, A. and {Dey}, A. and {Dhungana}, G. and {Fanning}, K. and {Font-Ribera}, A. and {Gontcho}, S. Gontcho A. and {Honscheid}, K. and {Ishak}, Mustapha and {Kisner}, T. and {Kov{\'a}cs}, A. and {Kremin}, A. and {Landriau}, M. and {Levi}, Michael E. and {Magneville}, C. and {Manera}, Marc and {Martini}, P. and {Meisner}, Aaron M. and {Miquel}, R. and {Moustakas}, J. and {Myers}, Adam D. and {Nie}, Jundan and {Palanque-Delabrouille}, N. and {Percival}, W.~J. and {Poppett}, C. and {Prada}, F. and {Raichoor}, A. and {Ross}, A.~J. and {Schlafly}, E. and {Schlegel}, D. and {Schubnell}, M. and {Tarl{\'e}}, Gregory and {Weaver}, B.~A. and {Wechsler}, R.~H. and {Y{\'e}che}, Christophe and {Zhou}, Zhimin},
        title = "{Target Selection and Validation of DESI Luminous Red Galaxies}",
      journal = {\aj},
         year = 2023,
        month = feb,
       volume = {165},
       number = {2},
          eid = {58},
        pages = {58},
          doi = {10.3847/1538-3881/aca5fb},
archivePrefix = {arXiv},
       eprint = {2208.08515},
 primaryClass = {astro-ph.CO},
       adsurl = {https://ui.adsabs.harvard.edu/abs/2023AJ....165...58Z}
}

@ARTICLE{ELG.TS.Raichoor.2023,
       author = {{Raichoor}, A. and {Moustakas}, J. and {Newman}, Jeffrey A. and {Karim}, T. and {Ahlen}, S. and {Alam}, Shadab and {Bailey}, S. and {Brooks}, D. and {Dawson}, K. and {de la Macorra}, A. and {de Mattia}, A. and {Dey}, A. and {Dey}, Biprateep and {Dhungana}, G. and {Eftekharzadeh}, S. and {Eisenstein}, D.~J. and {Fanning}, K. and {Font-Ribera}, A. and {Garc{\'\i}a-Bellido}, J. and {Gazta{\~n}aga}, E. and {A Gontcho}, S. Gontcho and {Guy}, J. and {Honscheid}, K. and {Ishak}, M. and {Kehoe}, R. and {Kisner}, T. and {Kremin}, Anthony and {Lan}, Ting-Wen and {Landriau}, M. and {Le Guillou}, L. and {Levi}, Michael E. and {Magneville}, C. and {Manera}, M. and {Martini}, P. and {Meisner}, Aaron M. and {Myers}, Adam D. and {Nie}, Jundan and {Palanque-Delabrouille}, N. and {Percival}, W.~J. and {Poppett}, C. and {Prada}, F. and {Ross}, A.~J. and {Ruhlmann-Kleider}, V. and {Sabiu}, C.~G. and {Schlafly}, E.~F. and {Schlegel}, D. and {Tarl{\'e}}, Gregory and {Weaver}, B.~A. and {Y{\`e}che}, Christophe and {Zhou}, Rongpu and {Zhou}, Zhimin and {Zou}, H.},
        title = "{Target Selection and Validation of DESI Emission Line Galaxies}",
      journal = {\aj},
         year = 2023,
        month = mar,
       volume = {165},
       number = {3},
          eid = {126},
        pages = {126},
          doi = {10.3847/1538-3881/acb213},
archivePrefix = {arXiv},
       eprint = {2208.08513},
 primaryClass = {astro-ph.CO},
       adsurl = {https://ui.adsabs.harvard.edu/abs/2023AJ....165..126R}
}

@ARTICLE{QSO.TS.Chaussidon.2023,
       author = {{Chaussidon}, Edmond and {Y{\`e}che}, Christophe and {Palanque-Delabrouille}, Nathalie and {Alexander}, David M. and {Yang}, Jinyi and {Ahlen}, Steven and {Bailey}, Stephen and {Brooks}, David and {Cai}, Zheng and {Chabanier}, Sol{\`e}ne and {Davis}, Tamara M. and {Dawson}, Kyle and {de laMacorra}, Axel and {Dey}, Arjun and {Dey}, Biprateep and {Eftekharzadeh}, Sarah and {Eisenstein}, Daniel J. and {Fanning}, Kevin and {Font-Ribera}, Andreu and {Gazta{\~n}aga}, Enrique and {A Gontcho}, Satya Gontcho and {Gonzalez-Morales}, Alma X. and {Guy}, Julien and {Herrera-Alcantar}, Hiram K. and {Honscheid}, Klaus and {Ishak}, Mustapha and {Jiang}, Linhua and {Juneau}, Stephanie and {Kehoe}, Robert and {Kisner}, Theodore and {Kov{\'a}cs}, Andras and {Kremin}, Anthony and {Lan}, Ting-Wen and {Landriau}, Martin and {Le Guillou}, Laurent and {Levi}, Michael E. and {Magneville}, Christophe and {Martini}, Paul and {Meisner}, Aaron M. and {Moustakas}, John and {Mu{\~n}oz-Guti{\'e}rrez}, Andrea and {Myers}, Adam D. and {Newman}, Jeffrey A. and {Nie}, Jundan and {Percival}, Will J. and {Poppett}, Claire and {Prada}, Francisco and {Raichoor}, Anand and {Ravoux}, Corentin and {Ross}, Ashley J. and {Schlafly}, Edward and {Schlegel}, David and {Tan}, Ting and {Tarl{\'e}}, Gregory and {Zhou}, Rongpu and {Zhou}, Zhimin and {Zou}, Hu},
        title = "{Target Selection and Validation of DESI Quasars}",
      journal = {\apj},
         year = 2023,
        month = feb,
       volume = {944},
       number = {1},
          eid = {107},
        pages = {107},
          doi = {10.3847/1538-4357/acb3c2},
archivePrefix = {arXiv},
       eprint = {2208.08511},
 primaryClass = {astro-ph.CO},
       adsurl = {https://ui.adsabs.harvard.edu/abs/2023ApJ...944..107C}
}

@ARTICLE{DESI.DR2.DR2,
  title = {DESI DR2 results. II. Measurements of baryon acoustic oscillations and cosmological constraints},
  author = {Abdul Karim, M. and Aguilar, J. and Ahlen, S. and Alam, S. and Allen, L. and Prieto, C. Allende and Alves, O. and Anand, A. and Andrade, U. and Armengaud, E. and Aviles, A. and Bailey, S. and Baltay, C. and Bansal, P. and Bault, A. and Behera, J. and BenZvi, S. and Bianchi, D. and Blake, C. and Brieden, S. and Brodzeller, A. and Brooks, D. and Buckley-Geer, E. and Burtin, E. and Calderon, R. and Canning, R. and Rosell, A. Carnero and Carrilho, P. and Casas, L. and Castander, F. J. and Charles, M. and Chaussidon, E. and Chaves-Montero, J. and Chebat, D. and Chen, X. and Claybaugh, T. and Cole, S. and Cooper, A. P. and Cuceu, A. and Dawson, K. S. and de la Macorra, A. and de Mattia, A. and Deiosso, N. and Della Costa, J. and Demina, R. and Dey, A. and Dey, B. and Ding, Z. and Doel, P. and Edelstein, J. and Eisenstein, D. J. and Elbers, W. and Fagrelius, P. and Fanning, K. and Fern\'andez-Garc\'{\i}a, E. and Ferraro, S. and Font-Ribera, A. and Forero-Romero, J. E. and Frenk, C. S. and Garcia-Quintero, C. and Garrison, L. H. and Gazta\~naga, E. and Gil-Mar\'{\i}n, H. and Gontcho, S. Gontcho A. and Gonzalez, D. and Gonzalez-Morales, A. X. and Gordon, C. and Green, D. and Gutierrez, G. and Guy, J. and Hadzhiyska, B. and Hahn, C. and He, S. and Herbold, M. and Herrera-Alcantar, H. K. and Ho, M.-F. and Honscheid, K. and Howlett, C. and Huterer, D. and Ishak, M. and Juneau, S. and Kamble, N. V. and Kara\ifmmode \mbox{\c{c}}\else \c{c}\fi{}ayl��, N. G. and Kehoe, R. and Kent, S. and Kim, A. G. and Kirkby, D. and Kisner, T. and Koposov, S. E. and Kremin, A. and Krolewski, A. and Lahav, O. and Lamman, C. and Landriau, M. and Lang, D. and Lasker, J. and Le Goff, J. M. and Le Guillou, L. and Leauthaud, A. and Levi, M. E. and Li, Q. and Li, T. S. and Lodha, K. and Lokken, M. and Lozano-Rodr\'{\i}guez, F. and Magneville, C. and Manera, M. and Martini, P. and Matthewson, W. L. and Meisner, A. and Mena-Fern\'andez, J. and Menegas, A. and Mergulh\~ao, T. and Miquel, R. and Moustakas, J. and Mu\~noz-Guti\'errez, A. and Mu\~noz-Santos, D. and Myers, A. D. and Nadathur, S. and Naidoo, K. and Napolitano, L. and Newman, J. A. and Niz, G. and Noriega, H. E. and Paillas, E. and Palanque-Delabrouille, N. and Pan, J. and Peacock, J. A. and Ibanez, M. P. and Percival, W. J. and P\'erez-Fern\'andez, A. and P\'erez-R\`afols, I. and Pieri, M. M. and Poppett, C. and Prada, F. and Rabinowitz, D. and Raichoor, A. and Ram\'{\i}rez-P\'erez, C. and Rashkovetskyi, M. and Ravoux, C. and Rich, J. and Rocher, A. and Rockosi, C. and Rohlf, J. and Rom\'an-Herrera, J. O. and Ross, A. J. and Rossi, G. and Ruggeri, R. and Ruhlmann-Kleider, V. and Samushia, L. and Sanchez, E. and Sanders, N. and Schlegel, D. and Schubnell, M. and Seo, H. and Shafieloo, A. and Sharples, R. and Silber, J. and Sinigaglia, F. and Sprayberry, D. and Tan, T. and Tarl\'e, G. and Taylor, P. and Turner, W. and Ure\~na-L\'opez, L. A. and Vaisakh, R. and Valdes, F. and Valogiannis, G. and Vargas-Maga\~na, M. and Verde, L. and Walther, M. and Weaver, B. A. and Weinberg, D. H. and White, M. and Wolfson, M. and Y\`eche, C. and Yu, J. and Zaborowski, E. A. and Zarrouk, P. and Zhai, Z. and Zhang, H. and Zhao, C. and Zhao, G. B. and Zhou, R. and Zou, H.},
  collaboration = {DESI},
  journal = {Phys. Rev. D},
  volume = {112},
  issue = {8},
  pages = {083515},
  numpages = {40},
  year = {2025},
  month = {Oct},
  publisher = {American Physical Society},
  doi = {10.1103/tr6y-kpc6},
  url = {https://link.aps.org/doi/10.1103/tr6y-kpc6},
  archivePrefix = {arXiv},
  eprint = {2503.14738},
  primaryClass = {astro-ph.CO}
}

@ARTICLE{DESI.DR2.BAO.lya,
       author = {{DESI Collaboration} and {Abdul-Karim}, M  and {Aguilar}, J. and {Ahlen}, S. and {Allende Prieto}, C. and {Alves}, O. and {Anand}, A. and {Andrade}, U. and {Armengaud}, E. and {Aviles}, A. and {Bailey}, S. and {Bault}, A. and {BenZvi}, S. and {Bianchi}, D. and {Blake}, C. and {Brodzeller}, A. and {Brooks}, D. and {Buckley-Geer}, E. and {Burtin}, E. and {Calderon}, R. and {Canning}, R. and {Carnero Rosell}, A. and {Carrilho}, P. and {Casas}, L. and {Castander}, F.~J. and {Cereskaite}, R. and {Charles}, M. and {Chaussidon}, E. and {Chaves-Montero}, J. and {Chebat}, D. and {Claybaugh}, T. and {Cole}, S. and {Cooper}, A.~P. and {Cuceu}, A. and {Dawson}, K.~S. and {de Belsunce}, R. and {de la Macorra}, A. and {de Mattia}, A. and {Deiosso}, N. and {Della Costa}, J. and {Dey}, A. and {Dey}, B. and {Ding}, Z. and {Doel}, P. and {Edelstein}, J. and {Eisenstein}, D.~J. and {Elbers}, W. and {Fagrelius}, P. and {Fanning}, K. and {Ferraro}, S. and {Font-Ribera}, A. and {Forero-Romero}, J.~E. and {Garcia-Quintero}, C. and {Garrison}, L.~H. and {Gazta{\~n}aga}, E. and {Gil-Mar{\'\i}n}, H. and {Gontcho}, S. Gontcho A and {Gonzalez-Morales}, A.~X. and {Gordon}, C. and {Green}, D. and {Gutierrez}, G. and {Guy}, J. and {Hahn}, C. and {Herbold}, M. and {Herrera-Alcantar}, H.~K. and {Ho}, M. and {Honscheid}, K. and {Howlett}, C. and {Huterer}, D. and {Ishak}, M. and {Juneau}, S. and {Kara{\c{c}}ayl{\i}}, N.~G. and {Kehoe}, R. and {Kent}, S. and {Kirkby}, D. and {Kisner}, T. and {Kitaura}, F. -S. and {Koposov}, S.~E. and {Kremin}, A. and {Lahav}, O. and {Lamman}, C. and {Landriau}, M. and {Lang}, D. and {Lasker}, J. and {Le Goff}, J.~M. and {Le Guillou}, L. and {Leauthaud}, A. and {Levi}, M.~E. and {Li}, Q. and {Li}, T.~S. and {Lodha}, K. and {Lokken}, M. and {Magneville}, C. and {Manera}, M. and {Martini}, P. and {Matthewson}, W. and {McDonald}, P. and {Meisner}, A. and {Mena-Fern{\'a}ndez}, J. and {Miquel}, R. and {Moustakas}, J. and {Mu{\~n}oz-Guti{\'e}rrez}, A. and {Mu{\~n}oz-Santos}, D. and {Myers}, A.~D. and {Newman}, J.~A. and {Niz}, G. and {Noriega}, H.~E. and {Paillas}, E. and {Palanque-Delabrouille}, N. and {Pan}, J. and {Percival}, W.~J. and {P{\'e}rez-R{\`a}fols}, I. and {Pieri}, M.~M. and {Poppett}, C. and {Prada}, F. and {Rabinowitz}, D. and {Raichoor}, A. and {Ram{\'\i}rez-P{\'e}rez}, C. and {Rashkovetskyi}, M. and {Ravoux}, C. and {Rich}, J. and {Rockosi}, C. and {Ross}, A.~J. and {Rossi}, G. and {Ruhlmann-Kleider}, V. and {Sanchez}, E. and {Sanders}, N. and {Satyavolu}, S. and {Schlegel}, D. and {Schubnell}, M. and {Seo}, H. and {Shafieloo}, A. and {Sharples}, R. and {Silber}, J. and {Sinigaglia}, F. and {Sprayberry}, D. and {Tan}, T. and {Tarl{\'e}}, G. and {Taylor}, P. and {Turner}, W. and {Valdes}, F. and {Vargas-Maga{\~n}a}, M. and {Walther}, M. and {Weaver}, B.~A. and {Wolfson}, M. and {Y{\`e}che}, C. and {Zarrouk}, P. and {Zhou}, R. and {Zou}, H.},
        title = "{DESI DR2 Results I: Baryon Acoustic Oscillations from the Lyman Alpha Forest}",
      journal = {arXiv e-prints},
         year = 2025,
        month = mar,
          eid = {arXiv:2503.14739},
        pages = {arXiv:2503.14739},
          doi = {10.48550/arXiv.2503.14739},
archivePrefix = {arXiv},
       eprint = {2503.14739},
 primaryClass = {astro-ph.CO},
       adsurl = {https://ui.adsabs.harvard.edu/abs/2025arXiv250314739D}
}

@ARTICLE{Y3.clust-s1.Andrade.2025,
       author = {{Andrade}, U. and {Paillas}, E. and {Mena-Fern{\'a}ndez}, J. and {Li}, Q. and {Ross}, A.~J. and {Nadathur}, S. and {Rashkovetskyi}, M. and {P{\'e}rez-Fern{\'a}ndez}, A. and {Seo}, H. and {Sanders}, N. and {Alves}, O. and {Chen}, X. and {Deiosso}, N. and {Abdul-Karim}, M  and {Ahlen}, S. and {Armengaud}, E. and {Aviles}, A. and {Bianchi}, D. and {Brieden}, S. and {Brodzeller}, A. and {Brooks}, D. and {Burtin}, E. and {Calderon}, R. and {Canning}, R. and {Carnero Rosell}, A. and {Casas}, L. and {Castander}, F.~J. and {Charles}, M. and {Chaussidon}, E. and {Chaves-Montero}, J. and {Claybaugh}, T. and {Cole}, S. and {Cuceu}, A. and {Dawson}, K.~S. and {de la Macorra}, A. and {de Mattia}, A. and {Della Costa}, J. and {Dey}, A. and {Dey}, B. and {Ding}, Z. and {Doel}, P. and {Eisenstein}, D.~J. and {Elbers}, W. and {Fern{\'a}ndez-Garc{\'\i}a}, E. and {Ferraro}, S. and {Font-Ribera}, A. and {Forero-Romero}, J.~E. and {Garcia-Quintero}, C. and {Garrison}, L.~H. and {Gazta{\~n}aga}, E. and {Gil-Mar{\'\i}n}, H. and {Gontcho}, S. Gontcho A and {Gonzalez-Morales}, A.~X. and {Gordon}, C. and {Gutierrez}, G. and {Guy}, J. and {Hahn}, C. and {He}, S. and {Herrera-Alcantar}, H.~K. and {Honscheid}, K. and {Howlett}, C. and {Huterer}, D. and {Ishak}, M. and {Juneau}, S. and {Kehoe}, R. and {Kirkby}, D. and {Kisner}, T. and {Kremin}, A. and {Lahav}, O. and {Lamman}, C. and {Landriau}, M. and {Le Guillou}, L. and {Leauthaud}, A. and {Levi}, M.~E. and {Magneville}, C. and {Manera}, M. and {Martini}, P. and {Matthewson}, W. and {Meisner}, A. and {Miquel}, R. and {Moustakas}, J. and {Mu{\~n}oz-Guti{\'e}rrez}, A. and {Mu{\~n}oz-Santos}, D. and {Myers}, A.~D. and {Napolitano}, L. and {Newman}, J.~A. and {Noriega}, H.~E. and {Palanque-Delabrouille}, N. and {Pan}, J. and {Percival}, W.~J. and {P{\'e}rez-R{\`a}fols}, I. and {Poppett}, C. and {Prada}, F. and {Raichoor}, A. and {Ram{\'\i}rez-P{\'e}rez}, C. and {Ravoux}, C. and {Rossi}, G. and {Ruggeri}, R. and {Samushia}, L. and {Sanchez}, E. and {Schlegel}, D. and {Schubnell}, M. and {Sinigaglia}, F. and {Sprayberry}, D. and {Tan}, T. and {Tarl{\'e}}, G. and {Taylor}, P. and {Turner}, W. and {Vaisakh}, R. and {Vargas-Maga{\~n}a}, M. and {Walther}, M. and {Weaver}, B.~A. and {White}, M. and {Wolfson}, M. and {Yu}, J. and {Y{\`e}che}, C. and {Zarrouk}, P. and {Zhou}, R. and {Zou}, H.},
        title = "{Validation of the DESI DR2 Measurements of Baryon Acoustic Oscillations from Galaxies and Quasars}",
      journal = {arXiv e-prints},
         year = 2025,
        month = mar,
          eid = {arXiv:2503.14742},
        pages = {arXiv:2503.14742},
          doi = {10.48550/arXiv.2503.14742},
archivePrefix = {arXiv},
       eprint = {2503.14742},
 primaryClass = {astro-ph.CO},
       adsurl = {https://ui.adsabs.harvard.edu/abs/2025arXiv250314742D}
}

@ARTICLE{Planck-2018-overview,
       author = {{Planck Collaboration} and {Aghanim}, N. and {Akrami}, Y. and {Arroja}, F. and {Ashdown}, M. and {Aumont}, J. and {Baccigalupi}, C. and {Ballardini}, M. and {Banday}, A.~J. and {Barreiro}, R.~B. and {Bartolo}, N. and {Basak}, S. and {Battye}, R. and {Benabed}, K. and {Bernard}, J. -P. and {Bersanelli}, M. and {Bielewicz}, P. and {Bock}, J.~J. and {Bond}, J.~R. and {Borrill}, J. and {Bouchet}, F.~R. and {Boulanger}, F. and {Bucher}, M. and {Burigana}, C. and {Butler}, R.~C. and {Calabrese}, E. and {Cardoso}, J. -F. and {Carron}, J. and {Casaponsa}, B. and {Challinor}, A. and {Chiang}, H.~C. and {Colombo}, L.~P.~L. and {Combet}, C. and {Contreras}, D. and {Crill}, B.~P. and {Cuttaia}, F. and {de Bernardis}, P. and {de Zotti}, G. and {Delabrouille}, J. and {Delouis}, J. -M. and {D{\'e}sert}, F. -X. and {Di Valentino}, E. and {Dickinson}, C. and {Diego}, J.~M. and {Donzelli}, S. and {Dor{\'e}}, O. and {Douspis}, M. and {Ducout}, A. and {Dupac}, X. and {Efstathiou}, G. and {Elsner}, F. and {En{\ss}lin}, T.~A. and {Eriksen}, H.~K. and {Falgarone}, E. and {Fantaye}, Y. and {Fergusson}, J. and {Fernandez-Cobos}, R. and {Finelli}, F. and {Forastieri}, F. and {Frailis}, M. and {Franceschi}, E. and {Frolov}, A. and {Galeotta}, S. and {Galli}, S. and {Ganga}, K. and {G{\'e}nova-Santos}, R.~T. and {Gerbino}, M. and {Ghosh}, T. and {Gonz{\'a}lez-Nuevo}, J. and {G{\'o}rski}, K.~M. and {Gratton}, S. and {Gruppuso}, A. and {Gudmundsson}, J.~E. and {Hamann}, J. and {Handley}, W. and {Hansen}, F.~K. and {Helou}, G. and {Herranz}, D. and {Hildebrandt}, S.~R. and {Hivon}, E. and {Huang}, Z. and {Jaffe}, A.~H. and {Jones}, W.~C. and {Karakci}, A. and {Keih{\"a}nen}, E. and {Keskitalo}, R. and {Kiiveri}, K. and {Kim}, J. and {Kisner}, T.~S. and {Knox}, L. and {Krachmalnicoff}, N. and {Kunz}, M. and {Kurki-Suonio}, H. and {Lagache}, G. and {Lamarre}, J. -M. and {Langer}, M. and {Lasenby}, A. and {Lattanzi}, M. and {Lawrence}, C.~R. and {Le Jeune}, M. and {Leahy}, J.~P. and {Lesgourgues}, J. and {Levrier}, F. and {Lewis}, A. and {Liguori}, M. and {Lilje}, P.~B. and {Lilley}, M. and {Lindholm}, V. and {L{\'o}pez-Caniego}, M. and {Lubin}, P.~M. and {Ma}, Y. -Z. and {Mac{\'\i}as-P{\'e}rez}, J.~F. and {Maggio}, G. and {Maino}, D. and {Mandolesi}, N. and {Mangilli}, A. and {Marcos-Caballero}, A. and {Maris}, M. and {Martin}, P.~G. and {Martinelli}, M. and {Mart{\'\i}nez-Gonz{\'a}lez}, E. and {Matarrese}, S. and {Mauri}, N. and {McEwen}, J.~D. and {Meerburg}, P.~D. and {Meinhold}, P.~R. and {Melchiorri}, A. and {Mennella}, A. and {Migliaccio}, M. and {Millea}, M. and {Mitra}, S. and {Miville-Desch{\^e}nes}, M. -A. and {Molinari}, D. and {Moneti}, A. and {Montier}, L. and {Morgante}, G. and {Moss}, A. and {Mottet}, S. and {M{\"u}nchmeyer}, M. and {Natoli}, P. and {N{\o}rgaard-Nielsen}, H.~U. and {Oxborrow}, C.~A. and {Pagano}, L. and {Paoletti}, D. and {Partridge}, B. and {Patanchon}, G. and {Pearson}, T.~J. and {Peel}, M. and {Peiris}, H.~V. and {Perrotta}, F. and {Pettorino}, V. and {Piacentini}, F. and {Polastri}, L. and {Polenta}, G. and {Puget}, J. -L. and {Rachen}, J.~P. and {Reinecke}, M. and {Remazeilles}, M. and {Renault}, C. and {Renzi}, A. and {Rocha}, G. and {Rosset}, C. and {Roudier}, G. and {Rubi{\~n}o-Mart{\'\i}n}, J.~A. and {Ruiz-Granados}, B. and {Salvati}, L. and {Sandri}, M. and {Savelainen}, M. and {Scott}, D. and {Shellard}, E.~P.~S. and {Shiraishi}, M. and {Sirignano}, C. and {Sirri}, G. and {Spencer}, L.~D. and {Sunyaev}, R. and {Suur-Uski}, A. -S. and {Tauber}, J.~A. and {Tavagnacco}, D. and {Tenti}, M. and {Terenzi}, L. and {Toffolatti}, L. and {Tomasi}, M. and {Trombetti}, T. and {Valiviita}, J. and {Van Tent}, B. and {Vibert}, L. and {Vielva}, P. and {Villa}, F. and {Vittorio}, N. and {Wandelt}, B.~D. and {Wehus}, I.~K. and {White}, M. and {White}, S.~D.~M. and {Zacchei}, A. and {Zonca}, A.},
        title = "{Planck 2018 results. I. Overview and the cosmological legacy of Planck}",
      journal = {\aap},
         year = 2020,
        month = sep,
       volume = {641},
          eid = {A1},
        pages = {A1},
          doi = {10.1051/0004-6361/201833880},
archivePrefix = {arXiv},
       eprint = {1807.06205},
 primaryClass = {astro-ph.CO},
       adsurl = {https://ui.adsabs.harvard.edu/abs/2020A&A...641A...1P}
}

@ARTICLE{Planck-2018-likelihoods,
       author = {{Planck Collaboration} and {Aghanim}, N. and {Akrami}, Y. and {Ashdown}, M. and {Aumont}, J. and {Baccigalupi}, C. and {Ballardini}, M. and {Banday}, A.~J. and {Barreiro}, R.~B. and {Bartolo}, N. and {Basak}, S. and {Benabed}, K. and {Bernard}, J. -P. and {Bersanelli}, M. and {Bielewicz}, P. and {Bock}, J.~J. and {Bond}, J.~R. and {Borrill}, J. and {Bouchet}, F.~R. and {Boulanger}, F. and {Bucher}, M. and {Burigana}, C. and {Butler}, R.~C. and {Calabrese}, E. and {Cardoso}, J. -F. and {Carron}, J. and {Casaponsa}, B. and {Challinor}, A. and {Chiang}, H.~C. and {Colombo}, L.~P.~L. and {Combet}, C. and {Crill}, B.~P. and {Cuttaia}, F. and {de Bernardis}, P. and {de Rosa}, A. and {de Zotti}, G. and {Delabrouille}, J. and {Delouis}, J. -M. and {Di Valentino}, E. and {Diego}, J.~M. and {Dor{\'e}}, O. and {Douspis}, M. and {Ducout}, A. and {Dupac}, X. and {Dusini}, S. and {Efstathiou}, G. and {Elsner}, F. and {En{\ss}lin}, T.~A. and {Eriksen}, H.~K. and {Fantaye}, Y. and {Fernandez-Cobos}, R. and {Finelli}, F. and {Frailis}, M. and {Fraisse}, A.~A. and {Franceschi}, E. and {Frolov}, A. and {Galeotta}, S. and {Galli}, S. and {Ganga}, K. and {G{\'e}nova-Santos}, R.~T. and {Gerbino}, M. and {Ghosh}, T. and {Giraud-H{\'e}raud}, Y. and {Gonz{\'a}lez-Nuevo}, J. and {G{\'o}rski}, K.~M. and {Gratton}, S. and {Gruppuso}, A. and {Gudmundsson}, J.~E. and {Hamann}, J. and {Handley}, W. and {Hansen}, F.~K. and {Herranz}, D. and {Hivon}, E. and {Huang}, Z. and {Jaffe}, A.~H. and {Jones}, W.~C. and {Keih{\"a}nen}, E. and {Keskitalo}, R. and {Kiiveri}, K. and {Kim}, J. and {Kisner}, T.~S. and {Krachmalnicoff}, N. and {Kunz}, M. and {Kurki-Suonio}, H. and {Lagache}, G. and {Lamarre}, J. -M. and {Lasenby}, A. and {Lattanzi}, M. and {Lawrence}, C.~R. and {Le Jeune}, M. and {Levrier}, F. and {Lewis}, A. and {Liguori}, M. and {Lilje}, P.~B. and {Lilley}, M. and {Lindholm}, V. and {L{\'o}pez-Caniego}, M. and {Lubin}, P.~M. and {Ma}, Y. -Z. and {Mac{\'\i}as-P{\'e}rez}, J.~F. and {Maggio}, G. and {Maino}, D. and {Mandolesi}, N. and {Mangilli}, A. and {Marcos-Caballero}, A. and {Maris}, M. and {Martin}, P.~G. and {Mart{\'\i}nez-Gonz{\'a}lez}, E. and {Matarrese}, S. and {Mauri}, N. and {McEwen}, J.~D. and {Meinhold}, P.~R. and {Melchiorri}, A. and {Mennella}, A. and {Migliaccio}, M. and {Millea}, M. and {Miville-Desch{\^e}nes}, M. -A. and {Molinari}, D. and {Moneti}, A. and {Montier}, L. and {Morgante}, G. and {Moss}, A. and {Natoli}, P. and {N{\o}rgaard-Nielsen}, H.~U. and {Pagano}, L. and {Paoletti}, D. and {Partridge}, B. and {Patanchon}, G. and {Peiris}, H.~V. and {Perrotta}, F. and {Pettorino}, V. and {Piacentini}, F. and {Polenta}, G. and {Puget}, J. -L. and {Rachen}, J.~P. and {Reinecke}, M. and {Remazeilles}, M. and {Renzi}, A. and {Rocha}, G. and {Rosset}, C. and {Roudier}, G. and {Rubi{\~n}o-Mart{\'\i}n}, J.~A. and {Ruiz-Granados}, B. and {Salvati}, L. and {Sandri}, M. and {Savelainen}, M. and {Scott}, D. and {Shellard}, E.~P.~S. and {Sirignano}, C. and {Sirri}, G. and {Spencer}, L.~D. and {Sunyaev}, R. and {Suur-Uski}, A. -S. and {Tauber}, J.~A. and {Tavagnacco}, D. and {Tenti}, M. and {Toffolatti}, L. and {Tomasi}, M. and {Trombetti}, T. and {Valiviita}, J. and {Van Tent}, B. and {Vielva}, P. and {Villa}, F. and {Vittorio}, N. and {Wandelt}, B.~D. and {Wehus}, I.~K. and {Zacchei}, A. and {Zonca}, A.},
        title = "{Planck 2018 results. V. CMB power spectra and likelihoods}",
      journal = {\aap},
         year = 2020,
        month = sep,
       volume = {641},
          eid = {A5},
        pages = {A5},
          doi = {10.1051/0004-6361/201936386},
archivePrefix = {arXiv},
       eprint = {1907.12875},
 primaryClass = {astro-ph.CO},
       adsurl = {https://ui.adsabs.harvard.edu/abs/2020A&A...641A...5P}
}

@ARTICLE{Torrado:2021,
       author = {{Torrado}, Jes{\'u}s and {Lewis}, Antony},
        title = "{Cobaya: code for Bayesian analysis of hierarchical physical models}",
      journal = {\jcap},
         year = 2021,
        month = may,
       volume = {2021},
       number = {5},
          eid = {057},
        pages = {057},
          doi = {10.1088/1475-7516/2021/05/057},
archivePrefix = {arXiv},
       eprint = {2005.05290},
 primaryClass = {astro-ph.IM},
       adsurl = {https://ui.adsabs.harvard.edu/abs/2021JCAP...05..057T}
}

@MISC{Torrado:2019,
       author = {{Torrado}, Jes{\'u}s and {Lewis}, Antony},
        title = "{Cobaya: Bayesian analysis in cosmology}",
 howpublished = {Astrophysics Source Code Library, record ascl:1910.019},
         year = 2019,
        month = oct,
          eid = {ascl:1910.019},
        pages = {ascl:1910.019},
archivePrefix = {ascl},
       eprint = {1910.019},
       adsurl = {https://ui.adsabs.harvard.edu/abs/2019ascl.soft10019T}
}

@ARTICLE{LewisCAMB:2000,
       author = {{Lewis}, Antony and {Challinor}, Anthony and {Lasenby}, Anthony},
        title = "{Efficient Computation of Cosmic Microwave Background Anisotropies in Closed Friedmann-Robertson-Walker Models}",
      journal = {\apj},
         year = 2000,
        month = aug,
       volume = {538},
       number = {2},
        pages = {473-476},
          doi = {10.1086/309179},
archivePrefix = {arXiv},
       eprint = {astro-ph/9911177},
 primaryClass = {astro-ph},
       adsurl = {https://ui.adsabs.harvard.edu/abs/2000ApJ...538..473L}
}

@article{Lewis:2019xzd,
       author = {{Lewis}, Antony},
        title = "{GetDist: a Python package for analysing Monte Carlo samples}",
      journal = {arXiv e-prints},
         year = 2019,
        month = oct,
          eid = {arXiv:1910.13970},
        pages = {arXiv:1910.13970},
          doi = {10.48550/arXiv.1910.13970},
archivePrefix = {arXiv},
       eprint = {1910.13970},
 primaryClass = {astro-ph.IM},
       adsurl = {https://ui.adsabs.harvard.edu/abs/2019arXiv191013970L}
}

@ARTICLE{Schoeneberg:2024,
       author = {{Sch{\"o}neberg}, Nils},
        title = "{The 2024 BBN baryon abundance update}",
      journal = {arXiv e-prints},
         year = 2024,
        month = jan,
          eid = {arXiv:2401.15054},
        pages = {arXiv:2401.15054},
archivePrefix = {arXiv},
       eprint = {2401.15054},
 primaryClass = {astro-ph.CO},
       adsurl = {https://ui.adsabs.harvard.edu/abs/2024arXiv240115054S}
}

@ARTICLE{Madhavacheril:ACT-DR6,
       author = {{Madhavacheril}, Mathew S. and {Qu}, Frank J. and {Sherwin}, Blake D. and {MacCrann}, Niall and {Li}, Yaqiong and {Abril-Cabezas}, Irene and {Ade}, Peter A.~R. and {Aiola}, Simone and {Alford}, Tommy and {Amiri}, Mandana and {Amodeo}, Stefania and {An}, Rui and {Atkins}, Zachary and {Austermann}, Jason E. and {Battaglia}, Nicholas and {Battistelli}, Elia Stefano and {Beall}, James A. and {Bean}, Rachel and {Beringue}, Benjamin and {Bhandarkar}, Tanay and {Biermann}, Emily and {Bolliet}, Boris and {Bond}, J. Richard and {Cai}, Hongbo and {Calabrese}, Erminia and {Calafut}, Victoria and {Capalbo}, Valentina and {Carrero}, Felipe and {Challinor}, Anthony and {Chesmore}, Grace E. and {Cho}, Hsiao-mei and {Choi}, Steve K. and {Clark}, Susan E. and {C{\'o}rdova Rosado}, Rodrigo and {Cothard}, Nicholas F. and {Coughlin}, Kevin and {Coulton}, William and {Crowley}, Kevin T. and {Dalal}, Roohi and {Darwish}, Omar and {Devlin}, Mark J. and {Dicker}, Simon and {Doze}, Peter and {Duell}, Cody J. and {Duff}, Shannon M. and {Duivenvoorden}, Adriaan J. and {Dunkley}, Jo and {D{\"u}nner}, Rolando and {Fanfani}, Valentina and {Fankhanel}, Max and {Farren}, Gerrit and {Ferraro}, Simone and {Freundt}, Rodrigo and {Fuzia}, Brittany and {Gallardo}, Patricio A. and {Garrido}, Xavier and {Givans}, Jahmour and {Gluscevic}, Vera and {Golec}, Joseph E. and {Guan}, Yilun and {Hall}, Kirsten R. and {Halpern}, Mark and {Han}, Dongwon and {Harrison}, Ian and {Hasselfield}, Matthew and {Healy}, Erin and {Henderson}, Shawn and {Hensley}, Brandon and {Herv{\'\i}as-Caimapo}, Carlos and {Hill}, J. Colin and {Hilton}, Gene C. and {Hilton}, Matt and {Hincks}, Adam D. and {Hlo{\v{z}}ek}, Ren{\'e}e and {Ho}, Shuay-Pwu Patty and {Huber}, Zachary B. and {Hubmayr}, Johannes and {Huffenberger}, Kevin M. and {Hughes}, John P. and {Irwin}, Kent and {Isopi}, Giovanni and {Jense}, Hidde T. and {Keller}, Ben and {Kim}, Joshua and {Knowles}, Kenda and {Koopman}, Brian J. and {Kosowsky}, Arthur and {Kramer}, Darby and {Kusiak}, Aleksandra and {La Posta}, Adrien and {Lague}, Alex and {Lakey}, Victoria and {Lee}, Eunseong and {Li}, Zack and {Limon}, Michele and {Lokken}, Martine and {Louis}, Thibaut and {Lungu}, Marius and {MacInnis}, Amanda and {Maldonado}, Diego and {Maldonado}, Felipe and {Mallaby-Kay}, Maya and {Marques}, Gabriela A. and {McMahon}, Jeff and {Mehta}, Yogesh and {Menanteau}, Felipe and {Moodley}, Kavilan and {Morris}, Thomas W. and {Mroczkowski}, Tony and {Naess}, Sigurd and {Namikawa}, Toshiya and {Nati}, Federico and {Newburgh}, Laura and {Nicola}, Andrina and {Niemack}, Michael D. and {Nolta}, Michael R. and {Orlowski-Scherer}, John and {Page}, Lyman A. and {Pandey}, Shivam and {Partridge}, Bruce and {Prince}, Heather and {Puddu}, Roberto and {Radiconi}, Federico and {Robertson}, Naomi and {Rojas}, Felipe and {Sakuma}, Tai and {Salatino}, Maria and {Schaan}, Emmanuel and {Schmitt}, Benjamin L. and {Sehgal}, Neelima and {Shaikh}, Shabbir and {Sierra}, Carlos and {Sievers}, Jon and {Sif{\'o}n}, Crist{\'o}bal and {Simon}, Sara and {Sonka}, Rita and {Spergel}, David N. and {Staggs}, Suzanne T. and {Storer}, Emilie and {Switzer}, Eric R. and {Tampier}, Niklas and {Thornton}, Robert and {Trac}, Hy and {Treu}, Jesse and {Tucker}, Carole and {Ullom}, Joel and {Vale}, Leila R. and {Van Engelen}, Alexander and {Van Lanen}, Jeff and {van Marrewijk}, Joshiwa and {Vargas}, Cristian and {Vavagiakis}, Eve M. and {Wagoner}, Kasey and {Wang}, Yuhan and {Wenzl}, Lukas and {Wollack}, Edward J. and {Xu}, Zhilei and {Zago}, Fernando and {Zheng}, Kaiwen},
        title = "{The Atacama Cosmology Telescope: DR6 Gravitational Lensing Map and Cosmological Parameters}",
      journal = {\apj},
         year = 2024,
        month = feb,
       volume = {962},
       number = {2},
          eid = {113},
        pages = {113},
          doi = {10.3847/1538-4357/acff5f},
archivePrefix = {arXiv},
       eprint = {2304.05203},
 primaryClass = {astro-ph.CO},
       adsurl = {https://ui.adsabs.harvard.edu/abs/2024ApJ...962..113M}
}

@ARTICLE{2007ApJ...664..675E,
       author = {{Eisenstein}, Daniel J. and {Seo}, Hee-Jong and {Sirko}, Edwin and {Spergel}, David N.},
        title = "{Improving Cosmological Distance Measurements by Reconstruction of the Baryon Acoustic Peak}",
      journal = {\apj},
         year = 2007,
        month = aug,
       volume = {664},
       number = {2},
        pages = {675-679},
          doi = {10.1086/518712},
archivePrefix = {arXiv},
       eprint = {astro-ph/0604362},
 primaryClass = {astro-ph},
       adsurl = {https://ui.adsabs.harvard.edu/abs/2007ApJ...664..675E}
}

@ARTICLE{AbacusSummit,
       author = {{Maksimova}, Nina A. and {Garrison}, Lehman H. and {Eisenstein}, Daniel J. and {Hadzhiyska}, Boryana and {Bose}, Sownak and {Satterthwaite}, Thomas P.},
        title = "{ABACUSSUMMIT: a massive set of high-accuracy, high-resolution N-body simulations}",
      journal = {\mnras},
         year = 2021,
        month = dec,
       volume = {508},
       number = {3},
        pages = {4017-4037},
          doi = {10.1093/mnras/stab2484},
archivePrefix = {arXiv},
       eprint = {2110.11398},
 primaryClass = {astro-ph.CO},
       adsurl = {https://ui.adsabs.harvard.edu/abs/2021MNRAS.508.4017M}
}

@ARTICLE{Eisenstein2007:astro-ph/0604362v1,
       author = {{Eisenstein}, Daniel J. and {Seo}, Hee-Jong and {Sirko}, Edwin and
         {Spergel}, David N.},
        title = "{Improving Cosmological Distance Measurements by Reconstruction of the Baryon Acoustic Peak}",
      journal = {\apj},
         year = "2007",
        month = "Aug",
       volume = {664},
       number = {2},
        pages = {675-679},
          doi = {10.1086/518712},
archivePrefix = {arXiv},
       eprint = {astro-ph/0604362},
 primaryClass = {astro-ph},
       adsurl = {https://ui.adsabs.harvard.edu/abs/2007ApJ...664..675E}
}

@ARTICLE{Landy1993,
       author = {{Landy}, Stephen D. and {Szalay}, Alexander S.},
        title = "{Bias and Variance of Angular Correlation Functions}",
      journal = {\apj},
         year = 1993,
        month = jul,
       volume = {412},
        pages = {64},
          doi = {10.1086/172900},
       adsurl = {https://ui.adsabs.harvard.edu/abs/1993ApJ...412...64L}
}

@ARTICLE{rascal,
       author = {{O'Connell}, Ross and {Eisenstein}, Daniel and {Vargas}, Mariana and
         {Ho}, Shirley and {Padmanabhan}, Nikhil},
        title = "{Large covariance matrices: smooth models from the two-point correlation function}",
      journal = {\mnras},
         year = "2016",
        month = "Nov",
       volume = {462},
       number = {3},
        pages = {2681-2694},
          doi = {10.1093/mnras/stw1821},
archivePrefix = {arXiv},
       eprint = {1510.01740},
 primaryClass = {astro-ph.CO},
       adsurl = {https://ui.adsabs.harvard.edu/abs/2016MNRAS.462.2681O}
}

@ARTICLE{rascalC,
       author = {{Philcox}, Oliver H.~E. and {Eisenstein}, Daniel J. and {O'Connell}, Ross and {Wiegand}, Alexander},
        title = "{RASCALC: a jackknife approach to estimating single- and multitracer galaxy covariance matrices}",
      journal = {\mnras},
         year = 2020,
        month = jan,
       volume = {491},
       number = {3},
        pages = {3290-3317},
          doi = {10.1093/mnras/stz3218},
archivePrefix = {arXiv},
       eprint = {1904.11070},
 primaryClass = {astro-ph.CO},
       adsurl = {https://ui.adsabs.harvard.edu/abs/2020MNRAS.491.3290P}
}

@software{pycorr,
       author = {{de Mattia}, Arnaud and {Rashkovetskyi}, Michael and {Sinha}, Manodeep and {Garrison}, Lehman H.},
        title = "{pycorr: Two-point correlation function estimation}",
 howpublished = {Astrophysics Source Code Library, record ascl:2403.009},
         year = 2024,
        month = mar,
          eid = {ascl:2403.009},
       adsurl = {https://ui.adsabs.harvard.edu/abs/2024ascl.soft03009D}
}

@article{EDR_HOD_LRGQSO2023,
  author        = {{Yuan}, Sihan and {Zhang}, Hanyu and {Ross}, Ashley J. and {Donald-McCann}, Jamie and {Hadzhiyska}, Boryana and {Wechsler}, Risa H. and {Zheng}, Zheng and {Alam}, Shadab and {Gonzalez-Perez}, Violeta and {Aguilar}, Jessica Nicole and {Ahlen}, Steven and {Bianchi}, Davide and {Brooks}, David and {de la Macorra}, Axel and {Fanning}, Kevin and {Forero-Romero}, Jaime E. and {Honscheid}, Klaus and {Ishak}, Mustapha and {Kehoe}, Robert and {Lasker}, James and {Landriau}, Martin and {Manera}, Marc and {Martini}, Paul and {Meisner}, Aaron and {Miquel}, Ramon and {Moustakas}, John and {Nadathur}, Seshadri and {Newman}, Jeffrey A. and {Nie}, Jundan and {Percival}, Will and {Poppett}, Claire and {Rocher}, Antoine and {Rossi}, Graziano and {Sanchez}, Eusebio and {Samushia}, Lado and {Schubnell}, Michael and {Seo}, Hee-Jong and {Tarle}, Gregory and {Weaver}, Benjamin Alan and {Yu}, Jiaxi and {Zhou}, Zhimin and {Zou}, Hu},
  title         = "{The DESI One-Percent Survey: Exploring the Halo Occupation Distribution of Luminous Red Galaxies and Quasi-Stellar Objects with AbacusSummit}",
  journal       = {arXiv e-prints},
  year          = 2023,
  month         = jun,
  eid           = {arXiv:2306.06314},
  pages         = {arXiv:2306.06314},
  doi           = {10.48550/arXiv.2306.06314},
  archiveprefix = {arXiv},
  eprint        = {2306.06314},
  primaryclass  = {astro-ph.CO},
  adsurl        = {https://ui.adsabs.harvard.edu/abs/2023arXiv230606314Y}
}

@article{EDR_HOD_ELG2023,
  author        = {{Rocher}, Antoine and {Ruhlmann-Kleider}, Vanina and {Burtin}, Etienne and {Yuan}, Sihan and {de Mattia}, Arnaud and {Ross}, Ashley J. and {Aguilar}, Jessica and {Ahlen}, Steven and {Alam}, Shadab and {Bianchi}, Davide and {Brooks}, David and {Cole}, Shaun and {Dawson}, Kyle and {de la Macorra}, Axel and {Doel}, Peter and {Eisenstein}, Daniel J. and {Fanning}, Kevin and {Forero-Romero}, Jaime E. and {Garrison}, Lehman H. and {Gontcho A Gontcho}, Satya and {Gonzalez-Perez}, Violeta and {Guy}, Julien and {Hadzhiyska}, Boryana and {Hahn}, ChangHoon and {Honscheid}, Klaus and {Kisner}, Theodore and {Landriau}, Martin and {Lasker}, James and {E. Levi}, Michael and {Manera}, Marc and {Meisner}, Aaron and {Miquel}, Ramon and {Moustakas}, John and {Mueller}, Eva-Maria and {Newman}, Jeffrey A. and {Nie}, Jundan and {Percival}, Will J. and {Poppett}, Claire and {Qin}, Fei and {Rossi}, Graziano and {Samushia}, Lado and {Sanchez}, Eusebio and {Schlegel}, David and {Schubnell}, Michael and {Seo}, Hee-Jong and {Tarl{\'e}}, Gregory and {Vargas-Maga{\~n}a}, Mariana and {Weaver}, Benjamin A. and {Yu}, Jiaxi and {Zhang}, Hanyu and {Zheng}, Zheng and {Zhou}, Zhimin and {Zou}, Hu},
  title         = "{The DESI One-Percent survey: exploring the Halo Occupation Distribution of Emission Line Galaxies with ABACUSSUMMIT simulations}",
  journal       = {\jcap},
  year          = 2023,
  month         = oct,
  volume        = {2023},
  number        = {10},
  eid           = {016},
  pages         = {016},
  doi           = {10.1088/1475-7516/2023/10/016},
  archiveprefix = {arXiv},
  eprint        = {2306.06319},
  primaryclass  = {astro-ph.CO},
  adsurl        = {https://ui.adsabs.harvard.edu/abs/2023JCAP...10..016R}
}

@ARTICLE{FKP1994,
       author = {{Feldman}, Hume A. and {Kaiser}, Nick and {Peacock}, John A.},
        title = "{Power-Spectrum Analysis of Three-dimensional Redshift Surveys}",
      journal = {\apj},
         year = 1994,
        month = may,
       volume = {426},
        pages = {23},
          doi = {10.1086/174036},
archivePrefix = {arXiv},
       eprint = {astro-ph/9304022},
 primaryClass = {astro-ph},
       adsurl = {https://ui.adsabs.harvard.edu/abs/1994ApJ...426...23F}
}

@article{CHANIOTIS2004253,
title = {High order interpolation and differentiation using B-splines},
journal = {Journal of Computational Physics},
volume = {197},
number = {1},
pages = {253-274},
year = {2004},
issn = {0021-9991},
doi = {https://doi.org/10.1016/j.jcp.2003.11.026},
url = {https://www.sciencedirect.com/science/article/pii/S0021999103006296},
author = {A.K. Chaniotis and D. Poulikakos},
}

@article{Dalal.RelV_2010,
doi = {10.1088/1475-7516/2010/11/007},
url = {https://doi.org/10.1088/1475-7516/2010/11/007},
year = {2010},
month = {nov},
publisher = {},
volume = {2010},
number = {11},
pages = {007},
author = {Neal Dalal and Ue-Li Pen and Uros Seljak},
title = {Large-scale BAO signatures of the smallest galaxies},
journal = {Journal of Cosmology and Astroparticle Physics},
}

@ARTICLE{Beutler.RelV_2017,
       author = {{Beutler}, Florian and {Seljak}, Uro{\v{s}} and {Vlah}, Zvonimir},
        title = "{Constraining the relative velocity effect using the Baryon Oscillation Spectroscopic Survey}",
      journal = {\mnras},
         year = 2017,
        month = sep,
       volume = {470},
       number = {3},
        pages = {2723-2735},
          doi = {10.1093/mnras/stx1196},
archivePrefix = {arXiv},
       eprint = {1612.04720},
 primaryClass = {astro-ph.CO},
       adsurl = {https://ui.adsabs.harvard.edu/abs/2017MNRAS.470.2723B}
}

@ARTICLE{class-approximation-schemes,
       author = {{Blas}, Diego and {Lesgourgues}, Julien and {Tram}, Thomas},
        title = "{The Cosmic Linear Anisotropy Solving System (CLASS). Part II: Approximation schemes}",
      journal = {\jcap},
         year = 2011,
        month = jul,
       volume = {2011},
       number = {7},
          eid = {034},
        pages = {034},
          doi = {10.1088/1475-7516/2011/07/034},
archivePrefix = {arXiv},
       eprint = {1104.2933},
 primaryClass = {astro-ph.CO},
       adsurl = {https://ui.adsabs.harvard.edu/abs/2011JCAP...07..034B}
}

@software{data-zenodo,
  author       = {Nick Sanders},
  title        = {Supplemental data for: A Unified Tracer Analysis of DESI DR2 Baryon Acoustic Oscillations},
  month        = aug,
  year         = 2026,
  publisher    = {Zenodo},
  doi          = {10.5281/zenodo.22135162},
  url          = {https://doi.org/10.5281/zenodo.22135162},
}

@article{DES-Dovekie,
    author = {Popovic, B and Shah, P and Kenworthy, W D and Kessler, R and Davis, T M and Goobar, A and Scolnic, D and Vincenzi, M and Wiseman, P and Chen, R and Charleton, E and Acevedo, M and Armstrong, P and Boyd, B M and Brout, D and Camilleri, R and Frieman, J and Galbany, L and Grayling, M and Kelsey, L and Rose, B and Sánchez, B and Lee, J and Möller, A and Smith, M and Sullivan, M and Shiamtanis, N and Alarcon, A and Allam, S S and Andrade-Oliveira, F and Avila, S and Bacon, D and Blazek, J and Bocquet, S and Brooks, D and Burke, D L and Rosell, A Carnero and Carretero, J and Cawthon, R and da Costa, L N and Pereira, M E da Silva and Diehl, H T and Dodelson, S and Doel, P and Everett, S and Frohmaier, C and García-Bellido, J and Gruen, D and Gutierrez, G and Herner, K and Hinton, S R and Hollowood, D L and Honscheid, K and Huterer, D and James, D J and Jeffrey, N and Kuehn, K and Lahav, O and Lee, S and Lidman, C and Marshall, J L and Mena-Fernández, J and Menanteau, F and Miquel, R and Muir, J and Myles, J and Ogando, R L C and Paterno, M and Malagón, A A Plazas and Porredon, A and Prat, J and Nichol, R C and Romer, A K and Roodman, A and Sanchez, E and Cid, D Sanchez and Sevilla-Noarbe, I and Suchyta, E and Swanson, M E C and To, C and Tucker, D L and Walker, A R and Weaverdyck, N and Aguena, M, The DES Collaboration},
    title = {The Dark Energy Survey supernova program: a reanalysis of cosmology results and evidence for evolving dark energy with an updated Type Ia supernova calibration},
    journal = {Monthly Notices of the Royal Astronomical Society},
    volume = {548},
    number = {4},
    pages = {stag632},
    year = {2026},
    month = {06},
    issn = {0035-8711},
    doi = {10.1093/mnras/stag632},
    url = {https://doi.org/10.1093/mnras/stag632},
    eprint = {2511.07517}
}

@article{SPT-3G,
  title = {SPT-3G D1: CMB temperature and polarization power spectra and cosmology from 2019 and 2020 observations of the SPT-3G main field},
  author = {Camphuis, E. and Quan, W. and Balkenhol, L. and Khalife, A. R. and Ge, F. and Guidi, F. and Huang, N. and Lynch, G. P. and Omori, Y. and Trendafilova, C. and Anderson, A. J. and Ansarinejad, B. and Archipley, M. and Barry, P. S. and Benabed, K. and Bender, A. N. and Benson, B. A. and Bianchini, F. and Bleem, L. E. and Bouchet, F. R. and Bryant, L. and Campitiello, M. G. and Carlstrom, J. E. and Chang, C. L. and Chaubal, P. and Chichura, P. M. and Chokshi, A. and Chou, T.-L. and Coerver, A. and Crawford, T. M. and Daley, C. and de Haan, T. and Dibert, K. R. and Dobbs, M. A. and Doohan, M. and Doussot, A. and Dutcher, D. and Everett, W. and Feng, C. and Ferguson, K. R. and Fichman, K. and Foster, A. and Galli, S. and Gambrel, A. E. and Gardner, R. W. and Goeckner-Wald, N. and Gualtieri, R. and Guns, S. and Halverson, N. W. and Hivon, E. and Holder, G. P. and Holzapfel, W. L. and Hood, J. C. and Hryciuk, A. and K{\'e}ruzor{\'e}, F. and Knox, L. and Korman, M. and Kornoelje, K. and Kuo, C.-L. and Levy, K. and Lowitz, A. E. and Lu, C. and Maniyar, A. and Martsen, E. S. and Menanteau, F. and Millea, M. and Montgomery, J. and Nakato, Y. and Natoli, T. and Noble, G. I. and Ouellette, A. and Pan, Z. and Paschos, P. and Phadke, K. A. and Pollak, A. W. and Prabhu, K. and Raghunathan, S. and Rahimi, M. and Rahlin, A. and Reichardt, C. L. and Rouble, M. and Ruhl, J. E. and Schiappucci, E. and Simpson, A. and Sobrin, J. A. and Stark, A. A. and Stephen, J. and Tandoi, C. and Thorne, B. and Umilta, C. and Vieira, J. D. and Vitrier, A. and Wan, Y. and Whitehorn, N. and Wu, W. L. K. and Young, M. R. and Zebrowski, J. A.},
  collaboration = {SPT-3G Collaboration},
  journal = {Phys. Rev. D},
  volume = {113},
  issue = {8},
  pages = {083504},
  numpages = {81},
  year = {2026},
  month = {Apr},
  publisher = {American Physical Society},
  doi = {10.1103/7wt3-9v2y},
  url = {https://link.aps.org/doi/10.1103/7wt3-9v2y}
}

@article{Wilks:1938dza,
    author = "Wilks, S. S.",
    title = "{The Large-Sample Distribution of the Likelihood Ratio for Testing Composite Hypotheses}",
    doi = "10.1214/aoms/1177732360",
    journal = "Annals Math. Statist.",
    volume = "9",
    number = "1",
    pages = "60--62",
    year = "1938"
}

\appendix

\section{Broadband of \eqo} \label{sec:broad-eq1}

\begin{figure*}
    \subfigure{
        \centering
        \includegraphics[width=0.45\linewidth]{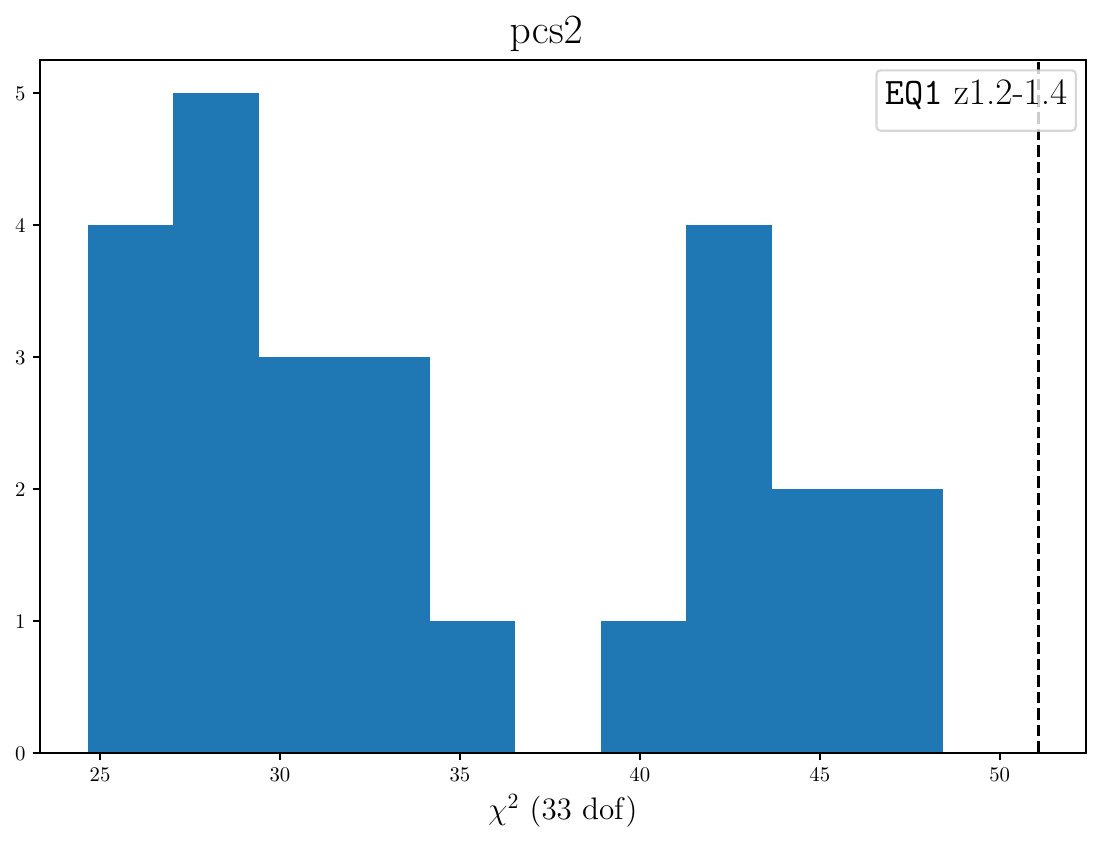}
        }
    \subfigure{
        \centering
        \includegraphics[width=0.45\linewidth]{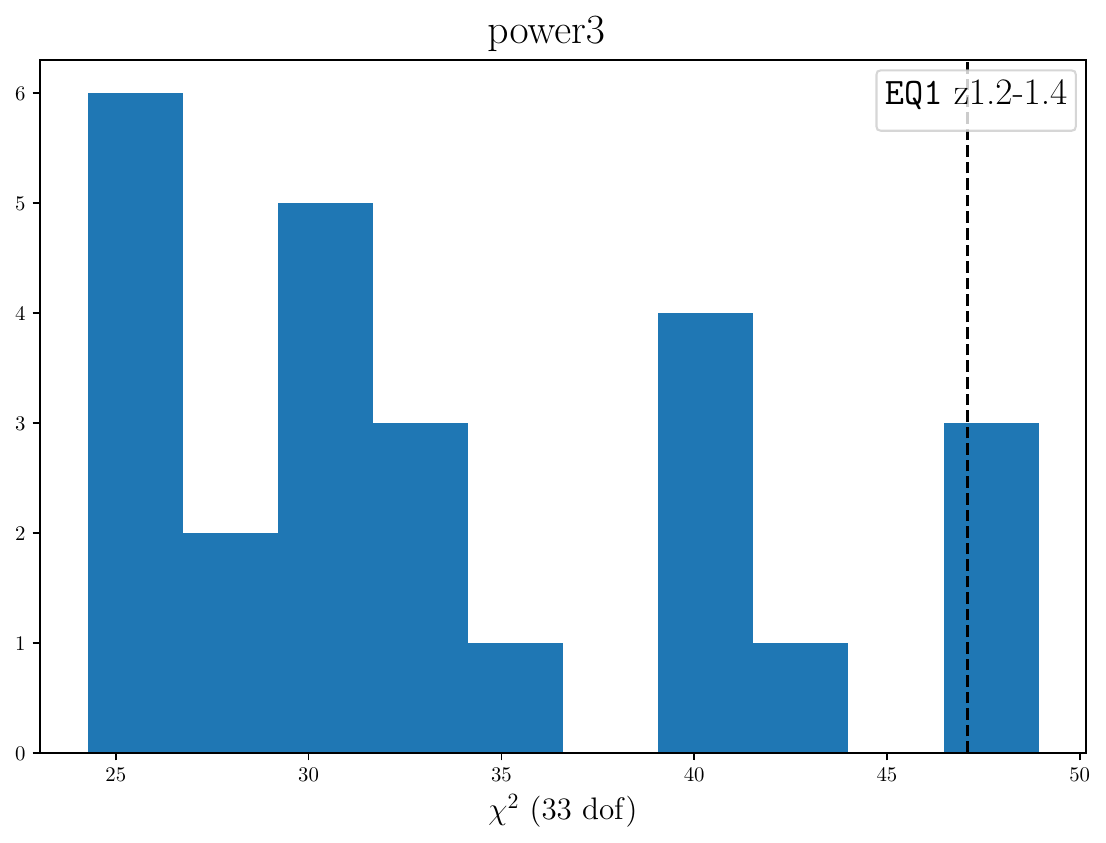}
        }
    \caption{Distributions of $\chi^2$ in 25 mocks for \eqo. The left panel corresponds to fits with the broadband term \pcs, and the right \pow. The dashed line is the $\chi^2$ in DR2 data. The data falls out of the distribution of mocks for \pcs, but within for \pow.}
    \label{fig:chi2-eq1}
\end{figure*}

\begin{figure}
    \centering
    \includegraphics[width=\linewidth]{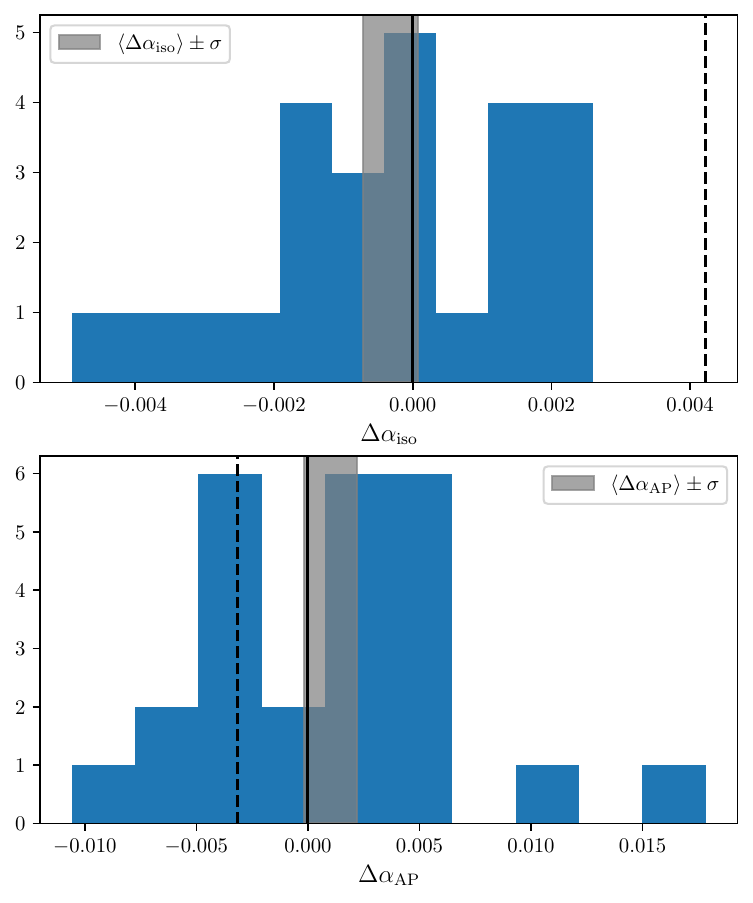}
    \caption{Differences in \eqo\ best fit $\aiso$ (top) and $\alap$ (bottom) between \pcs\ and \pow. The histogram shows 25 mocks, the grey band is the mean and $1\sigma$ dispersion (divided by $\sqrt{25}$), the dashed line is the DR2 data, and the solid line is 0. Mocks show no expectation of bias between the two choices. The standard deviation of mock differences are 0.002 and 0.006 for $\aiso$ and $\alap$ respectively, both only $\sim 15\%$ of the uncertainty in the data. Despite appearing to fall out of the distribution, the difference in data for $\aiso$ is only $\sim 2\sigma$ from 0, and so is not considered discrepant.}
    \label{fig:alpha-diff-eq1}
\end{figure}

\begin{figure}
    \centering
    \includegraphics[width=\linewidth]{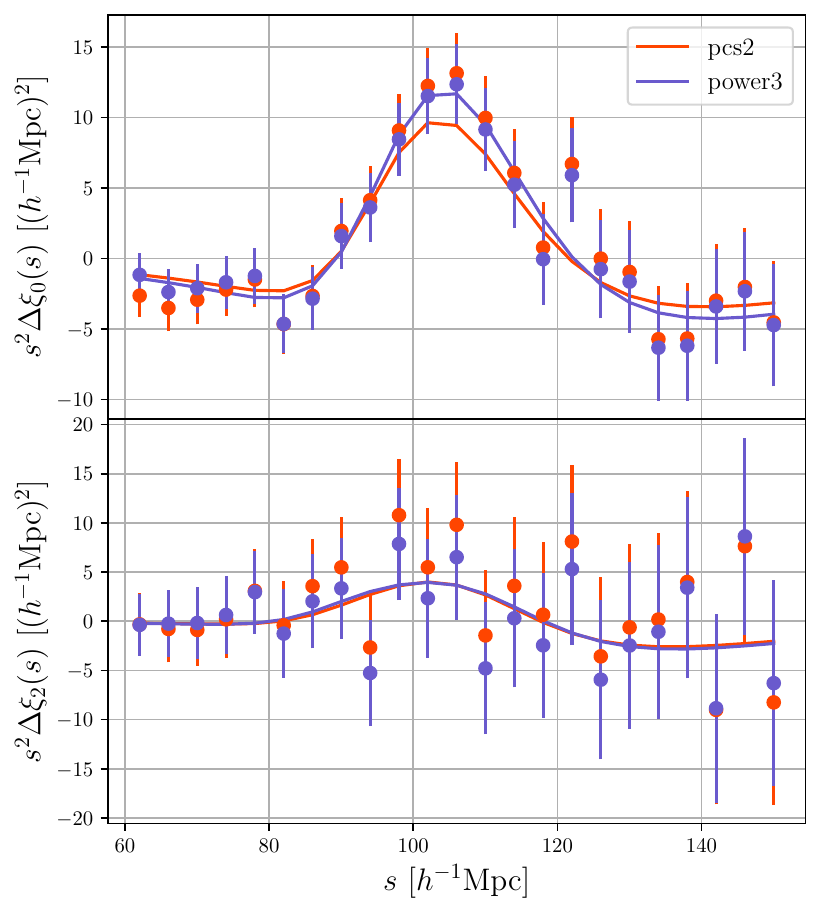}
    \caption{Best fits to the BAO feature in the monopole (top) and quadrupole (bottom) with the broadband parameterized as both \pcs\ and \pow.}
    \label{fig:bao-pcs-pow}
\end{figure}

As a validation test, we checked that the BAO fits for all bins have $\chi^2$ that fall within the distribution of 25 mocks, mimicking a test for data unblinding in the baseline analysis in \cite{Y3.clust-s1.Andrade.2025}. All unified tracer bins pass this test, with the exception of \eqo, as demonstrated in the left panel of \cref{fig:chi2-eq1}. As a check, we change the modeling of the broadband term to instead be a polynomial in separation $s$ rather than as the cubic spline in $k$ of \cref{eqn:spline-D}. We refer to the polynomial parameterization as \pow, and the cubic spline as \pcs. We find that the $\chi^2$ issue is resolved when modeling the broadband term as \pow, as shown in the right panel of \cref{fig:chi2-eq1}.

Here, we validate that there is a negligible difference between the best-fit BAO of \pcs\ and \pow. \cref{fig:alpha-diff-eq1} shows the difference between \pcs\ and \pow\ best fit $\aiso$ and $\alap$ in mocks and data. We see that the mean of the difference is consistent with zero within $1\sigma$ dispersion divided by $\sqrt{25}$, and the data falls within the symmetrized distribution of absolute differences in mocks, albeit only just. 
Only in 1 of the 25 mocks is there a more extreme result than the data for $\aiso$. Speculatively, these results suggest that there are broadband artifacts due to combining two very different samples on the clustering of the {\tt ELG+QSO} sample that are not well-captured by the default $\mathcal{D}(k)$ parameterization that warrant future investigation.
The BAO best fits with \pcs\ and \pow\ are compared in \cref{fig:bao-pcs-pow}. The \pow\ fits seem to better fit the monopole than \pcs\ in terms of $\chi2$ ($51.1/33$ versus $47.1/33$ for \pcs\ and \pow\ respectively) as well as visually, while the effect seems minimal in the quadrupole. This could be because, in configuration space, \pcs\ has one fewer broadband parameter than \pow\ for the monopole. As detailed in \citep{KP4s2-Chen}, when Hankel-transformed, the piecewise cubic spline terms for the broadband fitting are localized on smaller scales than the BAO scale except for the very low-order terms. Therefore, in configuration space, \pcs\ only adopts 2 $\mathcal{D}$-term parameters for the quadrupole for the low-order spline terms and an additional $2\times 2$ nuisance parameters for the monopole and the quadrupole to account for any potential, uncontrolled large-scale data systematics affecting $k<0.02\ihMpc$. On the other hand, \pow\ has 3 for each, meaning \pow\ has an additional parameter to better fit the monopole. Regardless, since we find no significant difference in the best-fit BAO scaling parameters, we present the \pcs\ results in the main text for consistency in the methodology, despite the failure of the $\chi^2$ validation test. 

\section{Additional Systematic Tests} \label{sec:add-systematics}

\begin{figure*}
    \centering
    \includegraphics[width=\linewidth]{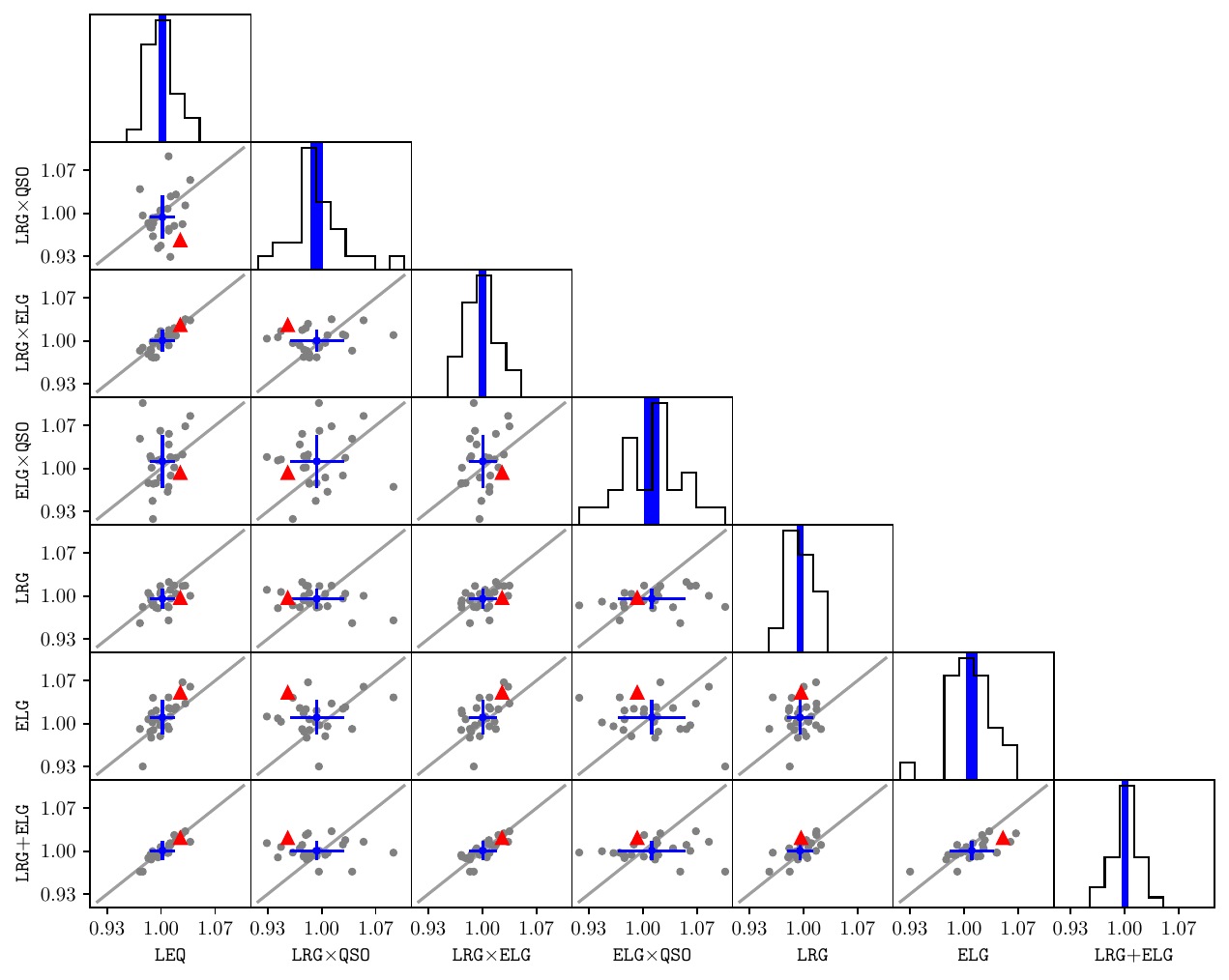}
    \caption{Testing for tracer-dependent systematics of $\alpha_{\rm AP}$ in redshift $0.8<z<1.1$, matching  $\alpha_{\rm iso}$ of \cref{fig:qiso-scatter}. No tracer-dependent systematics are detected from the mocks and from the data.}
    \label{fig:qap-scatter}
\end{figure*}

Here we present more tests of tracer-dependent systematics following the same procedure as in \cref{subsubsec:mock-systematics}. \cref{fig:qap-scatter} shows the same systematics test in the redshift bin $0.8 < z < 1.1$ on the fits to $\alap$. We find the same conclusion as for $\aiso$, that the deviations between tracers fall within the distribution of mocks, and therefore no tracer-dependent systematics are detected from the mocks and from the data.

\begin{figure*}
    \subfigure{
        \centering
        \includegraphics[width=0.45\linewidth]{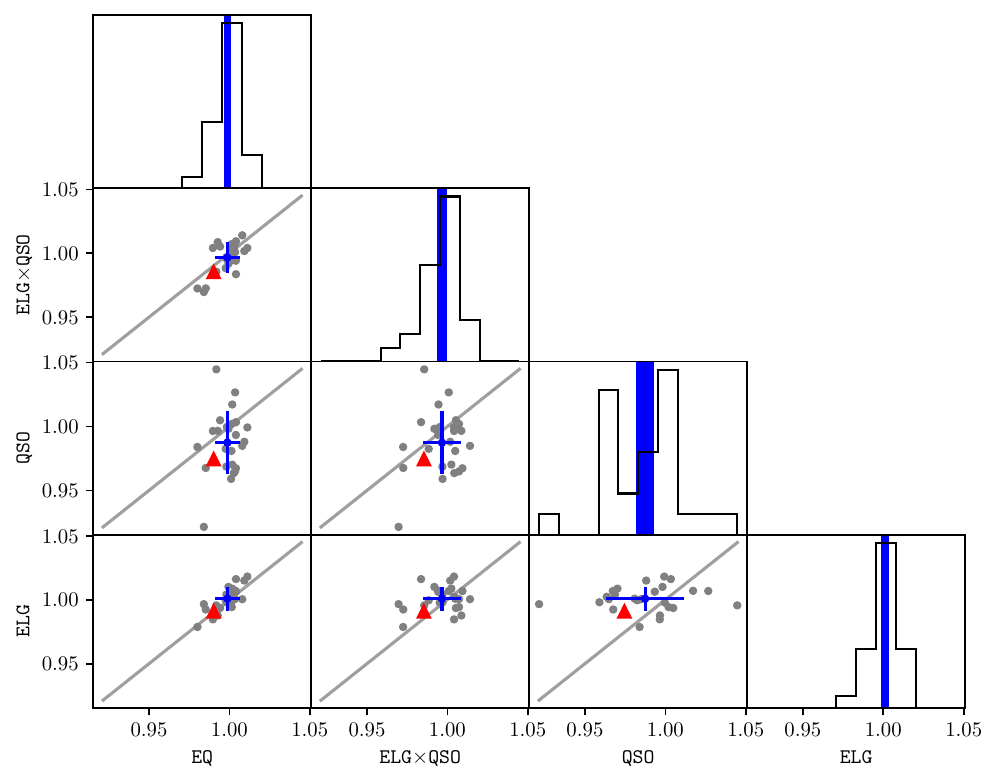}
        }
    \subfigure{
        \centering
        \includegraphics[width=0.45\linewidth]{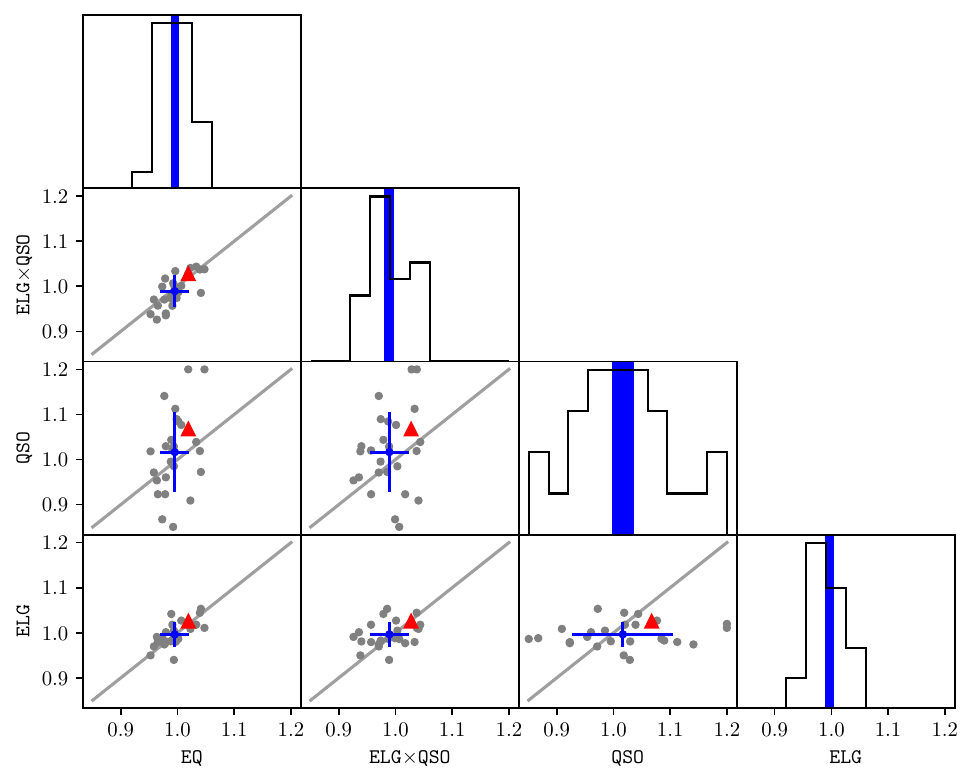}
        }
    \caption{Testing for tracer-dependent systematics of $\aiso$ (left panel) and $\alap$ (right panel) in redshift $1.1<z<1.6$. No tracer-dependent systematics are detected from the mocks and from the data.}
    \label{fig:tracer-sys-1.1-1.6}
\end{figure*}

We also repeat this test in the redshift bin $1.1 < z < 1.6$, with results shown in \cref{fig:tracer-sys-1.1-1.6}. We make the same conclusion of no detection of tracer-dependent systematics in both cases. \cref{tab:all-diffs-0.8-1.1} shows the quantitative measurements from the data, in comparison to the mocks, which supplements \cref{tab:diff-subset}.

\begin{table*}
    \centering
    \begin{tabular}{c|c|c|c|c|c|c|c}
        \hline
        \hline
        \multicolumn{7}{c}{$\Delta\alpha/\sigma_{\Delta\alpha}$} \\
        \hline
        Tracer & \tleq & \lrgxqso & \lrgxelg & \elgxqso & \lrg & \elg & \lrgelg \\
        \hline

        \tleq & --- & $0.85 \sigma$ & $0.84 \sigma$ & $1.10 \sigma$ & $2.15 \sigma$ & $0.29 \sigma$ & $0.68 \sigma$ \\
        \lrgxqso & $1.96 \sigma$ & --- & $0.59 \sigma$ & $0.56 \sigma$ & $0.30 \sigma$ & $0.62 \sigma$ & $0.70 \sigma$ \\
        \lrgxelg & $0.14 \sigma$ & $1.86 \sigma$ & --- & $0.93 \sigma$ & $0.98 \sigma$ & $0.07 \sigma$ & $0.47 \sigma$ \\
        \elgxqso & $0.75 \sigma$ & $0.69 \sigma$ & $0.73 \sigma$ & --- & $0.71 \sigma$ & $0.92 \sigma$ & $1.00 \sigma$ \\
        \lrg & $1.65 \sigma$ & $0.93 \sigma$ & $1.43 \sigma$ & $0.10 \sigma$ & --- & $0.52 \sigma$ & $1.82 \sigma$ \\
        \elg & $1.24 \sigma$ & $2.37 \sigma$ & $1.10 \sigma$ & $1.02 \sigma$ & $1.72 \sigma$ & --- & $0.12 \sigma$ \\
        \lrgelg & $0.45 \sigma$ & $1.68 \sigma$ & $0.17 \sigma$ & $0.63 \sigma$ & $1.88 \sigma$ & $1.39 \sigma$ & --- \\
        \hline
        \hline
    \end{tabular}
    \caption{Discrepancies from zero of differences in $\aiso$ (above diagonal) and $\alap$ (below diagonal) in DR2 data between different tracers in redshift bin $0.8<z<1.1$. Quantified by the standard deviation $\sigma$ of the difference measured in 25 mocks, as discussed in \cref{subsec:data-results}. No pair of tracers show a significant deviation from the null hypothesis of no tracer-dependent systematics.}
    \label{tab:all-diffs-0.8-1.1}
\end{table*}

\section{Additional Measurements} \label{sec:add-measures}

We present in \cref{tab:add-measures} additional measurements of BAO for other choices within the analysis pipeline that were not chosen in the main analysis of this work. These were used for validation tests, and are presented here for completeness.

\begin{table*}
    \centering
    \begin{tabular}{l|l|c|c|c|r|c}
        \hline\hline
        Choice & Tracer & Redshift Range & $\aiso$ & $\alap$ & \multicolumn{1}{c|}{$r$} & $\chi^2 / {\rm dof}$ \\
        \hline
        Default zbin & \tleq & $0.8<z<1.1$ & $0.9950 \pm 0.0071$ & $1.0172 \pm 0.0255$ & $-0.045$ & $33.6/33$ \\
        pre-recon  &   \eq & $1.1<z<1.6$ & $0.9866 \pm 0.0079$ & $1.0324 \pm 0.0313$ & $0.285$ & $52.4/33$ \\
        & \qsoo & $1.6<z<2.1$ & $0.9932 \pm 0.0181$ & $0.9631 \pm 0.0666$ & $-0.203$ & $24.7/33$ \\
        \hline\hline
        &  \tleqo & $0.8<z<1.0$ & $0.9932 \pm 0.0088$ & $1.0116 \pm 0.0334$ & $0.266$ & $26.0/33$ \\
        Rebinned & \tleqt & $1.0<z<1.2$ & $0.9929 \pm 0.0110$ & $0.9900 \pm 0.0377$ & $-0.260$ & $41.0/33$ \\
        pre-recon & \eqo & $1.2<z<1.4$ & $0.9773 \pm 0.0151$ & $0.9785 \pm 0.0561$ & $0.222$ & $37.6/33$ \\
        & \eqt & $1.4<z<1.6$ & $0.9885 \pm 0.0141$ & $1.1065 \pm 0.0648$ & $0.416$ & $41.2/33$ \\
        \hline\hline
        & {\tt LE1} & $0.8<z<1.0$ & $0.9889 \pm 0.0055$ & $1.0212 \pm 0.0193$ & $-0.041$ & $21.2/33$ \\
        Rebin w/o \qso &  {\tt LE2} & $1.0<z<1.2$ & $0.9896 \pm 0.0083$ & $1.0309 \pm 0.0303$ & $-0.111$ & $50.4/33$ \\ 
        & {\tt E1} & $1.2<z<1.4$ & $0.9954 \pm 0.0133$ & $0.9928 \pm 0.0435$ & $-0.079$ & $30.9/33$ \\ 
        &  {\tt E2} & $1.4<z<1.6$ & $0.9792 \pm 0.0123$ & $1.0776 \pm 0.0517$ & $0.207$ & $47.0/33$ \\
        \hline\hline
        Default $b(z)$\&$f(z)$ & \lrgelg & $0.8<z<1.1$ & $0.9876 \pm 0.0046$ & $1.0289 \pm 0.0158$ & $-0.090$ & $ 34.0/33$ \\
        \hline\hline
    \end{tabular}
    \caption{Additional BAO measurements provided for reference. Include measurements of both binnings before reconstruction, the rebinned analysis not including \qso, and the baseline \lrgelg\ with the redshift-dependent bias and growth rate reconstruction.}
    \label{tab:add-measures}
\end{table*}

\section{Optimal Weighting} \label{sec:alt-weights}

\begin{figure}
    \centering
    \includegraphics[width=\linewidth]{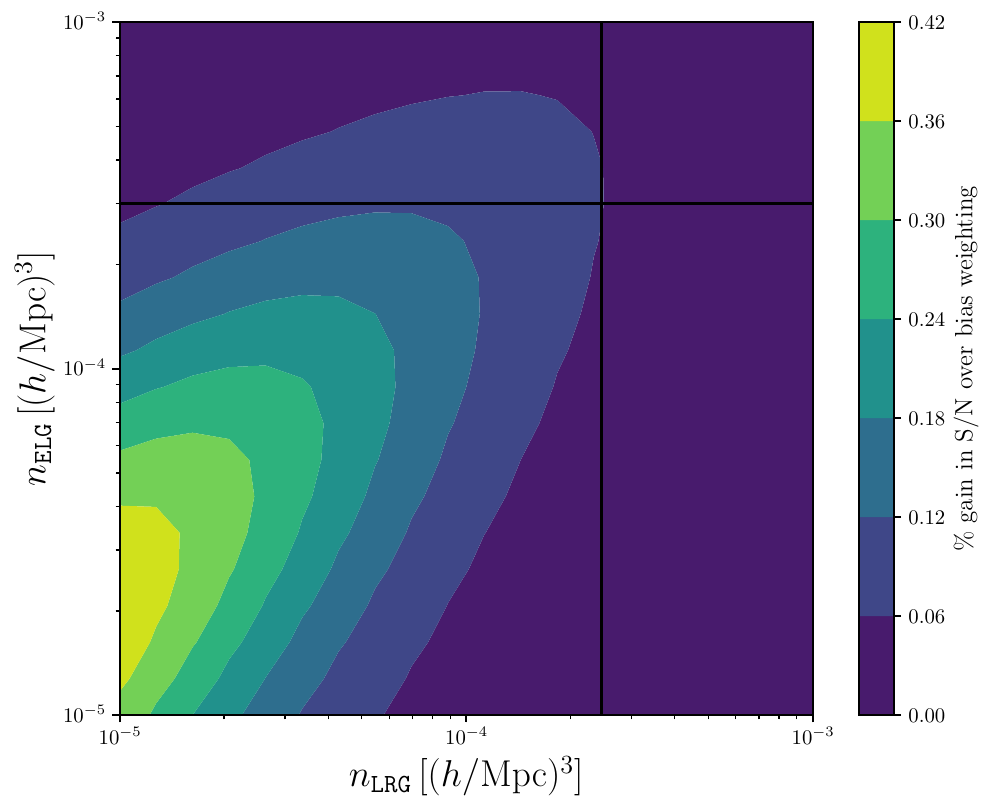}
    \caption{Expected gain in power spectrum S/N at BAO scale of full angular weighting as compared to simple bias weighting as a function of 2 tracer densities. The black lines indicate the densities of DR2 {\lrgs\ and \elgs.}}
    \label{fig:gain-contour}
\end{figure}

Weighting each catalog by the bias is only approximately optimal, where we assumed the bias matrix in the likelihood is identity (see Appendix A of \citep{KP4s5-Valcin}). Relaxing this assumption and including redshift-space distortions, the maximum-likelihood weighting is an anisotropic weight
\begin{equation} \label{eqn:optimal-weight}
    W_{\rm ML} = b(1 + \beta\mu^2).
\end{equation}
\noindent
Implementing this weighting is nontrivial, as one cannot simply reweight catalogs and concatenate them owing to the dependence on the angle between any given pair. 
In order to approximately estimate the expected gain from such a procedure, we forecast the signal-to-noise ratio of the power spectrum at the BAO scale ($k=0.14 \ihMpc$) when combining two tracers with the biases of \lrg\ and \elg:
\begin{equation}
    \left( \frac{S}{N} \right)^2 \propto \int_0^1 d\mu \left(\frac{[w_1b_1 + w_2b_2]^2}{[w_1b_1 + w_2b_2]^2 + \frac{w_1^2}{\bar{n}_1P_{\delta}} + \frac{w_1^2}{\bar{n}_2P_{\delta}}} \right).
\end{equation}
\noindent
Here, $w$ can be any arbitrary weight. We choose to compare the bias weighting to the optimal weight \cref{eqn:optimal-weight}.
\cref{fig:gain-contour} shows the expected gain in the signal-to-noise ratio of this anisotropic weighting as compared to bias weighting as a function of the two tracer densities. We find that at DESI DR2 densities, we expect a gain on the order of a tenth of a percent. Additionally, there are diminishing returns as the densities increase, so future data releases stand to gain even less. We conclude that this marginal gain is not worth the effort required to implement this weighting.

\section{Validating the single-tracer treatment for the combined tracer covariance matrices}
\label{sec:cov-as-multi}

\subsection{Motivation and theory}
It is very important to estimate the shot noise contributions to the covariance correctly, because the tracer combination pipeline optimizes the shot noise contribution to the variance (covariance) of clustering statistics.
Indeed, the derivation in Appendix A of \cite{KP4s5-Valcin} gives an (approximately) optimal estimate of the underlying matter density (in our realization of the Universe) as an inverse-variance-weighted average of the matter overdensity estimates $\delta_i/b_i$, with inverse variance as Poisson shot noise (independent between tracers) enhanced by linear bias, $b_i^2 \bar n_i$.
The matter density remains subject to the volume-dependent cosmic variance, but the shot noise variance is reduced: its inverse is the sum over the tracers:
\begin{equation}
    \sum_i b_i^2 \bar n_i = b_{\rm eff} \sum_i b_i \bar n_i, \label{eqn:shot-noise-factor-correct}
\end{equation}
where
\begin{equation}
    b_{\rm eff} (z) = \frac{\sum_i b_i^2 \bar{n}_i}{\sum_i b_i \bar{n}_i} \label{eqn:beff-noz}
\end{equation}
analogously to \cref{eqn:beff}, just without redshift dependence.

The concern here is that we risk underestimating the benefit of the combined tracer if we do not capture the shot noise reduction accurately.
Therefore, let us consider in more detail how the \rascalc{} covariance framework estimates shot noise.

For a single tracer, \rascalc{} takes the empirical correlation function, enhanced by $b^2$.
The 4-point term contains an integral of products of these correlation functions, so it scales as $b^4$, and also does not depend on the number density \citep{rascalC} — as expected from the (bias-upscaled) intrinsic covariance of matter 2PCF.
Part of the 2-point term doesn't contain correlation functions, so it scales as $b^0/\bar n^2$, providing a relative variance component scaling as $1/(b^2 \bar n)^2$, which seems reasonable.
There is also another part of the 2-point term containing one correlation function and thus scaling as $b^2/\bar n^2$.
The 3-point term also contains one correlation function, and scales as $b^2/\bar n$; its relative contribution is $1/(b^2 \bar n)$ after dividing by the overall $b^4$.
The number (density) is unweighted and taken from the data.

Treating the combined tracer covariance as a single-tracer one gives
$b \rightarrow b_{\rm eff}$ (\cref{eqn:beff-noz}) as the upscaling of the 2PCF across the given redshift range, not redshift-dependent, and
$\bar n \rightarrow \sum \bar n_i$ as the sum of unweighted densities averaged over the given redshift range, again not redshift-dependent.
This gives the main shot-noise factor of
\begin{equation}
    b_{\rm eff}^2 \sum_i \bar n_i = \frac{(\sum_i b_i^2 \bar{n}_i)^2 \sum_i \bar n_i}{(\sum_i b_i \bar{n}_i)^2}, \label{eqn:shot-noise-factor-rascalc-single}
\end{equation}
which is generally different from the correct value given by \cref{eqn:shot-noise-factor-correct}.

The mismatch between \cref{eqn:shot-noise-factor-correct,eqn:shot-noise-factor-rascalc-single} may be fixed to some extent by the shot-noise rescaling, optimized to fit the jackknife covariance reference.
To remind, the shot-noise rescaling $\alpha_{\rm SN}$ rescales the 2-point term by $\alpha_{\rm SN}^2$ and the 3-point one by $\alpha_{\rm SN}$ (and the 4-point term is not scaled).
On the other hand, the shot-noise rescaling already mimics other effects like intrinsic non-Gaussian 4/3-point correlations \citep{rascal}, reconstruction artifacts (when applicable) \citep{2023MNRAS.524.3894R} and fiber assignment \citep{KP4s7-Rashkovetskyi}, so additional shot noise mismatch may become a breaking point.

Running the combined tracer covariance as a multi-tracer one (using the combined catalog weights and all the different auto- and cross-correlations measured with them) seems theoretically better to deal with shot noise, $b_i(z)$, $\bar n_i(z)$, keeping track of same- and different-tracer pairs and different auto- and cross-correlations, and giving more freedom with multiple shot-noise rescaling values (one per tracer).
On the other hand, this may lead to some loss of flexibility due to shot-noise rescaling not affecting some of the cross-covariance terms (see Appendix A of \cite{KP4s7-Rashkovetskyi}).
Moreover, the shot-noise rescaling values are currently calibrated separately for each tracer on its auto-covariance.

\subsection{Obtaining the combined tracer covariance from multi-tracer}

A multi-tracer covariance can be collapsed into a single combined tracer covariance in a way analogous to the combination of correlation functions in different regions described in Appendix B of \cite{KP4s7-Rashkovetskyi}.
Considering $s$ bins $a,b$ and $|\mu|$ bins $c,d$, all the weighted binned counts ($DD$, $DR$, $RR$) for the combined tracer are simply the sums of the weighted auto- and cross-counts (provided the weights remain the same in the combined and separated catalogs):
\begin{align*}
    (Q^{X+Y} Q^{X+Y})_a^c = & (Q^X Q^X)_a^c + (Q^X Q^Y)_a^c \\
    & + (Q^Y Q^X)_a^c + (Q^Y Q^Y)_a^c,
\end{align*}
where each $Q$ can be $D$ or $R$, data or randoms\footnote{I.e. $Q^XQ^Y$ encompasses $D^XD^Y$, $D^XR^Y$, $R^XD^Y$ and $R^XR^Y$ count arrays if standard BAO reconstruction is not used. After reconstruction, we also have $S$ -- shifted randoms, and the counts relevant to the Landy-Szalay estimator are $D^XD^Y$, $D^XS^Y$, $S^XD^Y$, $S^XS^Y$ and $R^XR^Y$.}.

Then, from the form of the Landy-Szalay estimator (\cref{eqn:LS-estimator}), it follows that the total correlation function is the average of the auto- and cross-correlations weighted by the $RR$ counts:\footnote{The counts must not be normalized before this, in contrast to the note in Appendix B.1 of \cite{KP4s7-Rashkovetskyi}, because the region combination logic is different in this aspect. However, there is a subtlety because the {\tt wnorm} values are added, they may cause artifacts if the data-to-random weighted ratio is not the same in a way that might not be possible to handle on the \rascalc{} side. Fortunately, the data-to-random weighted ratio is forced to be the same in the combination pipeline since \cite{KP4s5-Valcin}.}
\begin{align}
    (\xi^{X+Y})_a^c &= \sum_{t_1,t_2} (W^{t_1t_2})_a^c (\xi^{t_1t_2})_a^c, \\
    (W^{t_1t_2})_a^c &\equiv \frac{(R^{t_1}R^{t_2})_a^c}{\sum_{t_3,t_4} (R^{t_3}R^{t_4})_a^c}.
\end{align}
At this point, in $|\mu|$ bins, $XY$ and $YX$ are the same, so for two tracers, there is a single cross-correlation function $\xi^{XY}$ that appears twice with the same $R^X R^Y$ cross-counts.
Then the covariance matrix for the combined region is simply
\begin{equation}
    (C^{X+Y})_{ab}^{cd} = \sum_{t_1,t_2,t_3,t_4} (C^{t_1t_2,t_3t_4})_{ab}^{cd} (W^{t_1t_2})_a^c (W^{t_3t_4})_b^d.
\end{equation}

For Legendre multipoles in $s$ bins, we can follow Appendix B.2 of \cite{KP4s7-Rashkovetskyi}, approximately converting from Legendre moments to $|\mu|$ bins before the combination and converting back after it:
\begin{align}
    (\xi^{XY})^\ell_a &= \sum_c (\xi^{XY})_a^c F^\ell_c, \\
    L_\ell^c &\equiv \frac1{\Delta\mu_c} \int_{\Delta\mu_c} d\mu\, L_\ell(\mu) = \frac{F^\ell_c}{(2\ell + 1) \Delta\mu_c} \\ 
    (\xi^{XY})_a^c &\approx \sum_\ell (\xi^{XY})^\ell_a L_\ell^c. 
\end{align}
Then we can work out the partial derivatives:
\begin{equation}
    \frac{\partial (\xi^{X+Y})^\ell_a}{\partial (\xi^{t_1t_2}_G)^{\ell_1}_a} \approx \sum_c F_c^\ell (W^{t_1t_2})_a^c L^c_{\ell_1} \equiv (W_G^{t_1t_2})^{\ell,\ell_1}_a;
\end{equation}
the different separation bins stay independent.

Then the covariance matrix for the combined tracer 2PCF (Legendre moments or multipoles) is
\begin{align}
    & (C^{X+Y})_{ab}^{\ell\ell^\prime} \approx \nonumber \\
    & \sum_{t_1,t_2,t_3,t_4} \sum_{\ell_1,\ell^\prime_1} (C^{t_1 t_2,t_3 t_4})_{ab}^{\ell_1 \ell^\prime_1} (W^{t_1 t_2})_a^{\ell,\ell_1} (W^{t_3 t_4})_b^{\ell^\prime,\ell^\prime_1}.
\end{align}

\subsection{Test setup and results}

Having this machinery, we have run a test on \lrgelg\ $0.8<z<1.1$.
We separated the combined catalog back into \lrg\ and \elg\ (keeping the weights) and computed the two autocorrelation functions and the cross-correlation function.
We provided the separated catalogs and the three correlation functions to \rascalc{}.
We compare the main fit results (goodness of fit and the BAO scale parameters) with the two covariance variants in \cref{tab:cov-as-multi-lrgelg}, and find minimal differences.
Therefore, we conclude that it seems safe to treat \lrgelg\ as a single tracer for the covariance computation after all.

\begin{table}[htb!]
    \centering
    \begin{tabular}{|c|c|c|}
        \hline
         & Single tracer (default) & Two tracers \\
        \hline
        $\chi^2$/dof & 38.44 / 33 & 36.90 / 33 \\
        $\aiso$ & $0.9887 \pm 0.0045$ & $0.9887 \pm 0.0046$ \\
        $\alap$ & $1.0214 \pm 0.0152$ & $1.0215 \pm 0.0155$ \\
        \hline
    \end{tabular}
    \caption{Comparison of the fit results for \lrgelg\ $0.8<z<1.1$ (GCcomb post-recon) with different covariance treatments.
    There is a slight difference in $\chi^2$, and the $\aiso$ and $\alap$ are nearly identical.}
    \label{tab:cov-as-multi-lrgelg}
\end{table}

\begin{table}[htb!]
    \centering
    \begin{tabular}{|c|c|c|}
        \hline
         & Single tracer (default) & Two tracers \\
        \hline
        $\chi^2$/dof & 33.86 / 33 & 33.62 / 33 \\
        $\aiso$ & $0.9875 \pm 0.0045$ & $0.9874 \pm 0.0046$ \\
        $\alap$ & $1.025 \pm 0.015$ & $1.025 \pm 0.015$ \\
        \hline
    \end{tabular}
    \caption{Comparison of the fit results for \tleq\ $0.8<z<1.1$ (GCcomb post-recon) with different covariance treatments.
    \tleq\ is a combination of three tracers, but we separate \lrgelg\ and \qso\ as two because the \rascalc{} framework does not support more than 2 tracers at a time, and we have shown that \lrgelg\ behaves well as a single tracer in \cref{tab:cov-as-multi-lrgelg}.
    There is a slight difference in $\chi^2$, and the $\aiso$ and $\alap$ are nearly identical.}
    \label{tab:cov-as-multi-leq}
\end{table}

We have conducted an additional test involving \qso{}, using \tleq\ $0.8<z<1.1$.
A technical challenge is that \tleq\ is a combination of three tracers, whereas the \rascalc{} implementation does not support more than two at the same time\footnote{A 3-tracer covariance computation also may become much longer, as it will probably need 27 independent (4-point term) computations (instead of 7 for 2 tracers and 1 for 1).}.
To overcome it, we treated \lrgelg\ as a single tracer in this test, having shown that it is adequate in the previous one.
Then, we provided the two separated catalogs and the three correlation functions computed from them to \rascalc{}.
We compare the fits with the two \tleq\ covariance options in \cref{tab:cov-as-multi-leq}, and find even smaller changes.
The conclusion is that the single-tracer covariance treatment for the combined tracer is sufficiently precise, at least for BAO analyses.

\end{document}